\documentclass[12pt,journal,onecolumn]{IEEEtran}
\usepackage[hang]{subfigure}
\usepackage{epsfig,rotating,setspace,latexsym,amsmath,epsf,amssymb,amsfonts,bm,theorem,epstopdf}
\usepackage[normalem]{ulem}
\usepackage{cite,authblk}
\usepackage{algorithm}
\usepackage[noend]{algpseudocode}
\usepackage{setspace}
\usepackage{color}
\usepackage{graphicx}
\usepackage{mathrsfs}
\ifCLASSOPTIONcompsoc
\usepackage[caption=false, font=normalsize, labelfont=sf, textfont=sf]{subfig}
\else
\usepackage[caption=false, font=footnotesize]{subfig}
\fi

\newtheorem{theorem}{Theorem}
\newtheorem{remark}{Remark}
\newtheorem{lemma}{Lemma}
\newtheorem{corollary}{Corollary}
\newtheorem{example}{Example}
\newtheorem{definition}{Definition}
\newtheorem{prop}{{Proposition}}
\newenvironment{Proof}[1]{\medskip\par\noindent{\bf Proof:\,}\,#1}{{\mbox{\,$\blacksquare$}\par}}

\allowdisplaybreaks

\definecolor{green1}{rgb}{0.2,0.7,0.2}
\definecolor{brown}{rgb}{1,0.5,0.2}
\DeclareMathOperator*{\argmin}{\arg\!\min}

\newcommand{\rknote}[1]{\noindent{\leavevmode\color{red}[Raj -- #1]}}

\usepackage[colorinlistoftodos,textwidth=\marginparwidth]{todonotes}
\usepackage{comment}
\begin{document}

\title{Value of Information in Games with Multiple Strategic Information Providers\thanks{Research of the authors was supported in part by the ARO MURI Grant AG285 and in part by ARL-DCIST Grant AK868.}}
\author{Raj Kiriti Velicheti \qquad  Melih Bastopcu \qquad  Tamer Ba{\c s}ar\\
\normalsize Coordinated Science Laboratory\\
	\normalsize University of Illinois Urbana-Champaign, Urbana, IL, 61801\\
	\normalsize  \texttt{{\small (rkv4,bastopcu,basar1)@illinois.edu}}}

\maketitle
\begin{abstract}
In the classical communication setting multiple senders having access to the same source of information and transmitting it over channel(s) to a receiver in general leads to a decrease in estimation error at the receiver as compared with the single sender case. However, if the objectives of the information providers are different from that of the estimator, this might result in interesting strategic interactions and outcomes. In this work, we consider a hierarchical signaling game between multiple senders (information designers) and a single receiver (decision maker) each having their own, possibly misaligned, objectives. The senders lead the game by committing to individual information disclosure policies simultaneously, within the framework of a non-cooperative Nash game among themselves. This is followed by the receiver's action decision.
With Gaussian information structure and quadratic objectives (which depend on underlying state and receiver's action) for all the players, we show that in general the equilibrium is not unique. We hence identify a set of equilibria and further show that linear noiseless policies can achieve a minimal element of this set. 
Additionally, we show that competition among the senders is beneficial to the receiver, as compared with cooperation among the senders. Further, we extend our analysis to a dynamic signaling game of finite horizon with Markovian information evolution. We show that linear memoryless policies can achieve equilibrium in this dynamic game. We also consider an extension to a game with multiple receivers having coupled objectives.
We provide algorithms to compute the equilibrium strategies in all these cases. Finally, via extensive simulations, we analyze the effects of multiple senders in varying degrees of alignment among their objectives.
\end{abstract}
\newpage
\section{Introduction}\label{Sect:Intro}


With increasing ease of communication and emergence of remote estimation, questions about the incentives of the data sources take the center stage. Consider, for example, the community notes feature rolled out recently by Twitter to combat misinformation. This feature lets volunteered users to post a \textit{note} on a Tweet that might possibly be incorrect. However, since this may lead to strategic reporting, Twitter further aggregates data from a group of volunteers with diverse views. This brings us to the following two questions: How diverse should the views be? Is there a systematic way to understand the amount of information revealed given a prior on possible personal biases that people can have? As another example, consider a robot learning to perform a task based on human demonstrations. In order to perform the task efficiently a large number of human demonstrations need to be collected but since humans might have their own biases, this causes a possible mismatch in objectives\cite{kostrikov2018discriminator}. 
Learning the underlying task amidst these mismatched (possibly strategic) objectives is still a problem of active interest\cite{arora2021survey}. Such an estimation with strategic sources occurs in multiple situations including estimation from strategic sensors in cyber-physical systems, Federated Learning, and information theory \cite{bhagoji2019analyzing,kashyap2007quantized,milgrom1986relying}. We aim to 
model these interactions as a communication game and analyze the resulting outcome.

A model of strategic communication was introduced in the seminal work \cite{crawford1982strategic} which considers a game between a sender and a receiver with simultaneous strategy commitment. Different from the traditional communication setting, their work considers a sender and a receiver with mismatched objectives and noise-free channels, which makes the problem a communication game. Under few regularity assumptions, they have shown that all Nash Equilibria are quantization based mappings.
 Following \cite{crawford1982strategic}, several extensions to the model have been considered in the literature such as multiple senders\cite{ambrus2008multi,mcgee2013cheap}, adding of a strategic mediation channel\cite{ivanov2010communication}, considering dynamic information disclosure\cite{ivanov2015dynamic}, and multidimensional settings\cite{saritacs2016quadratic}, all under Nash equilibrium.

An alternate line of work pioneered in \cite{kamenica2011bayesian,rayo2010optimal} (termed as Bayesian Persuasion), however, considers instead a hierarchical signaling game where the receiver responds to a strategy committed by the sender. This commitment power adds benefit to the sender, and hence can be used to characterize the optimal utility that a sender can derive in such a communication game. The key idea to solve for the optimal sender strategy in such a game is to pose an equivalent optimization problem in the space of posteriors using the \textit{splitting lemma} \cite{aumann1995repeated}. Due to the natural hierarchical structure occurring in many scenarios, (such as the ones described earlier), the last decade has seen a surge of interest in this model from various disciplines including control theory\cite{farokhi2016estimation,saritacs2016quadratic,saritacs2020dynamic}, machine learning \cite{chen2022selling}, information theory \cite{akyol2016information}, and mechanism design\cite{bergemann2019information}. 
A follow-up work \cite{dughmi2016algorithmic} identifies that the optimal sender strategy can be computed by solving a linear program. Extensions to the dynamic case with discrete states evolving as a Markov chain are studied in \cite{ely2017beeps,renault2017optimal}. \cite{kamenica2019bayesian} provides a more detailed review regarding recent works along this line. Additionally, outside the communication context, games with mismatched objectives between multiple principals (players who are similar to senders in this work) and a social optimizer are considered broadly under the name of common agent games in \cite{fickinger2020multi,martimort2002revelation}. 

In spirit, our work adds on to the growing literature of multi-sender hierarchical communication games. References \cite{gentzkow2017bayesian,gentzkow2016competition} solve for Nash equilibrium (NE) among senders, assuming access to arbitrarily correlated signaling policies, by utilizing concavification. They show that there can be more than one equilibrium with full revelation by all senders always being an equilibrium. Such intuitive results also extend to our setting although proving them requires different mathematical tools which we will formalize in the following sections. Studies in \cite{li2021sequential,wu2021sequential} consider a slightly different game where senders can commit sequentially and prove the uniqueness of equilibrium in such a game. On the other hand, works \cite{wang2013bayesian,arieli2019private,babichenko2016computational} analyze a game consisting of multiple receivers and investigate the effects of public and private signaling but confine attention to a single sender. All these works mentioned above consider finite state and action spaces where state is a one-dimensional variable.

In contrast, in this work we consider a more general infinite multidimensional state and action spaces while restricting our attention to a Gaussian prior and quadratic utilities. Multidimensional signaling games require a significantly different set of tools than those described previously. In one of the initial works in this direction, \cite{tamura2018bayesian} utilizes semi-definite programming (SDP) to solve for a lower bound on sender's cost\footnote{In line with the control literature, most of the works discussed here consider the scenario where each player wants to minimize their individual cost function rather than maximize their own utility function.} and prove that for Gaussian information structures, linear policies can be used to achieve this lower bound. Reference \cite{sayin2019hierarchical} extends this by considering a dynamic sender receiver game and shows that equilibrium posteriors follow a recursive structure which can still be achieved by using linear policies. \cite{9222530} shows that noisy linear policies are sufficient to identify an SDP which is equivalent to sender's optimization problem. Reference \cite{sayin2021bayesian} goes beyond Gaussian distribution and utilizes techniques from co-positive programming to identify lower bounds for arbitrary prior distributions when players have quadratic objectives. Finally, \cite{tamura2018bayesian} mentions a game with multiple senders as an extension to their earlier work, and  presents a necessary condition for existence of a partially revealing equilibrium for the underlying game. However, \cite{tamura2018bayesian} does not characterize equilibrium posteriors nor do they specify Nash equilibrium strategies for the senders. Although \cite{farokhi2016estimation} considers the multi-sender communication game, it restricts attention to symmetric sender objectives while allowing for private realization of state for each sender and solving for a herding equilibrium. 


Coming to the specifics of this paper, our contributions are multi-fold. Firstly, we pose a strategic communication game between multiple senders and a single receiver. We then solve for a hierarchical equilibrium where all senders commit to an information disclosure strategy\footnote{In this work, we use the terms strategy and policy interchangeably.} simultaneously (and thus, play a Nash game among themselves) followed by the receiver taking a decision (and as a result, play a Stackelberg game between senders and the receiver)\cite{bacsar1998dynamic}. Due to the  presence of multiple senders, each sender faces an equilibrium computation problem which can be significantly different from the optimization problem in a single sender game discussed in \cite{tamura2018bayesian,sayin2019hierarchical}.In particular, each sender needs to compute a strategy accounting for the presence of other senders with possibly misaligned utilities. Further, as will be illustrated in the following sections, the game under consideration is a general sum game but is not convex in each players action. 
To ease computation of the equilibrium, we pose the problem in space of posterior covariance, propose a notion of stable posterior and design an algorithm to identify such posteriors. We show that the equilibrium might not be unique and identify necessary conditions that equilibrium policies should satisfy. Additionally, we propose an equilibrium refinement, termed as partially informative equilibrium, to identify equilibria which are robust against more informative deviation by all the senders. Although noiseless signaling policies restrict the space of possible posteriors\cite{tamura2018bayesian}, we show that these equilibria can be achieved by linear noiseless signaling policies. For ease of exposition, we start by considering a game with two senders and a single receiver and then extend it to a game with more than two senders. Next, we formulate a dynamic game and extend the result of \cite{sayin2019hierarchical} to the multiple-sender setting. We show that greedy equilibrium computation for a single stage game results in posteriors that are stable for the dynamic game, and thus we identify a recursive structure for equilibrium posteriors. Further, similar to the single-stage case, we also show that these equilibrium posteriors can be achieved by linear noiseless signaling policies. We also provide extension to a game with multiple receivers with possibly coupled objectives. Finally, we provide extensive numerical results to demonstrate the effect of competition in information revelation. A key takeaway from our work, apart from the technical tools developed, is that in line with \cite{gentzkow2017bayesian}, even in multidimensional infinite state spaces with quadratic cost\footnote{In this work we use the terms objective and cost interchangeably.} functions, competition among multiple senders is beneficial to the receiver. However, unlike in \cite{gentzkow2017bayesian}, we do not require arbitrary correlation between senders' signaling policies, and thus overcome limitations of the work in \cite{gentzkow2017bayesian}.

The paper is organized as follows: In Section~\ref{Sect:system_model}, we introduce our system model with a single-stage and formulate the strategic information disclosure problem for an $m$-sender and a single receiver game. In Section~\ref{Sect:Strategic_com}, we first consider a sub-class with 2 senders and identify the equilibrium policies and prove the optimality of linear policies. Then, we extend this result to a more general case with $m>2$ senders. In Section~\ref{Sect:Dynamic}, we formulate the multi-stage dynamic strategic information disclosure problem with multiple senders. Here, we show that applying the same solution technique developed for the single-stage problem greedily to the multi-stage game achieves the equilibrium, which can also be obtained by using linear signaling strategies. In Section~\ref{Sec: Multireceiver}, we extend our results to a game with multiple receivers. In Section~\ref{Sect:Num_result}, we provide extensive numerical results to understand the effect of multiple senders and the achieved equilibrium policies in various scenarios. Section~\ref{Sect:Discussion} provides insights on the commitment structure assumed in this work and discusses other possible equilibrium definitions for multi-sender games. Finally, we conclude the paper with some major highlights of our work and potential future directions in Section~\ref{Sect:Conclusion}.

\subsubsection*{Notations}
We denote by $[m]$ the set $\{1,\dots,m\}$. $\mathbb{N}(0,\Sigma)$ denotes a Gaussian distribution with mean 0 and co-variance $\Sigma$ where $\Sigma$ is a positive semi-definite (psd) matrix. For a psd matrix $S$, we use $S^{\frac{1}{2}}$ to denote its unique psd square root. We use $\mathcal{N}(.)$ to denote the null set of a matrix. $\bar{\mathcal{N}}$ denotes a set of possible ordering. $Tr(.)$ denotes the trace of a matrix. We denote vectors with bold lower symbols. For a given matrix $A$ and a vector $\boldsymbol{y}$, $A^\top$ and $\boldsymbol{y}^\top$ denote the transposes of that matrix and the vector, respectively. We use $B^\dagger$ to denote the pseudo-inverse of matrix $B$. The identity and zero matrices are denoted by $I$ and $O$, respectively. For two symmetric matrices $A$ and $B$, $A\succ B$ ($A\succeq B$) means that $A-B$ is a positive (semi-)definite matrix. For a set $\boldsymbol{X}$, we use $\mathcal{B}(\boldsymbol{X})$ to denote the Borel set of $\boldsymbol{X}$.
\begin{figure}[t]
	\centerline{\includegraphics[width=0.65\columnwidth]{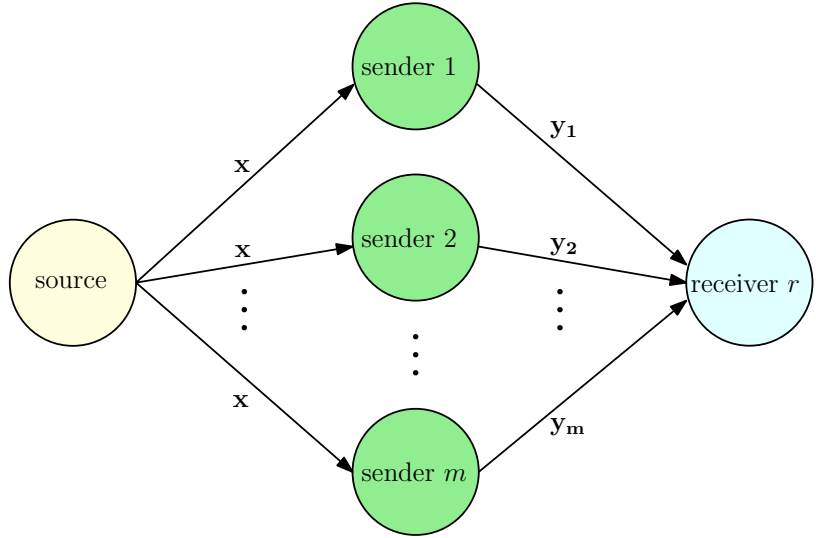}}
	\caption{The strategic communication system consisting of $m$ senders denoted by $i =1,\dots,m$ and a receiver $r$. Each sender $i$ observes the state of the world $\boldsymbol{x}$ and designs its signaling policy so as to minimize its cost $J_{i}$ while the receiver estimates the state of the world based on the signals that it receives to minimize its cost $J_r$. }
	\vspace{-0.35cm}
	\label{Fig:system_model}
	\vspace{-0.3cm}
\end{figure}

\section{System Model and Problem Formulation}\label{Sect:system_model}
Consider a non-cooperative game among  a group of $m+1$ players, comprised of $m$ senders denoted by $i =1,\dots,m$, and a receiver denoted by $r$. The senders have access to a random state variable $\boldsymbol{x}\in \boldsymbol{X}\subseteq  \mathbb{R}^p$, which is sampled from a zero-mean Gaussian distribution with positive-definite covariance $\Sigma_x$, i.e., $ \boldsymbol{x}\sim \mathbb{N}(0,\Sigma_x)$ with $\Sigma_x\succ O$. After observing the state realization $ \boldsymbol{x}$, 
the senders send signals $\{\boldsymbol{y}_i\}_{i\in[m]}$ to the receiver. After observing these signals, 
 the receiver forms an estimate of $\boldsymbol{x}$  and takes an action $\boldsymbol{u}$. Thus, let the action of sender $i$ be $\boldsymbol{y}_i\in\boldsymbol{Y}_i$ and that of the receiver be $\boldsymbol{u}\in\boldsymbol{U}$, where  $\boldsymbol{Y}_i\subseteq \mathbb{R}^p$ and  $\mathbf{U}\subseteq \mathbb{R}^t$ are the action spaces of sender $i$ and the receiver, respectively. 
 
 All players have their own individual loss functions $L_i(\boldsymbol{x};\boldsymbol{u})$ for $i=1,\hdots,m,r$, mapping from $\boldsymbol{X}\times\boldsymbol{U}\rightarrow\mathbb{R}$. Let $I_i\subseteq \mathcal{B}(\boldsymbol{X}\times \boldsymbol{Y}_1\times\hdots\boldsymbol{Y}_m)$ denote the information structure of player $i=1,\hdots,m,r$.  Formally, we have $I_i=\{\boldsymbol{x}\}$ for all the senders $i\in[m]$ and $I_r=\{\boldsymbol{y}_1,\hdots,\boldsymbol{y}_m\}$ as shown in Fig.~\ref{Fig:system_model}. We specifically consider quadratic loss functions which are of the form:
\begin{align}\label{eqn:loss_fnc}
L_i(\boldsymbol{x};\boldsymbol{u})=\|Q_i\boldsymbol{x}+R_i\boldsymbol{u}\|^2,\quad i=1,\hdots,m,r.
\end{align}
Let the strategy of sender $i$ be $\eta_i:I_i\rightarrow \boldsymbol{Y}_i$ such that $\boldsymbol{y}_i=\eta_i(\boldsymbol{x})$ is transmitted to the receiver. We denote by $\Omega$ the policy space of each sender (i.e., $\eta_i\in\Omega$ for $i\in[m]$) which we consider to be the space of all Borel-measurable functions from $\mathbb{R}^p$ to $\mathbb{R}^p$.\footnote{We note that this is the most general case that provides maximum flexibility to the sender. As to be shown through simulations in Section~\ref{Sect:Num_result} a message space $\mathbb{R}^s$ with $s<p$ suffices in specific instances.} Let the strategy of the receiver be $\gamma:I_r\rightarrow \boldsymbol{U}$ such that the receiver selects an action $\mathbf{u}=\gamma(\{\boldsymbol{y}_i\}_{i\in[m]})$ based on the policy $\gamma\in\Gamma$ where $\Gamma$ is a set of all Borel measurable functions from $\mathbb{R}^{mp}$ to $\mathbb{R}^t$. Finally, the expected cost function of each player $i=1,\hdots,m,r$ is given by
\begin{align}  J_i(\eta_1,\hdots,\eta_m;\gamma)=\mathbb{E}[L_i(\boldsymbol{x};\boldsymbol{u}=\gamma(I_r))],
\end{align}
where the dependence on $\eta_i$ appears through the information structure of the receiver, $I_r$.

We consider a hierarchical commitment structure where the senders announce, and thus commit to their policies followed by the receiver's reaction optimizing its expected cost. This structure leads to a Nash game played between the senders, followed by a Stackelberg game between the senders and the receiver\cite{bacsar1998dynamic}. Due to this commitment structure, the receiver's strategy $\gamma$ can depend on senders' signaling policies, i.e., $\gamma(\{\boldsymbol{y}_i\}_{i\in[m]})=\gamma(\{\eta_i\}_{i\in[m]})(\{\boldsymbol{y}_i\}_{i\in[m]})$. Let $\Gamma^*(\{\eta_i\}_{i\in[m]})$ denote receiver $r$'s best response set for given tuple of senders' policies $(\{\eta_i\}_{i\in[m]})$. This is a subset of $\Gamma$ and is given by the following expression where we assume that a minimum exists (which turns out to be the case as to be shown later):
\begin{align*}
    \Gamma^*(\{\eta_i\}_{i\in[m]})=\argmin_{\gamma \in \Gamma}  J_{r}( \{\eta_i\}_{i\in[m]}, \gamma(\{\eta_i\}_{i\in[m]})).
\end{align*}
Under certain assumptions such as the convexity of the receiver's utility, which is true in our setting due to the quadratic cost structure as given by (\ref{eqn:loss_fnc}), the receiver's best response set $\Gamma^*$ forms an equivalence class, and hence results in the same random variable $\boldsymbol{u}$ for any $\gamma^*\in\Gamma^*$ almost surely. Therefore, the set of signaling rules $(\{\eta^*_i\}_{i\in[m]},\gamma^*)$ are said to achieve equilibrium if
\begin{align}\label{eq: Nash}\begin{split}
    J_{i}(\eta_i^*,\{\eta_{-i}^*\},\gamma^*(\eta_i^*,\{\eta_{-i}^*\}))
    &\leq J_{i}(\eta_i,\{\eta_{-i}^*\},\gamma^*(\eta_i,\{\eta_{-i}^*\})), \quad\forall\eta_{i}\in\Omega,\forall i\in[m],\\
    \gamma^*(\{\eta_i\}_{i\in[m]}) &= \argmin_{\gamma \in \Gamma}  J_{r}( \{\eta_i\}_{i\in[m]}, \gamma(\{\eta_i\}_{i\in[m]}),
    \end{split}
\end{align}
where $\eta_{-i}^*$ denotes the equilibrium strategies of all senders other than sender $i$, and we assumed that the best response set of the receiver, $\Gamma^*$, forms an equivalence class, and thus leads to a unique $\boldsymbol{u}$ almost surely.
Note that in the equilibrium formulation provided in (\ref{eq: Nash}), all senders select their strategies independent of the signal realizations. Thus each sender is assumed to minimize its expected cost. In the following section we analyze equilibrium strategies for all players considering quadratic costs.
\section{Multi Sender Strategic Information Disclosure}\label{Sect:Strategic_com}
In this section, we first start by considering a three-player communication game consisting of two senders, to be referred to as sender $1$ and sender $2$, and one receiver $r$, and then extend the analysis to a general $m>2$ sender communication game. As illustrated in Section~\ref{Sect:system_model}, we consider expected cost functions for each player, given by

\begin{align}\label{eqn:quad_cost}   J_i(\eta_1,\eta_2;\gamma)=\mathbb{E}[L_i(\boldsymbol{x};\boldsymbol{u}=\gamma(I_r))]=\mathbb{E}[||Q_i\boldsymbol{x}+R_i\boldsymbol{u}||^2],
\end{align}
where $i=1,2,r$; $Q_i\in\mathbb{R}^{q\times p}$, $R_i\in\mathbb{R}^{q\times t}$ such that $t\leq q$ and $R_r^\top R_r$ is invertible; $\boldsymbol{u}=\gamma(\eta_1,\eta_2)(\boldsymbol{y}_1, \boldsymbol{y}_2)$, and $L_i(\boldsymbol{x};\boldsymbol{u}=\gamma(I_r))$ is given in (\ref{eqn:loss_fnc}). 
By using first order optimality conditions, the receiver’s (essentially unique) best response is given by
\begin{align}\label{opt_est_receiver}
    \gamma^*(\boldsymbol{y_1},\boldsymbol{y_2})=-(R_r^\top R_r)^{-1}R_r^\top Q_r\hat{\boldsymbol{x}}(\eta_1,\eta_2).
\end{align}
where, $\boldsymbol{\hat{x}}(\eta_1,\eta_2)=\mathbb{E}[\boldsymbol{x}|\boldsymbol{y}_1,\boldsymbol{y}_2]$ denotes the explicit dependence of the receiver's posterior on senders' strategies. Notice that for a given pair of sender strategies $(\eta_1,\eta_2)$, the receiver's optimum response is unique which is in line with the assumption made while defining equilibrium in (\ref{eq: Nash}). Incorporating this into the senders' objective functions, lets us obtain 
the best response correspondence maps \cite{bacsar1998dynamic} which we utilize to compute Nash equilibrium policies $(\eta_1^*,\eta_2^*)$. Particularly, the optimal response of sender $i$ for $i=1,2$ to a fixed strategy $\eta_{-i}$ of the other sender can be obtained by solving the following optimization problem 
\begin{align}
    \min_{\eta_i\in\Omega} \mathbb{E}[||Q_i\boldsymbol{x}-R_i(R_r^\top R_r)^{-1}R_r^\top Q_r\hat{\boldsymbol{x}}(\eta_i,\eta_{-i})||^2].\label{eq: senobj}
\end{align}
The problem in (\ref{eq: senobj}) is a functional optimization problem for each sender on Borel measurable functions in $\Omega$ which 
can be intractable. 
Thus, in order to find these functions following the approach of \cite{sayin2019hierarchical,tamura2018bayesian}, we first relax the optimization problem of each sender in (\ref{eq: senobj}) to a semi-definite program (SDP). Then, we provide an optimal solution to the SDP formulation which lower bounds the original optimization problem in (\ref{eq: senobj}) for each sender. Finally, we find the functions $\eta_i\in\Omega$ that can achieve this lower-bound, and thus provide the optimal solution to the problem in (\ref{eq: senobj}) for each $i=1,2$.

First, we note that the objective function in (\ref{eq: senobj}) is quadratic in $\boldsymbol{x}$, and, $\hat{\boldsymbol{x}}$ and senders' policies can only affect the cost through influencing $\hat{\boldsymbol{x}}$. Thus, we can simplify the senders' optimization problem in (\ref{eq: senobj}) as
\begin{align}\label{eq:obj}
    \mathbb{E}[||Q_i\boldsymbol{x}-R_i(R_r^\top R_r)^{-1}R_r^\top Q_r\hat{\boldsymbol{x}}||^2]=\mathbb{E}[\boldsymbol{x}^\top Q_i^\top Q_i\boldsymbol{x}]-2\mathbb{E}[\hat{\boldsymbol{x}}^\top\Lambda_i^\top Q_i\boldsymbol{x}]+\mathbb{E}[\hat{\boldsymbol{x}}^\top\Lambda_i^\top\Lambda_i\hat{\boldsymbol{x}}],
\end{align}
where $\Lambda_i=R_i(R_r^\top R_r)^{-1}R_r^\top Q_r$. The first term in (\ref{eq:obj}) does not depend on senders' strategies. By using the law of iterated expectation, the second term in (\ref{eq:obj}) can be shown to be equal to
\begin{align*}
    \mathbb{E}[\hat{\boldsymbol{x}}^\top\Lambda_i^\top Q_i\boldsymbol{x}]=\mathbb{E}[\mathbb{E}[\hat{\boldsymbol{x}}^\top\Lambda_i^\top Q_i\boldsymbol{x}|\boldsymbol{y}_1,\boldsymbol{y}_2]]=\mathbb{E}[\hat{\boldsymbol{x}}^\top\Lambda_i^\top Q_i\mathbb{E}[\boldsymbol{x}|\boldsymbol{y}_1,\boldsymbol{y}_2]]=\mathbb{E}[\hat{\boldsymbol{x}}^\top\Lambda_i^\top Q_i\hat{\boldsymbol{x}}].
\end{align*}
Then, we can rewrite the optimization problem in (\ref{eq: senobj}) equivalently as
\begin{align}\label{eq:senobj_v2}
    \min_{\eta_i\in \Omega(\eta_{-i})}\mathbb{E}[\hat{\boldsymbol{x}}^\top V_i\hat{\boldsymbol{x}}],
\end{align}
where
\begin{align}\label{eqn:Vi}
V_i=\Lambda_i^\top\Lambda_i-\Lambda_i^\top Q_i-Q_i^\top\Lambda_i.
\end{align}
Here, we use $\Omega(\eta_{-i})$, or equivalently $\Omega(\Sigma_{-i})$, to denote the space of all Borel measurable policies of sender $i$ that induce a posterior covariance $\Sigma$ such that $\Sigma_x\succeq \Sigma \succeq \Sigma_{-i}$.\footnote{Although sender $i$ can commit to any strategy in $\Omega$, we note that sender $i$ cannot induce a posterior at the receiver that is less informative than the posterior already induced by the other senders, i.e., $\Sigma_{-i}\succeq\Sigma \succeq O$ where $\Sigma_{-i}$ denotes the posterior of the receiver induced by other senders.} Note that the first moment of $\hat{\boldsymbol{x}}$ is equal to $\mathbb{E}[\boldsymbol{x}]$ which we take to be 0 without loss of generality. Thus, the posterior covariance denoted by $\Sigma$ is given by $\mathbb{E}[||\hat{\boldsymbol{x}}-\mathbb{E}[\hat{\boldsymbol{x}}]||^2]=\mathbb{E}[\hat{\boldsymbol{x}}\hat{\boldsymbol{x}}^\top]$. 

To illustrate the above simplification, we provide an example below where the state of the world $\boldsymbol{x}\in \mathbb{R}^3$ consists of 3 scalar random variables $z$, $\theta_A$, and $\theta_B$. 
\begin{example}\label{ex1}
Consider senders $1$ and $2$ where each sender wants to deceive the receiver about a variable of interest. We consider the state of the world as $\boldsymbol{x}=\begin{bmatrix}z & \theta_A &\theta_B\end{bmatrix}^\top$. For this example, while the receiver is only interested in $z$, the senders, being self interested, want the receiver to track a biased version of the variable of interest. Formally, 
let sender $1$ have the objective $J_{1}=\mathbb{E}[(z+\theta_A-u)^2]$ and sender $2$ have the objective function $J_{2}=\mathbb{E}[(z+\theta_B-u)^2]$. On the other hand, the receiver is only interested in $z$, and thus has the objective function  $J_{r} = \mathbb{E}[(z-u)^2]$. With these objective functions, we have the following matrices:
\begin{align*}
    Q_1=\begin{bmatrix}1 &1 &0\end{bmatrix},\quad R_1=-1,\quad Q_2=\begin{bmatrix}1 &0 &1\end{bmatrix},\quad R_2=-1,\quad Q_r=\begin{bmatrix}1 &0 &0\end{bmatrix},\quad R_r=-1.
\end{align*}
Thus, by using the definition of $V_i$ in (\ref{eqn:Vi}), we have
\begin{align*}
    V_1=\begin{bmatrix}-1 & -1 &0\\-1 & 0 &0\\0 &0 &0\end{bmatrix},\qquad V_2=\begin{bmatrix}-1 &0 &-1\\0 &0 &0\\-1 &0 &0\end{bmatrix}.
\end{align*}
\end{example}
Later in Section~\ref{Sect:Num_result}, we provide a complete solution to Example~\ref{ex1} after characterizing the overall solution for the general problem in (\ref{eqn:quad_cost}). In what follows, we first state and prove a few lemmas which will help identify the reaction set of sender $i$ for a fixed strategy of the other sender $-i$. 

\begin{lemma} \label{Lemma_1}
For a fixed strategy of the other sender $-i$, if the posterior covariance at the receiver is $\Sigma_{-i}$, sender $i$ can only induce posterior covariance $\Sigma$ in the interval of $\Sigma_x\succeq\Sigma\succeq\Sigma_{-i}$, holding for $i=1,2$.
\end{lemma}
\begin{Proof}
Since $\mathbb{E}[\hat{\boldsymbol{x}}^\top V_i\hat{\boldsymbol{x}}] = Tr(\mathbb{E}[V_i\hat{\boldsymbol{x}}\hat{\boldsymbol{x}}^\top])$, the optimization problem in (\ref{eq: senobj}) for sender $i$ for $i =1,2$, can be written as
\begin{align*}
    \min_{\eta_i\in\Omega(\eta_{-i})} Tr(\mathbb{E}[V_i\hat{\boldsymbol{x}}\hat{\boldsymbol{x}}^\top])=\min_{\eta_i\in\Omega(\eta_{-i})} Tr (V_i \Sigma)=\min_{\eta_i\in\Omega(\Sigma_{-i})} Tr(V_i \Sigma).
\end{align*}

Note that senders' strategies only affect the cost through the second moment of the posterior estimate. Further, each sender $i$'s strategy $\eta_i$ induces a posterior covariance matrix which by definition is positive semi-definite and symmetric. Thus, it follows that 
 \begin{align*}
     \mathbb{E}[(\boldsymbol{x}-\hat{\boldsymbol{x}})(\boldsymbol{x}-\hat{\boldsymbol{x}})^\top]=\Sigma_x-\Sigma\succeq O,
 \end{align*}
 which proves the upper-bound on $\Sigma$, i.e., $\Sigma_x\succeq\Sigma$. Next, we show the lower-bound on $\Sigma$. For that, we write 
 \begin{align}\label{eq:lem_1}
     \mathbb{E}[(\hat{\boldsymbol{x}}-\mathbb{E}[\boldsymbol{x}|\boldsymbol{y}_{-i}])(\hat{\boldsymbol{x}}-\mathbb{E}[\boldsymbol{x}|\boldsymbol{y}_{-i}])^\top]=&\mathbb{E}[\hat{\boldsymbol{x}}\hat{\boldsymbol{x}}^\top]-2\mathbb{E}[\mathbb{E}[\boldsymbol{x}|\boldsymbol{y}_{-i}]\mathbb{E}[\boldsymbol{x}|\boldsymbol{y}_{-i},\boldsymbol{y}_i]^\top]\nonumber\\&+\mathbb{E}[\mathbb{E}[\boldsymbol{x}|\boldsymbol{y}_{-i}]\mathbb{E}[\boldsymbol{x}|\boldsymbol{y}_{-i}]^\top].
 \end{align}
 By using the law of iterated expectations, the second term can be shown to be equal to
 \begin{align}\label{eq:lem_2}
     \mathbb{E}[\mathbb{E}[\boldsymbol{x}|\boldsymbol{y}_{-i}]\mathbb{E}[\boldsymbol{x}|\boldsymbol{y}_{-i},\boldsymbol{y}_i]^\top]=\mathbb{E}[\mathbb{E}[[\mathbb{E}[\boldsymbol{x}|\boldsymbol{y}_{-i}]\mathbb{E}[\boldsymbol{x}|\boldsymbol{y}_{-i},\boldsymbol{y}_i]^\top|\boldsymbol{y}_{-i}]]=\mathbb{E}[\mathbb{E}[\boldsymbol{x}|\boldsymbol{y}_{-i}]\mathbb{E}[\boldsymbol{x}|\boldsymbol{y}_{-i}]^\top].
 \end{align}
 Thus, by using (\ref{eq:lem_2}), we can write (\ref{eq:lem_1}) as
 \begin{align*}
     \mathbb{E}[(\hat{\boldsymbol{x}}-\mathbb{E}[\boldsymbol{x}|\boldsymbol{y}_{-i}])(\hat{\boldsymbol{x}}-\mathbb{E}[\boldsymbol{x}|\boldsymbol{y}_{-i}])^\top]=\Sigma-\Sigma_{-i}\succeq O.
 \end{align*}
 Finally, we obtain $\Sigma_x\succeq \Sigma\succeq \Sigma_{-i}$, which completes the proof.
\end{Proof}
Therefore the real, symmetric posterior covariance matrix $\Sigma$ induced by sender $i$'s policy $\eta_i$ should at least follow the condition provided in Lemma~\ref{Lemma_1}. This result in Lemma~\ref{Lemma_1} is intuitive. If sender $i$'s policy is to reveal the state of the world $\boldsymbol{x}$ entirely, then the posterior covariance induced by this strategy is $\Sigma = \Sigma_x$. On the other hand, if the sender $i$ chooses a policy with no information disclosure, then the receiver can always use the signal from the sender $-i$ and as a result, its posterior covariance becomes equal to $\Sigma = \Sigma_{-i}$. Thus, any partial information disclosure strategy from sender $i$ leads to the posterior covariance $\Sigma_x\succ \Sigma\succ \Sigma_{-i}$. 
In the following lemma, we relax the optimization problem in (\ref{eq: senobj}) by an SDP problem formulation. 
 \begin{lemma}\label{lem: bestresplbound}
 There exists an SDP problem formulation which lower bounds the optimization problem in (\ref{eq: senobj}) 
 and is given by
 \begin{align}
\min_{S\in\mathbb{S}^p} & \quad Tr(V_iS) \nonumber\\
\textrm{s.t.} & \quad 
\Sigma_x\succeq S\succeq \Sigma_{-i}\succeq O , \label{eq: SDPseni}
\end{align}
where $\mathbb{S}^p$ denotes the space of psd matrices with dimension $p\times p$. Moreover (\ref{eq: SDPseni}) is a convex optimization problem.  
 \end{lemma}
 \begin{Proof} 
Lemma \ref{Lemma_1} shows that the constraint in optimization problem  (\ref{eq: SDPseni}) is necessary.\footnote{Although \cite{9222530} proves that using linear plus noise strategies would make this an equivalent optimization problem, we are interested in purely linear  strategies (without any additive noise) for reasons that would be evident soon, and hence this forms a lower bound to the original optimization problem as in \cite{sayin2019hierarchical}.} 
Thus, the SDP problem (\ref{eq: SDPseni}) constitutes a lower bound to the original optimization problem of each sender in (\ref{eq: senobj}). 
 \end{Proof}
Next, we introduce a few quantities that build on the above two lemmas and will be useful in identifying the equilibrium posteriors.
\begin{definition}\label{Def_1}
A posterior covariance $\Sigma$ is said to be stable for sender $i$ if 
\begin{align*}
    \min_{\Sigma_x\succeq S\succeq\Sigma\succeq O}Tr(V_iS)=Tr(V_i\Sigma).
\end{align*}
\end{definition}
For the relaxed optimization problem in (\ref{eq: SDPseni}), at a stable equilibrium $\Sigma$, none of the players (senders) wants to reveal any more information. We note that since this definition is for the relaxed optimization problem, not every stable posterior covariance can be attained by the signaling policies $\eta_i \in \Omega(\Sigma)$ for a considered policy class $\Omega(\cdot)$ defined earlier. To address this, we provide the following definition.
\begin{definition}\label{Def_2}
A posterior covariance $\Sigma$ is said to be stable and achievable for sender $i$ if there exists a strategy $\eta_i\in\Omega$ such that $\mathbb{E}[\mathbb{E}[\boldsymbol{x}|\boldsymbol{y_{-i}}]\mathbb{E}[\boldsymbol{x}|\boldsymbol{y_{-i}}]^\top]=\mathbb{E}[\mathbb{E}[\boldsymbol{x}|\boldsymbol{y_i},\boldsymbol{y_{-i}}]\mathbb{E}[\boldsymbol{x}|\boldsymbol{y_i},\boldsymbol{y_{-i}}]^\top]=\Sigma$, where $\boldsymbol{y_i}=\eta_i(\boldsymbol{x})$.
\end{definition}
In other words, in Definition~\ref{Def_2}, we say that a stable posterior covariance $\Sigma$ is achievable if it can be induced by a strategy $\eta_i \in \Omega$ for sender $i$ in the considered policy class while sender $-i$ already induces the \textit{same} posterior covariance.

\begin{lemma}\label{lem: stabsig}
If a posterior covariance $\Sigma^*$ is stable and is achievable for both senders in the chosen class of sender policies ${\eta_i}\in\Omega$, $i=1,2$, then $\Sigma^*$ is an equilibrium posterior covariance. 
\end{lemma}
\begin{Proof}
Consider a posterior covariance $\Sigma^*$ which is stable and achievable for both senders. This implies that there exists a strategy $\eta_{-i}'$ for sender $-i$ which can induce the posterior $\Sigma^*$. If $\Sigma^*$ is a stable posterior for sender $i$ then, from Definition~\ref{Def_1}, we have
\begin{align*}
    \min_{\Sigma_x\succeq S\succeq\Sigma^*} Tr(V_iS)=Tr(V_i\Sigma^*).
\end{align*}
That is, given sender $-i$'s policy  $\eta_{-i}'$, sender $i$ also prefers the same posterior covariance. 
Furthermore, if $\Sigma^*$ is achievable by sender $i$, then there exists a strategy $\eta_i'\in\Omega$ for sender $i$ such that $\mathbb{E}[\mathbb{E}[\boldsymbol{x}|\boldsymbol{y_{-i}}]\mathbb{E}[\boldsymbol{x}|\boldsymbol{y_{-i}}]^\top]=\mathbb{E}[\mathbb{E}[\boldsymbol{x}|\boldsymbol{y_i},\boldsymbol{y_{-i}}]\mathbb{E}[\boldsymbol{x}|\boldsymbol{y_i},\boldsymbol{y_{-i}}]^\top]=\Sigma^*$, where $\boldsymbol{y_i}=\eta_i(\boldsymbol{x})$.
As a result, if $\Sigma^*$ is also stable and achievable posterior for sender $i$, then $\Sigma^*$ indeed constitutes a posterior that is the result of equilibrium strategies.
\end{Proof}


Thus, to identify the NE among the senders for the two-sender strategic communication game, it is sufficient to identify posteriors which are stable and achievable by both senders\footnote{We will see later that this property of NE between two senders is equally valid in the $m>2$ sender problem as well.}. To identify such posteriors, we reformulate the relaxed SDP problem given in (\ref{eq: SDPseni}) using the following lemma.

\begin{lemma}\label{lem: transf}
For any $\Sigma'$ such that $\Sigma_x\succeq\Sigma'\succeq O$, the optimization problem
\begin{align}\label{eqn:lemma_4}
\min_{S\in\mathbb{S}^p} & \quad Tr(V_iS) \nonumber\\
\textrm{s.t.} & \quad 
\Sigma_x\succeq S\succeq\Sigma',
\end{align}
can be equivalently written as
\begin{align}\label{eqn:lemma_4_1}
\min_{Z\in\mathbb{S}^p} & \quad Tr(W_iZ)+ Tr(V_i\Sigma') \nonumber\\
\textrm{s.t.} & \quad 
I\succeq Z\succeq O,
\end{align}
where $W_i=(\Sigma_x-\Sigma')^{\frac{1}{2}}V_i(\Sigma_x-\Sigma')^{\frac{1}{2}}$.
\end{lemma}
\begin{Proof}
Let $Z:=I-K^2$, where
\begin{align}
    K=\left((\Sigma_x-\Sigma')^{\frac{1}{2}}\right)^{\dagger}(\Sigma_x-S)^{\frac{1}{2}}, 
\end{align}
where $(.)^{\dagger}$ denotes the pseudo-inverse of a matrix and $S$ satisfies the constraint in (\ref{eqn:lemma_4}).\footnote{We note that since $\Sigma_x-\Sigma'$ might not be invertible, we use pseudo-inverse of a matrix which always exists.} Further, let $\lambda(K)$ 
denote the largest eigenvalue of $K$. 
Then $\lambda(K)\in[0,1]$ \cite{gross2001lowner}. By using the fact that any non-zero vector is an eigenvector of the identity matrix, it follows that $\lambda(Z)\in [0,1]$. Thus, we have $ I\succeq Z\succeq O$. 

Next, for ease of exposition, we let $ Q := \Sigma_x-S$. Then, we have $(\Sigma_x-\Sigma')^{\frac{1}{2}}K=Q^{\frac{1}{2}}$ and $Q$ can equivalently be written as 
\begin{align}\label{eqn:lem4}
 Q=(\Sigma_x-\Sigma')^{\frac{1}{2}}K^2(\Sigma_x-\Sigma')^{\frac{1}{2}}.
\end{align}
Since $Z=I-K^2$, $Q$ in (\ref{eqn:lem4}) is equal to 
\begin{align}\label{eqn:lem5}
Q=(\Sigma_x-\Sigma')-(\Sigma_x-\Sigma')^{\frac{1}{2}}Z(\Sigma_x-\Sigma')^{\frac{1}{2}}.
\end{align}
Then, we have $S=\Sigma'+(\Sigma_x-\Sigma')^{\frac{1}{2}}Z(\Sigma_x-\Sigma')^{\frac{1}{2}}$. Thus, the objective function $Tr(V_iS)$ in (\ref{eqn:lemma_4}) is equal to the objective function $Tr(W_iZ)+ Tr(V_i\Sigma')$ in (\ref{eqn:lemma_4_1}) where $W_i=(\Sigma_x-\Sigma')^{\frac{1}{2}}V_i(\Sigma_x-\Sigma')^{\frac{1}{2}}$, which completes the proof.
\end{Proof}

  
Based on Lemmas \ref{lem: stabsig} and \ref{lem: transf}, we have the following result:
\begin{prop}\label{rem: wpsd} If
$Tr(V_i\Sigma^*_i)\!=\!\min_{\Sigma_x\succeq S\succeq O} Tr(V_i S)$, then $W_i^*\!=\!(\Sigma_x-\Sigma_i^*)^{\frac{1}{2}}V_i(\Sigma_x-\Sigma_i^*)^{\frac{1}{2}}\!\succeq\! O.$ 
\end{prop}
\begin{Proof}
We note that if $Tr(V_i\Sigma^*_i)=\min_{\Sigma_x\succeq S\succeq O} Tr(V_i S)$, then we have 
\begin{align} \label{eqn:rmk1}
    Tr(V_i\Sigma^*_i)=\min_{\Sigma_x\succeq S\succeq \Sigma^*_i} Tr(V_i S).
\end{align}
Then, by using Lemma~\ref{lem: transf}, we rewrite (\ref{eqn:rmk1}) as 
\begin{align}\label{eqn:rmk2}
\min_{Z\in\mathbb{S}^p} & \quad Tr(W_i^*Z)+ Tr(V_i\Sigma^*_i) \nonumber\\
\textrm{s.t.} & \quad 
I\succeq Z\succeq O,
\end{align}
where $W_i^*=(\Sigma_x-\Sigma_i^*)^{\frac{1}{2}}V_i(\Sigma_x-\Sigma_i^*)^{\frac{1}{2}}$. Since the minimum value in (\ref{eqn:rmk2}) is equal to $Tr(V_i\Sigma^*_i)$, we must have $\min_{I\succeq Z\succeq O} Tr(W_iZ) = 0$ which happens when $W_i^*$ is a positive semi-definite matrix, i.e., $W_i^*=(\Sigma_x-\Sigma_i^*)^{\frac{1}{2}}V_i(\Sigma_x-\Sigma_i^*)^{\frac{1}{2}}\succeq O$. To show this, suppose that there exists a $Q \in \mathbb{R}^{p\times r}$ where $r$ is a positive integer such that $Z =QQ^\top$. Then, we have 
\begin{align}\label{eqn:rmk3}
    Tr(W_i^*Z) = Tr(Q^\top W_i^*Q) = \sum_{\ell =1}^{r} q_\ell^\top W_i^*q_\ell, 
\end{align}
where $q_\ell$ is the $\ell$th column of $Q$. We note that if $W_i^*$ is not a positive semi-definite matrix, then we can find a $Q$ matrix such that $ Tr(W_i^*Z)$ in (\ref{eqn:rmk3}) is less than $0$. However, since we have $\min_{I\succeq Z\succeq O} Tr(W_i^*Z) = 0$, we conclude that $W_i^*$ is a positive semi-definite matrix.
\end{Proof}

In Proposition~\ref{rem: wpsd}, we have shown that when there is no constraint on the posterior covariance in the relaxed optimization problem in (\ref{eq: SDPseni}), i.e., when $\Sigma_{-i} = O$, the overall minimum value achieved in (\ref{eq: SDPseni}) is given by $Tr(V_i\Sigma^*_i)$. Further, such a $\Sigma^*_i$ in Proposition~\ref{rem: wpsd} may not be unique. In other words, there can be multiple $\Sigma^*_i$s with $W_i^*\succeq O$ and achieve the same minimum value $Tr(V_i\Sigma^*_i)$. Moreover, we may not have a strict ordering between such $\Sigma^*_i$ posterior covariances. We utilize tools developed thus far to prove that any $\Sigma'\succeq\Sigma_i^*$ is stable for sender $i$.

\begin{lemma}\label{lem: greatersig}
Let $Tr(V_i\Sigma^*_i)=\min_{\Sigma_x\succeq S\succeq O} Tr(V_iS)$. Then, for any $\Sigma'$ such that $\Sigma_x\succeq\Sigma'\succeq\Sigma^*_i\succeq O$, we have
\begin{align}
    \min_{\Sigma_x\succeq S\succeq\Sigma'} Tr(V_iS)= Tr(V_i\Sigma').
\end{align}
\end{lemma}
\begin{Proof}
By using Lemma~\ref{lem: transf}, we have
\begin{align*}
    \min_{\Sigma_x\succeq S\succeq\Sigma'} Tr(V_iS)=\min_{I\succeq Z\succeq O} Tr(W_iZ)+ Tr(V_i\Sigma'),
\end{align*}
where $S=\Sigma'+(\Sigma_x-\Sigma')^{\frac{1}{2}}Z(\Sigma_x-\Sigma')^{\frac{1}{2}}$ and $W_i=(\Sigma_x-\Sigma')^{\frac{1}{2}}V_i(\Sigma_x-\Sigma')^{\frac{1}{2}}$. By definition, $\Sigma_x-\Sigma_i^*\succeq\Sigma_x-\Sigma'$. Hence, there exists a matrix $K'$ with $\lambda (K')\subseteq[0,1]$ such that $(\Sigma_x-\Sigma_i^*)^\frac{1}{2}K'=(\Sigma_x-\Sigma')^{\frac{1}{2}}\succeq O$ \cite{gross2001lowner}. This leads to 
\begin{align*}
    W_i=(\Sigma_x-\Sigma')^{\frac{1}{2}}V_i(\Sigma_x-\Sigma')^{\frac{1}{2}}=K'^\top(\Sigma_x-\Sigma_i^*)^{\frac{1}{2}}V_i(\Sigma_x-\Sigma_i^*)^{\frac{1}{2}}K'\succeq O,
\end{align*}
since $W_i^*=(\Sigma_x-\Sigma_i^*)^{\frac{1}{2}}V_i(\Sigma_x-\Sigma_i^*)^{\frac{1}{2}}\succeq O$ from Proposition~\ref{rem: wpsd}. Thus, $\min_{I\succeq Z\succeq O} Tr(W_iZ)$ is equal to $0$, and we have $ \min_{\Sigma_x\succeq S\succeq\Sigma'} Tr(V_iS) =Tr(V_i\Sigma')$. 
\end{Proof}

Lemma~\ref{lem: greatersig} hints at the possibility of existence of multiple equilibria. To formalize this, we first characterize the set of stable posteriors:
\begin{prop}\label{Remark_2}
The full set of posterior covariances that are stable for sender $i$ for $i=1,2$ is given by the set $\{\Sigma'\mid W_i=(\Sigma_x-\Sigma')^{\frac{1}{2}}V_i(\Sigma_x-\Sigma')^{\frac{1}{2}}\succeq O\}$.
\end{prop}
\begin{Proof}
For a given posterior covariance $\Sigma'$, sender $i$ solves the following optimization problem
\begin{align} 
    \min_{\Sigma_x\succeq S\succeq \Sigma'} Tr(V_i S),
\end{align}
which can equivalently be written, by using Lemma~\ref{lem: transf}, as
\begin{align}\label{eqn:rmk2_1}
    \min_{I\succeq Z\succeq O} Tr(W_iZ)+ Tr(V_i\Sigma').
\end{align}
Since $W_i=(\Sigma_x-\Sigma')^{\frac{1}{2}}V_i(\Sigma_x-\Sigma')^{\frac{1}{2}}\succeq O$, the minimum value of (\ref{eqn:rmk2_1}) is equal to $Tr(V_i\Sigma')$. Thus, every $\Sigma'$ that generates $W_i\succeq O$ is stable for sender $i$ in the sense that the sender $i$ does not reveal any more information to the receiver.
\end{Proof}

Proposition~\ref{Remark_2} is in line with the observation made in \cite{tamura2018bayesian} as a necessary and sufficient condition for existence of a partially revealing equilibrium for a game with multiple senders. In particular, by using Lemma~\ref{lem: greatersig} and Proposition~\ref{Remark_2}, it follows that any posterior covariance $\Sigma^* \in \cap_{i=1}^2 \mathcal{D}_i$ where  $\mathcal{D}_i = \{\Sigma'\mid\Sigma_x\succeq\Sigma'\succeq \Sigma_i^*\}$ is a stable posterior. 
However, only a subset of these stable posteriors can be achieved by using linear noiseless signaling policies considered in this work. In the following proposition, we show that full revelation is always an equilibrium.

\begin{prop}\label{Prop: Fullinf}
Full information revelation is always an equilibrium.
\end{prop}
\begin{Proof}

In the case of full information revelation the proof follows directly from noticing that $W_i=O\succeq O,$ $\forall i\in[m]$. Since $\Sigma_x$ is stable and achievable for all senders, it is also an equilibrium posterior covariance.
\end{Proof}
This is in line with the observations of \cite{gentzkow2017bayesian, sobel2013giving} for finite state and action spaces. It can be seen that even if all senders have identical preferences, full revelation is an equilibrium outcome although there might be other equilibria which are less informative and preferable by all the senders. This calls for a refinement to identify non-trivial equilibrium outcomes. We utilize the following definition to propose a refinement\cite{bacsar1998dynamic}.
\begin{definition}\label{defn_adm}
An equilibrium is said to be partially informative if there exists at least one sender for whom deviating to a more informative equilibrium does not lead to a better cost.
\end{definition}

Definition~\ref{defn_adm} says that when the senders are at a partially informative equilibrium posterior, there is no other more informative equilibrium that is preferred by all senders. 

\subsection{Sequential Optimization Technique to Find Equilibrium Posteriors}\label{subsect:3_A}
We now propose two algorithms to identify partially informative equilibrium posteriors for the two-sender game. The algorithms differ in the techniques utilized to solve the underlying SDP. The first algorithm uses an analytical approach similar to \cite{tamura2012theory} to solve the SDP explicitly whereas the second algorithm uses the standard SDP solver in \cite{grant2008cvx} with minor variations.


\begin{algorithm}[t]
\caption{Finding an Equilibrium Posterior via \cite{tamura2012theory} }\label{algorithm2}
\begin{algorithmic}[1]
\State{\textbf{Parameters:} $\Sigma_x$}
\State{\textbf{Compute} $V_i$ for $ i\in\{1,2\}$ by using (\ref{eqn:Vi})}
\State{\textbf{Variables:} $S_i\in\mathbb{S}^p$ with $S_i\succeq O$ $\forall i\in\{1,2\}$}
\State{\textbf{Compute} $W_1 = \Sigma_x^{\frac{1}{2}}V_{1}\Sigma_x^{\frac{1}{2}} $}
\State{\textbf{Define} $Q_1 =[q_1,\cdots, q_r]$ consisting of eigenvectors of negative eigenvalues of $W_1$ } 
\State{\textbf{Compute:} $ P_1^* = Q_1(Q_1^\top Q_1)^{-1}Q_1^\top$} and $ S_1 =\Sigma_x^{\frac{1}{2}}P_1^*\Sigma_x^{\frac{1}{2}} $
\State $ \Sigma_{-1} \gets S_{1}$
\State{\textbf{Compute} $W_2 = (\Sigma_x- \Sigma_{-1})^{\frac{1}{2}}V_{2}(\Sigma_x- \Sigma_{-1})^{\frac{1}{2}} $}
\State{\textbf{Define} $Q_2 =[q'_1,\cdots, q'_r]$ consisting of eigenvectors of negative eigenvalues of $W_2$ }
\State{\textbf{Compute:} $ P_2^* = Q_2(Q_2^\top Q_2)^{-1}Q_2^\top$} and $ S_2 =\Sigma_{-1} +(\Sigma_x- \Sigma_{-1})^{\frac{1}{2}}P_2^*(\Sigma_x- \Sigma_{-1})^{\frac{1}{2}} $
\State{\textbf{Return:} $\Sigma^* \gets  S_{2}$}
\end{algorithmic}
\end{algorithm}

More precisely, in Algorithm~\ref{algorithm2}, we use the fact that the following optimization problem 
\begin{align} \label{eqn:tamura}
    \min_{\Sigma_x\succeq S\succeq O} Tr(V_i S)
\end{align}
admits a solution of the form 
\begin{align}\label{eqn:soln_tamura}
    S_i^* =\Sigma_x^{\frac{1}{2}}P_i^*\Sigma_x^{\frac{1}{2}}, 
\end{align}
where $P_i^* = Q_i(Q_i^\top Q_i)^{-1}Q_i^\top$ is a projection matrix. Here, $ Q_i = [q_1,\cdots,q_r]$ is a matrix consisting of the eigenvectors $q_r$s that correspond to the non-positive eigenvalues of $W_i = \Sigma_x^{\frac{1}{2}}V_{i}\Sigma_x^{\frac{1}{2}}$.  Due to the sequential nature of the algorithm, for a two sender problem, we are able to find two equilibrium posteriors. Here, we emphasize that the sequential nature of the algorithm is just to find a partially informative equilibrium posterior but this does not imply anything about commitment orders for policy computation. We later prove that this posterior can be achieved by simultaneous commitment by the senders.

In Algorithm~\ref{alg: postcalc}, we propose another way to reach such equilibrium posteriors. 
Since $V_i$ can be rank deficient, the SDP problem (at line 4) in Algorithm~\ref{alg: postcalc} may have multiple equivalent (minimal) solutions.
Similar to Algorithm~\ref{algorithm2}, in order to provide the largest feasible posterior set 
among these solutions, we choose the one obtained from the following optimization problem   
\begin{align}\label{eqn:alg1}
\min_{\Sigma_x\succeq S'_{i}\succeq O} & \quad  ||\Sigma_x^{-\frac{1}{2}}S'_{i}\Sigma_x^{-\frac{1}{2}}||_* \nonumber\\
\textrm{s.t.} & \quad 
Tr(V_iS_i)\geq Tr(V_iS_i'),
\end{align}
where $||.||_*$ denotes the nuclear norm which is the summation of the singular values of a matrix. Since the objective function in (\ref{eqn:alg1}) is a psd matrix, the nuclear norm in this case is equal to the summation of the eigen-values of the matrix $ \Sigma_x^{-\frac{1}{2}}S'_{i}\Sigma_x^{-\frac{1}{2}}$ which has a similar form as $P_i^*$ in (\ref{eqn:soln_tamura}) with one difference, which is that the eigenvalues of $\Sigma_x^{-\frac{1}{2}}S'_{i}\Sigma_x^{-\frac{1}{2}}$ are in $[0,1]$. To be more precise, the eigen-values of $\Sigma_x^{-\frac{1}{2}}S'_{i}\Sigma_x^{-\frac{1}{2}}$ are equal to 1 corresponding to the negative eigen-values of $W_1=\Sigma_x^{\frac{1}{2}}V_{1}\Sigma_x^{\frac{1}{2}}$ and take arbitrary values in $[0,1]$ for the 0 eigen-values of $W_1$. The nuclear norm minimization problem in (\ref{eqn:alg1}) is a convex optimization problem \cite{recht2011simpler} which finds an equilibrium posterior where $ \Sigma_x^{-\frac{1}{2}}S'_{i}\Sigma_x^{-\frac{1}{2}}$ is a projection matrix. Thus, $S'_{i}$ obtained in Algorithm~\ref{alg: postcalc}  has the same form as the posteriors obtained in Algorithm~\ref{algorithm2}. 
Due to Lemma~\ref{lem: greatersig}, the posterior covariance $\Sigma^*$ obtained from both of these algorithms is stable for both senders, i.e., neither of the senders wants to reveal more information to the receiver. 

\begin{algorithm}[t]
\caption{Finding an Equilibrium Posterior via SDP}\label{alg: postcalc}
\begin{algorithmic}[1]
\State{\textbf{Parameters:} $\Sigma_x$}
\State{\textbf{Compute} $V_i$ $\forall i\in\{1,2\}$ by using (\ref{eqn:Vi})}
\State{\textbf{Variables:} $S_i, S_i'\in\mathbb{S}^p$ with $S_i,S_i'\succeq O$ $\forall i\in\{1,2\}$}
\State{\textbf{Solve:} $ \min_{\Sigma_x\succeq S_{1}\succeq O} Tr(V_1 S_{1}),$ by using CVX toolbox in MatLab}
\State{\textbf{Solve:} $ \min_{\Sigma_x\succeq S'_{1}\succeq O} ||\Sigma_x^{-\frac{1}{2}}S'_{1}\Sigma_x^{-\frac{1}{2}}||_*,$} 
\State{$\qquad \quad$s.t. $Tr(V_1S_1)\succeq Tr(V_1S_1')$ by using CVX toolbox in MatLab} 
\State $ \Sigma_{1}^* \gets S_{1}$
\State{\textbf{Solve:} $ \min_{\Sigma_x\succeq S_{2}\succeq \Sigma_{1}^*} Tr(V_2 S_{2}),$ by using CVX toolbox in MatLab}
\State{\textbf{Solve:} $ \min_{\Sigma_x\succeq S'_{2}\succeq \Sigma_{1}^*} ||\Sigma_x^{-\frac{1}{2}}S'_{2}\Sigma_x^{-\frac{1}{2}}||_*,$} 
\State{$\qquad \quad$s.t. $Tr(V_2S_{2})\succeq Tr(V_2S_{2}')$ by using CVX toolbox in MatLab} 
\State{\textbf{Return:} $\Sigma^* \gets  S'_{2}$}
\end{algorithmic}
\end{algorithm}

Next we analyze the structure of equilibrium posteriors computed by using
Algorithms~\ref{algorithm2} or \ref{alg: postcalc}, and find a linear strategy that can achieve this posterior. For ease of exposition, we follow the analysis in Algorithm~\ref{alg: postcalc}. 
In the following corollary, we state the form of the stable posterior covariance $\Sigma^*$.  

\begin{corollary}\label{cor: stabpost1}
Stable posteriors obtained from Algorithm~\ref{alg: postcalc} are of the form
\begin{align}\label{eqn:stabpost}
    \Sigma^*=\Sigma_i^*+(\Sigma_x-\Sigma_i^*)^{\frac{1}{2}}P(\Sigma_x-\Sigma_i^*)^{\frac{1}{2}},
\end{align}
for some projection matrix $P$, where $Tr(V_i\Sigma_i^*)=\min_{\Sigma_x\succeq S\succeq O}Tr(V_iS)$.
\end{corollary}
\begin{Proof}
For the sake of simplifying the notation in the proof, in Algorithm~\ref{alg: postcalc} we consider the SDP optimization steps in line 4 as $V_1$ and in line 8 as $V_2$. First note that the posterior $\Sigma_{1}^*$ obtained in line 7 of Algorithm~\ref{alg: postcalc} is of the form $\Sigma_{1}^*=\Sigma_x^{\frac{1}{2}}P_{1}^*\Sigma_x^\frac{1}{2}$, which follows from (\ref{eqn:soln_tamura}). The remaining part of Algorithm~\ref{alg: postcalc} solves for
\begin{align}\label{eqn:corr_1}
    \min_{\Sigma_x\succeq S\succeq\Sigma_{1}^*} Tr(V_{2}S).
\end{align}
By using Lemma \ref{lem: transf}, we can rewrite the optimization problem in (\ref{eqn:corr_1}) as 
\begin{align}\label{eqn:corr_1_2}
    \min_{I\succeq Z\succeq O} Tr(W_{2}Z)+ Tr(V_2\Sigma_{1}^*).
\end{align}
It has been shown in \cite{tamura2012theory,sayin2019hierarchical} that the solution to the optimization problem in (\ref{eqn:corr_1_2}) is of the form of a projection matrix, i.e., $Z=\Tilde{P}$. Thus, the posterior obtained from Algorithm~\ref{alg: postcalc} is of the form 
\begin{align}\label{eq: eqpost}
    \Sigma^*=\Sigma_1^*+(\Sigma_x-\Sigma_1^*)^{\frac{1}{2}}\Tilde{P}(\Sigma_x-\Sigma_1^*)^{\frac{1}{2}},
\end{align}
where $\Sigma_1^*=\Sigma_x^{\frac{1}{2}}P_1^*\Sigma_x^{\frac{1}{2}}$. A similar approach can be used to solve for the posterior structure with $V_2$ in line 4 and $V_1$ in line 8 of Algorithm~\ref{alg: postcalc}, which provides the general $ \Sigma^*$ expression in (\ref{eqn:stabpost}). 
\end{Proof}
We note that the posterior obtained from the proposed algorithm is stable for both the senders. This follows for sender $-i$ because $\Sigma^*\succeq\Sigma_{i}^*$ from line 8 of Algorithm~\ref{alg: postcalc}, and is true for sender $i$ because $\Sigma^*$ is the solution of the optimization problem in lines  9 and 10 of Algorithm~\ref{alg: postcalc}, which implies $W_i\succeq O$. Below, we provide policies that can achieve the computed equilibrium posterior.

\subsection{Achievability of Nash Equilibrium using Linear Strategies}\label{subsect:3_B}

Following Corollary~\ref{cor: stabpost1}, we still have to show that the stable posteriors can be achieved as a result of simultaneous policy commitment by the senders, that is that there exists a pair of policies that achieve the posterior covariance and constitute of a NE 
as given in (\ref{eq: Nash}). One way to achieve Nash equilibrium is to design strategies such that each sender achieves the same stable posterior. This is because from Definition \ref{Def_1}, the posterior is stable for each sender given the other sender's policy. Further, this satisfies Definition~\ref{Def_2} and hence such posterior covariances are achievable. In the following theorem we show that linear strategies can indeed achieve a stable equilibrium posterior of the form given in Corollary~\ref{cor: stabpost1}.
\begin{theorem}\label{th: Nasheq}
Let $\Sigma^*$ be a stable posterior covariance obtained from Algorithm \ref{alg: postcalc}. Then, policies $\eta_1(\boldsymbol{x})=G_1L\boldsymbol{x} $ and $\eta_2(\boldsymbol{x})=G_2L\boldsymbol{x}$, where $L^\top=\Sigma_x^{-\frac{1}{2}}U_{P'}\Lambda_{P'}$, $P'=\Sigma_x^{-\frac{1}{2}}\Sigma^*\Sigma_x^{-\frac{1}{2}}=U_{P'}\Lambda_{P'}U_{P'}^\top$, and $G_1$, $G_2$ are arbitrary full-rank matrices in $\mathbb{R}^{p\times p}$ with orthonormal columns, 
achieve Nash equilibrium, resulting in a posterior $\Sigma^*$ in (\ref{eqn:stabpost}). 
\end{theorem}
\begin{Proof}
As a result of Corollary~\ref{cor: stabpost1}, the equilibrium posterior is of the form
\begin{align}\label{eqn:equlib}
    \Sigma^*=\Sigma_i^*+(\Sigma_x-\Sigma_i^*)^{\frac{1}{2}}P(\Sigma_x-\Sigma_i^*)^{\frac{1}{2}}.
\end{align}
Since $\Sigma_x\succeq\Sigma_x-\Sigma^*_i$, we have $(\Sigma_x-\Sigma_i^*)^{\frac{1}{2}} = \Sigma_x^{\frac{1}{2}}K$ for some K with $\lambda(K)\subseteq[0,1]$. Then, $\Sigma^*$ in (\ref{eqn:equlib}) can be written as
\begin{align*}
    \Sigma^*=\Sigma_x^{\frac{1}{2}}(P^*_i+KPK^\top)\Sigma_x^{\frac{1}{2}}.
\end{align*}
Moreover, since $\Sigma_x-\Sigma_i^* =\Sigma_x^{\frac{1}{2}}KK^\top\Sigma_x^{\frac{1}{2}} $, we have $KK^\top = I-P_i^*$, and thus $KK^\top$ is a projection matrix. If we multiply $KK^\top$ with $P_i^*$, we have $P_i^*KK^\top = 0$ and also $P_i^*KK^\top P_i^* =0$. Thus, $P_i^*K=K^\top P_i^*=0$.

Further, the singular value decomposition (SVD) of $K$ is given by $K=U_k\Lambda_kV_k^\top$ which implies that $KK^\top=U_k\Lambda^2_kU_k^\top$. However, since $KK^\top$ is a projection matrix, it has eigenvalues $0$ or $1$, and thus $\Lambda_k$ is a diagonal matrix with entries either 0 or 1. Finally, as $K$ is a square matrix, $U_k$ and $V_k$ are unitary matrices which form an orthonormal basis.

Next, we prove that $KPK^\top$ is a projection matrix. By following Corollary \ref{cor: stabpost1} and noting that $W_{-i} =K^\top\Sigma_x^\frac{1}{2}V_{-i}\Sigma_x^{\frac{1}{2}}K $, the optimization problem for sender $-i$, i.e., $\min_{I\succeq P\succeq O} Tr(W_{-i}P)$, can equivalently be written as
\begin{align}\label{eqn:equlib_2}
    \min_{I\succeq P\succeq O} Tr(K^\top\Sigma_x^\frac{1}{2}V_{-i}\Sigma_x^{\frac{1}{2}}KP).
\end{align}
As given in (\ref{eqn:soln_tamura}), the solution to the SDP problem in (\ref{eqn:equlib_2}) is a projection matrix $P$ which is given by 
\begin{align}\label{eqn:projection_mat}
  P=Q_{-i}^{-}(Q_{-i}^{-\top} Q_{-i}^{-})^{-1}Q_{-i}^{-\top}, 
\end{align}
where $Q_{-i}^{-}$ is a matrix consisting of eigenvectors corresponding to the negative eigenvalues of $W_{-i}$. Note that $\mathcal{N}(K)\subseteq\mathcal{N}(W_{-i})$ where $\mathcal{N}(\cdot)$ denotes the null space of a matrix. Similarly, we have $\mathcal{R}(W_{-i})\subseteq\mathcal{R}(K)$, where $\mathcal{R}(\cdot)$ denotes the range space of a matrix. Thus, any eigenvector corresponding to the non-zero eigenvalues of $W_{-i}$ can be expressed as a linear combination of columns of $V_k$ that has non-zero singular values. Let us assume that the number of negative eigenvalues of $W_{-i}$ is $\ell$ (or equivalently $Q_{-i}^{-}$ has rank $\ell$). If we assume that $rank(\Lambda_k)=rank(I-P^*)=k'$, we would have $k'$ columns of $V_k$ which correspond to non-zero singular values. We note that by changing the columns of $U_k$ and $V_k$, it is possible to change the order of the singular values in $\Lambda_k$. Without loss of generality, we assume that $\Lambda_k$ has the following form  
\begin{align*}
   \Lambda_k=\begin{bmatrix}I & O\\
    O & O\end{bmatrix},
\end{align*}
where $I\in\mathbb{R}^{k'\times k'}$. Since $\mathcal{R}(W_{-i})\subseteq\mathcal{R}(K)$, it follows that $\ell\leq k'$. Let $P=C\Lambda_p C^\top$ be the eigen-decomposition of $P$ where $C$  consists of orthogonal eigenvectors of $P$. Similarly, by arranging the columns of $C$, we can bring $\Lambda_p$ to the form 
\begin{align*}
   \Lambda_p=\begin{bmatrix}I & O\\
    O & O\end{bmatrix},
\end{align*}
where $I\in\mathbb{R}^{p\times p}$. Since $P$ is a projection matrix, it defines a subspace. Any vector $\boldsymbol{v}$ chosen from this subspace is an eigenvector of $P$ corresponding to eigenvalue 1 as we have $P\boldsymbol{v}= \boldsymbol{v}$. Any vector $\boldsymbol{u}$ that is orthogonal to that subspace is an eigenvector with eigenvalue 0. From the above discussion it follows that there exists a matrix M such that $C=V_kM$ with
\begin{align*}
    M=\begin{bmatrix}E & O\\
    O & F\end{bmatrix},
\end{align*}
where $E\in\mathbb{R}^{\ell\times\ell}$. The block structure follows from the fact that there exists $\ell \leq k'$ vectors in $V_k$ (corresponding to the non-zero singular values) that can span the space of $Q_{-i}^{-}$, and the rest of the vectors in $V_k$ would span the space perpendicular to the space spanned by these $\ell$ vectors. The latter follows from the fact that the columns of $V_k$ are orthonormal. Further, since $C^\top C=V_k^\top V_k=I$, we have $M^\top M=I$. Thus, we have
\begin{align}\label{eqn:projection}
    KPK^\top KP K^\top&=U\Lambda_k V_k^\top C\Lambda_p C^\top V_k\Lambda_kU^\top U\Lambda_k V_k^\top C\Lambda_p C^\top V_k\Lambda_kU^\top\nonumber\\
    &\overset{(a)}=U\Lambda_kV_k^\top C\Lambda_pC^\top V_k\Lambda_k V_k^\top C\Lambda_p C^\top V_k\Lambda_kU^\top\nonumber\\
    &=U\Lambda_kV_k^\top C\Lambda_p M^\top\Lambda_k M \Lambda_p C^\top V_k\Lambda_kU^\top\nonumber\\
    &\overset{(b)}=U\Lambda_kV_k^\top CM^\top\Lambda_p\Lambda_k \Lambda_p MC^\top V_k\Lambda_kU^\top\nonumber\\
    &\overset{(c)}= U\Lambda_kV_k^\top CM^\top\Lambda_pMC^\top V_k\Lambda_kU^\top\nonumber\\
    &\overset{(d)}=U\Lambda_k V^\top_k C\Lambda_p C^\top V_k\Lambda_k U^\top\nonumber\\
    &= KPK^\top,
\end{align}
where (a) follows from the facts that $U^\top U=I$ and $\Lambda_k^2=\Lambda_k$, (b) follows since $M$ is a block diagonal matrix where block $E$ has size $\ell\times \ell$ and $\Lambda_p$ has also rank $\ell$, they commute with each other, (c) follows from the observation that $\Lambda_k$ and $\Lambda_p$ are diagonal matrices with entries as 0 or 1, and $rank(\Lambda_k)\geq rank(\Lambda_p)$, and finally (d) follows from the fact that $M^\top\Lambda_pM=\Lambda_pM^\top M=\Lambda_p$.

By using (\ref{eqn:projection}), we can finally show that $P_i^*+KPK^\top$ is a projection matrix, i.e.,
\begin{align*}
    (P_i^*+KPK^\top)(P_i^*+KPK^\top)=P_i^*+KPK^\top.
\end{align*}
This is true for any P which is a solution of (\ref{eqn:equlib_2}), and thus is true for $\tilde{P}$ in (\ref{eq: eqpost}). Therefore, $P'=P_i^*+K\Tilde{P}K^\top$ is a projection matrix. Now consider the linear policies for both senders given by $\eta_1(\boldsymbol{x})=\eta_2(\boldsymbol{x})=L\boldsymbol{x}$. Then the posterior induced by such policies is given by $\Sigma_xL^\top(L\Sigma_xL^\top)^{\dagger}L\Sigma_x$. The stable posterior in (\ref{eq: eqpost}) can be achieved by $L^\top=\Sigma_x^{-\frac{1}{2}}U_{P'}\Lambda_{P'}$ since  $\mathbb{E}[\mathbb{E}[\boldsymbol{x}|\boldsymbol{y_{-i}}]\mathbb{E}[\boldsymbol{x}|\boldsymbol{y_{-i}}]^\top]=\mathbb{E}[\mathbb{E}[\boldsymbol{x}|\boldsymbol{y_i},\boldsymbol{y_{-i}}]\mathbb{E}[\boldsymbol{x}|\boldsymbol{y_i},\boldsymbol{y_{-i}}]^\top]=\Sigma^*$, where $\boldsymbol{y_i}=L\boldsymbol{x}$.
Thus, the Nash equilibrium posterior can be achieved by using the linear strategies.

Further, consider $\boldsymbol{y_i} = \eta_i'(\boldsymbol{x}) =G_i L\boldsymbol{x}$. By just observing $\boldsymbol{y}_1$, corresponding posterior formed by the receiver, $\Sigma_1^*$, is given by
\begin{align}
    \Sigma_1^*= \Sigma_x(G_1 L)^\top((G_1 L)\Sigma_x(G_1 L)^\top)^{\dagger}(G_1 L)\Sigma_x= \Sigma_x  L^\top G_1^\top(G_1 L\Sigma_x L^\top G_1^\top)^{\dagger}G_1 L\Sigma_x.
\end{align}
We note that $(AB)^{\dagger} = B^{\dagger}A^{\dagger}$ if $A$ has orthonormal columns, or similarly, $B$ has orthonormal rows. Then, we have
\begin{align}
    \Sigma_1^*= \Sigma_x  L^\top G_1^\top (G_1^\top)^{\dagger}( L\Sigma_x L^\top )^{\dagger}G_1^{\dagger}G_1 L\Sigma_x.
\end{align}
Since $G_1^{\dagger}G_1= I$ and $G_1^\top (G_1^\top)^{\dagger}=I$, we obtain $\Sigma_1^*$ as
\begin{align}\label{eqn_sigma_1}
    \Sigma_1^*= \Sigma_x  L^\top ( L\Sigma_x L^\top )^{\dagger} L\Sigma_x,
\end{align}
which is the same Nash equilibrium posterior $\Sigma^*$. Additionally, the innovation in observing $\boldsymbol{y_2}$, after observing $\boldsymbol{y_1}$ becomes $\boldsymbol{\Tilde{y}_2}=\boldsymbol{y_2}-\mathbb{E}[\boldsymbol{y_2}|\boldsymbol{y_1}]=0$. 
Thus, we have $\mathbb{E}[\boldsymbol{x}|\boldsymbol{y_1},\boldsymbol{y_2}] =\mathbb{E}[\boldsymbol{x}|\boldsymbol{y_1}]$. As the posterior estimation after observing only $\boldsymbol{y_1}$, i.e., $\Sigma_1^*$ in (\ref{eqn_sigma_1}), is equal to Nash equilibrium posterior $\Sigma^*$, and the innovation of observing $\boldsymbol{y_2}$ is equal to 0, we conclude that the signaling policies $\eta_i'(\boldsymbol{x})=G_i L\boldsymbol{x}$ also form Nash equilibrium strategies.   
\end{Proof}
For ease of exposition, for the rest of this paper, among these different strategies, we choose the Nash equilibrium policy as the symmetric one $\eta_1(\boldsymbol{x})=\eta_2(\boldsymbol{x})=L\boldsymbol{x}$. Before we
generalize our results to the multiple sender case with $m>2$
in the following subsection, we consider a sequential single sender game.

\subsection{Sequentially adding a sender to the one-stage game}\label{subsect:3_C}

It might be of interest to study the effect of adding a sender to an existing single-sender one-stage game, and thus we present a solution to such a problem in this subsection and highlight its relation to the Nash Equilibrium obtained in the previous section. Consider a single-stage, base strategic communication game with a single sender and a single receiver. Following the notation described in Section~\ref{Sect:Strategic_com}, a policy pair $(\tilde{\eta}_1^*,\tilde{\gamma^*})$ is the Stackelberg equilibrium of this base game, with the sender as the leader, if
\begin{align}\label{eq: basestack} \tilde{\eta_1}^*&=\argmin_{\eta_1\in\Omega(\cdot)} J_{1}(\eta_1,\tilde{\gamma}^*(\eta_1)),\\
    \tilde{\gamma}^*&=\argmin_{\tilde{\gamma}\in\Tilde{\Gamma}} J_r(\eta_1,\tilde{\gamma}(\eta_1)),
\end{align}
where we assume uniqueness in the receiver's best response (which holds in our case). For this game with a single sender (denoted by sender 1), the equilibrium described in (\ref{eq: basestack}) can be obtained by solving the following trace minimization \cite{sayin2019hierarchical,tamura2018bayesian}:
\begin{align*} \min_{\Sigma_x\succeq S_1\succeq O}Tr(V_1S_1),
\end{align*}
i.e., line 4 of Algorithm~\ref{alg: postcalc}. Here, the equilibrium posterior covariance is of the form $\Sigma_1^*=\Sigma_x^{\frac{1}{2}}P_1^*\Sigma_x^{\frac{1}{2}}$ for some projection matrix $P_1^*$.\footnote{We assume that, if the sender has two policies which result in the same cost value, a policy which reveals only the required information (resulting in lower rank posterior covariance) is chosen.} Hence, linear strategies of the form $\tilde{\eta}_{1}^*(\boldsymbol{x})=A\boldsymbol{x}$ where $A^\top=\Sigma_x^{-\frac{1}{2}}U_{1}^*\Lambda_{1}^*$ and $P_1^*=U_1^*\Lambda_1^*U_1^{*\top}$ achieve the equilibrium.

Now suppose, after this commitment of $\tilde{\eta}_1^*$ by sender $1$, sender $2$ enters the game. For this fixed strategy ($\tilde{\eta}_1^*$) of sender 1, we can define a Stackelberg equilibrium of the game between sender $2$ and the receiver. A policy pair ($\eta_2^*,\gamma^*$) is said to be in equilibrium if
\begin{align}\label{Eq: stack}
    \eta_2^*(\tilde{\eta}_1^*)&=\argmin_{\eta_2\in\Omega(\tilde{\eta}_1^*)}J_{2}(\tilde{\eta}_1^*,\eta_2(\tilde{\eta}_1^*),\gamma^*),\\
    \gamma^*(\tilde{\eta}_1^*,\eta_2)&=\argmin_{\gamma\in\Gamma} J_r(\tilde{\eta}_1^*,\eta_2(\tilde{\eta}_1^*),\gamma(\tilde{\eta}_1^*,\eta_2)),\label{Eq: stack_2}
\end{align}
where we assumed uniqueness of the receiver's response as stated in (\ref{eq: Nash}). Equivalently, sender 2 faces a problem of Stackelberg equilibrium computation when the receiver has a side information (due to prior commitment of sender 1). Utilizing this equilibrium definition and the techniques developed in this section, the optimization problem of sender 2 considering the receiver's best response is lower bounded by the following SDP:
\begin{align*}
    \min_{\Sigma_x\succeq S_2\succeq\Sigma'}Tr(V_2S_2).
\end{align*}
Note that this is a single-agent optimization problem because the second sender enters the game after the first one who is already in the game and has determined its policy. In this scenario, we assume that the first sender \textit{does not} know the existence of the second sender while it commits to a policy. This is a natural situation that might occur in real-world markets where a player enters and disrupts an existing monopoly. We defer a discussion on the setting of the anticipated second player's entry to Section~\ref{Sect:Discussion}. Since the above optimization problem mimics a sequential approach described in Algorithms~\ref{algorithm2} and \ref{alg: postcalc}, the same arguments can be used to identify the posterior reached in this sequential setting. 
We now show that sender 2 can attain an equilibrium in such a game by using linear policies.
\begin{theorem}\label{th: seqcommit}
Consider the scenario where the first sender does not know the existence of the second sender while it commits to a policy. Then, let $\Sigma^*$ be the stable posterior covariance obtained from Algorithm~\ref{algorithm2} and $\Sigma_1^*$ be the equilibrium posterior covariance of the base game described in (\ref{eq: basestack}). Then, we have $\eta_{2}^*(\boldsymbol{x})=B\boldsymbol{x}$ with $B^\top=(\Sigma_x-\Sigma_{1}^*)^{\dagger\frac{1}{2}}U_{2}^*\Lambda_{2}^*$ where $\Tilde{P}_{2}=U_{2}^*\Lambda_{2}^*U_{2}^{*\top}$ is a projection matrix such that
$\Tilde{P}_{2}=(\Sigma_x-\Sigma_1^*)^{-\frac{1}{2}}(\Sigma^*-\Sigma_1^*)(\Sigma_x-\Sigma_1^*)^{-\frac{1}{2}}$ yields an equilibrium (as defined in (\ref{Eq: stack}) and (\ref{Eq: stack_2})) with posterior covariance $\Sigma^*$, in (\ref{eq: eqpost}) when we solve senders' optimization problems sequentially, with sender $1$ moving first within the general class of policies.
\end{theorem}
\begin{Proof}
The optimization problem faced by Sender 2 after the commitment of Sender 1 is
\begin{align*}
    \min_{\Sigma_x\succeq S_2\succeq\Sigma_1^*}Tr(V_2S_2).
\end{align*}
This is exactly the optimization problem solved in lines 8 and 9 of Algorithm~\ref{alg: postcalc}. Hence $\Sigma^*$ is the equilibrium posterior for this sequential game as defined in (\ref{Eq: stack}) and (\ref{Eq: stack_2}). Further, $\Sigma^*=\Sigma_1^*+(\Sigma_x-\Sigma_1^*)^{\frac{1}{2}}\tilde{P}(\Sigma_x-\Sigma_1^*)^{\frac{1}{2}}$ for some projection matrix $\tilde{P}$ as proved in Corollary~\ref{cor: stabpost1}. 

Now, consider $\eta_{2}^*(\boldsymbol{x})=B\boldsymbol{x}$ with $B^\top=(\Sigma_x-\Sigma_{1}^*)^{\dagger\frac{1}{2}}U_{2}^*\Lambda_{2}^*$ where $\Tilde{P}_{2}=U_{2}^*\Lambda_{2}^*U_{2}^{*\top}$ and the estimation process $\boldsymbol{x}=\mathbb{E}[\boldsymbol{x}|\boldsymbol{y}_1,\boldsymbol{y}_2]$. Viewing this as a sequential observation process of $\boldsymbol{y}_1$ and $\boldsymbol{y}_2$. After observing $\boldsymbol{y_1}$, the corresponding posterior formed by the receiver is $\boldsymbol{\hat{x}_1}=\mathbb{E}[\boldsymbol{x}|\boldsymbol{y_1}]$, which is given by
\begin{align}
    \boldsymbol{\hat{x}_1}=\mathbb{E}[\boldsymbol{x}|\boldsymbol{y_1}]= \Sigma_x A^\top(A\Sigma_x A^\top)^{\dagger}A\boldsymbol{x}.
\end{align}

Then, the receiver can estimate $\boldsymbol{\Tilde{x}}=\boldsymbol{x}-\boldsymbol{\hat{x}_1}$ using the \textit{innovation} in observing $\boldsymbol{y_2}$, i.e., $\boldsymbol{\Tilde{y}_2}=\boldsymbol{y_2}-\mathbb{E}[\boldsymbol{y_2}|\boldsymbol{y_1}]=\boldsymbol{y_2}-B\boldsymbol{\hat{x}_1}$. Since $\mathbb{E}[\boldsymbol{\Tilde{y}_2}]=0$, we have
\begin{align*}
    \mathbb{E}[\boldsymbol{x}-\boldsymbol{\hat{x}_1}|\boldsymbol{\Tilde{y}_2}]=\mathbb{E}[\boldsymbol{x}|\boldsymbol{y_1},\boldsymbol{y_2}]-\boldsymbol{\hat{x}_1}=C_{\boldsymbol{\Tilde{x}}\boldsymbol{\Tilde{y}_2}}C^{\dagger}_{\boldsymbol{\Tilde{y}_2}\boldsymbol{\Tilde{y}_2}}\boldsymbol{\Tilde{y}_2},
\end{align*}
where
\begin{align*}
    C_{\boldsymbol{\Tilde{y}_2}\boldsymbol{\Tilde{y}_2}}&=\mathbb{E}[(\boldsymbol{y_2}-B\boldsymbol{\hat{x}_1})(\boldsymbol{y_2}-B\boldsymbol{\hat{x}_1})^\top]=B\mathbb{E}[(\boldsymbol{x}-\boldsymbol{\hat{x}_1})(\boldsymbol{x}-\boldsymbol{\hat{x}_1})^\top]B^\top,\\
    C_{\boldsymbol{\Tilde{x}}\boldsymbol{\Tilde{y}_2}}&=\mathbb{E}[(\boldsymbol{x}-\boldsymbol{\hat{x}_1})(\boldsymbol{x}-\boldsymbol{\hat{x}_1})^\top]B^\top.
\end{align*}
Since $\mathbb{E}[\boldsymbol{x}\boldsymbol{\hat{x}_1}^\top]=\mathbb{E}[\mathbb{E}[\boldsymbol{x}\boldsymbol{\hat{x}_1}^\top|\boldsymbol{y_1}]]=\mathbb{E}[\boldsymbol{\hat{x}_1}\boldsymbol{\hat{x}_1}^\top]$, we have $\mathbb{E}[(\boldsymbol{x}-\boldsymbol{\hat{x}_1})(\boldsymbol{x}-\boldsymbol{\hat{x}_1})^\top]=\Sigma_x-\Sigma_1$ where $\Sigma_1=\mathbb{E}[\boldsymbol{\hat{x}_1}\boldsymbol{\hat{x}_1}^\top]$. Thus, we write $\mathbb{E}[\boldsymbol{x}|\boldsymbol{y_1},\boldsymbol{y_2}]$ as
\begin{align}\label{eq: postseqest}
    \mathbb{E}[\boldsymbol{x}|\boldsymbol{y_1},\boldsymbol{y_2}]=\boldsymbol{\hat{x}_1}+(\Sigma_x-\Sigma_1)B^\top(B(\Sigma_x-\Sigma_1)B^\top)^{\dagger}(\boldsymbol{y_2}-B\boldsymbol{\hat{x}_1}).
\end{align}
We use $\boldsymbol{\hat{x}} =\mathbb{E}[\boldsymbol{x}|\boldsymbol{y_1},\boldsymbol{y_2}]$ in (\ref{eq: postseqest}) to find the posterior $\mathbb{E}[\boldsymbol{\hat{x}}\boldsymbol{\hat{x}}^\top]$ attained by the linear strategies. Since $\mathbb{E}[\boldsymbol{\hat{x}_1}(\boldsymbol{y_2}-B\boldsymbol{\hat{x}_1})^\top]=\mathbb{E}[\mathbb{E}[\boldsymbol{\hat{x}_1}\boldsymbol{x}^\top|\boldsymbol{y_1}]]B^\top-\mathbb{E}[\boldsymbol{\hat{x}_1}\boldsymbol{\hat{x}_1}]B^\top=0$, we have
\begin{align}\label{eqn:thm_1}
    \mathbb{E}[\boldsymbol{\hat{x}}\boldsymbol{\hat{x}}^\top]=\Sigma_1+(\Sigma_x-\Sigma_1)B^\top(B(\Sigma_x-\Sigma_1)B^\top)^{\dagger}B(\Sigma_x-\Sigma_1).
\end{align}

Next, we define $\mathcal{C}:=(\Sigma_x-\Sigma_1)^\frac{1}{2}B^\top$. Then, $\mathbb{E}[\boldsymbol{\hat{x}}\boldsymbol{\hat{x}}^\top]$ in (\ref{eqn:thm_1}) becomes 
\begin{align*}
    \mathbb{E}[\boldsymbol{\hat{x}}\boldsymbol{\hat{x}}^\top]=\Sigma_1+(\Sigma_x-\Sigma_1)^{\frac{1}{2}}P'(\Sigma_x-\Sigma_1)^{\frac{1}{2}},
\end{align*}
where $P'=\mathcal{C}(\mathcal{C}^\top\mathcal{C})^{\dagger}\mathcal{C}^\top$ is a projection matrix. Thus, the posterior covariance obtained from linear strategies follows an equation similar to equilibrium posterior given by (\ref{eq: eqpost}). 
In particular, for the equilibrium posterior $\Sigma^*$ in (\ref{eq: eqpost}), we define $P_1^*=U_1^*\Lambda_1^*U_1^{*\top}$, and $\Tilde{P}=U_{2}^*\Lambda_{2}^*U_{2}^{*\top}$. Substituting $\mathcal{C}=U_{2}^*\Lambda_{2}^*$ we have $P'=U_{2}^*\Lambda_{2}^*U_{2}^{*\top}=\Tilde{P}$. Similarly, let $\mathcal{D}=U_1^*\Lambda_1^*=\Sigma_x^{\frac{1}{2}}A^\top$, then $\Sigma_1=\Sigma_x A^\top(A\Sigma_x A^\top)^{-\dagger}A\Sigma_x=\Sigma_x^\frac{1}{2}\mathcal{D}(\mathcal{D}^\top\mathcal{D})^{\dagger}\mathcal{D}^\top\Sigma_x^{\frac{1}{2}}=\Sigma_x^{\frac{1}{2}}P_1^*\Sigma_x^{\frac{1}{2}}$. Thus, for these choices of $A$ and $B$, we have
\begin{align*}
    \mathbb{E}[\boldsymbol{\hat{x}}\boldsymbol{\hat{x}}^\top]=\Sigma_x^{\frac{1}{2}}P_{1}^*\Sigma_x^{\frac{1}{2}}+(\Sigma_x-\Sigma_{1})^{\frac{1}{2}}\tilde{P}(\Sigma_x-\Sigma_{1})^{\frac{1}{2}},
\end{align*}
which is the same as (\ref{eq: eqpost}). Thus, we obtain the linear policies
$\eta_1(\boldsymbol{x})=A\boldsymbol{x}$ and $\eta_2(\boldsymbol{x})=B\boldsymbol{x}$ where $A^\top=\Sigma_x^{-\frac{1}{2}}U_{1}^*\Lambda_{1}^*$ and $B^\top=(\Sigma_x-\Sigma_{1})^{\dagger\frac{1}{2}}U_{2}\Lambda_{2}$ that yield the equilibrium posterior $\Sigma^*$ in (\ref{eq: eqpost}) within the general class of policies.

\end{Proof}
We note that a different order of commitment would lead to a different equilibrium posterior which can be obtained by interchanging $V_1$, $V_2$ in lines 4 and 8 of Algorithm~\ref{algorithm2}, but the analysis would follow similarly. Thus, the order of commitment is important in sequential games. In the next subsection, we consider the Nash equilibrium strategies for a general $m\geq 2$ sender game. 

\subsection{Multiple-Sender Strategic Communication}\label{Subsect:Multiple_senders}
We now extend the results from the 2-sender game discussed heretofore to an $m$-sender game, structured in a similar way, as an $m+1$ player game where $m$ players (referred to as senders), who have access to an underlying random state of the world $\boldsymbol{x}$, denoted by $i$ for $i=[m]$, compete to disclose information to a receiver, denoted by $r$, whose action affects the costs of all the players. Each player $i=1,\hdots, m,r$ has an expected cost function given by

\begin{align*}
J_i(\eta_1,\hdots,\eta_m;\gamma)=\mathbb{E}[L_i(\boldsymbol{x};\boldsymbol{u}=\gamma(I_r))]=\mathbb{E}[||Q_i\boldsymbol{x}+R_i\boldsymbol{u}||^2],
\end{align*}
where $\boldsymbol{u}$ is the receiver's action and $L_i(\boldsymbol{x};\boldsymbol{u}=\gamma(I_r))$ is given in  (\ref{eqn:loss_fnc}). Our goal is to find hierarchical equilibrium as defined in (\ref{eq: Nash}). We first start by identifying stable posteriors for the $m$-sender game.
\begin{lemma}\label{lem: nsenstable}
A posterior $\Sigma^{'}$ is stable for all senders $i\in[m]$ if and only if $W_i^*=(\Sigma_x-\Sigma^{'})^{\frac{1}{2}}V_i(\Sigma_x-\Sigma^{'})^{\frac{1}{2}}\succeq O$ for all $i\in[m]$.
\end{lemma}
The proof follows from extending Proposition \ref{Remark_2} to the $m$-sender case. For the 2-sender case, Lemma \ref{lem: nsenstable} showed that there can be multiple posteriors which are stable. This hints naturally at the possibility of multiple equilibria in the general $m$-sender communication game as well. Further, if $\Sigma^{'}$, $\Sigma^{''}$ are both stable and achievable posteriors (thus, equilibrium posteriors) for all senders $i\in[m]$ and can be ordered as $\Sigma^{''}\succeq\Sigma^{'}$, then all senders $i\in[m]$ prefer $\Sigma^{'}$ over $\Sigma^{''}$. Thus not all equilibrium posteriors are partially informative as defined earlier. We propose Algorithm \ref{alg: npostcalc} as a way to achieve a set of partially informative equilibria, which is proved using the following proposition.
\begin{algorithm}[t]
\caption{Finding an Equilibrium Posterior for the $m$-Sender Game via SDP}\label{alg: npostcalc}
\begin{algorithmic}[1]
\State{\textbf{Parameters:} $\Sigma_x$, $N=\{k_1,k_2,\dots,k_m\}\in \bar{\mathcal{N}}$}
\State{\textbf{Compute} $V_i$ $\forall i\in[m]$ by using (\ref{eqn:Vi})}
\State{\textbf{Variables:} $S_i\in\mathbb{S}^p$ with $S_i\succeq O$ $\forall i\in[m]$}
\State{$\Sigma_{-i}=O$}
\For{$j=1,\hdots, m$}
\State $i \gets k_j$
\State{\textbf{Solve:} $ \min_{\Sigma_x\succeq S_{i}\succeq \Sigma_{-i}} Tr(V_i S_{i}),$ by using CVX toolbox in MatLab}
\State{\textbf{Solve:} $ \min_{\Sigma_x\succeq S'_{i}\succeq \Sigma_{-i}} ||\Sigma_x^{-\frac{1}{2}}S'_{i}\Sigma_x^{-\frac{1}{2}}||_*,$} 
\State{$\qquad \quad$s.t. $Tr(V_iS_i)\succeq Tr(V_iS_i')$ by using CVX toolbox in MatLab} 
\State $ \Sigma_{-i} \gets S'_{i}$
\EndFor
\State{\textbf{Return:} $\Sigma^* \gets  S'_{m}$}
\end{algorithmic}
\end{algorithm}

\begin{prop}\label{Remark_3_alt}
Let the stable posterior obtained from Algorithm \ref{alg: npostcalc} be $\Sigma^*$. There does not exist a stable posterior $\Sigma'$ such that $ \Sigma' \succeq \Sigma^* $ and $\Sigma'$ is preferred by all senders compared to the posterior $\Sigma^*$.
\end{prop}
\begin{Proof}
We prove this by contradiction. Assume that there is a stable posterior $\Sigma'$ which is preferred by all senders compared to $\Sigma^*$ such that we have $\Sigma' \succeq \Sigma^*$. Let us consider the iteration $j\in \{1,\dots,m\}$ of Algorithm~\ref{alg: npostcalc} that gives $\Sigma^*$ as the solution for the first time. If there were such a $\Sigma'$ that is preferred by all senders compared to $\Sigma^*$, such a $\Sigma'$ would be a solution to the optimization problem at iteration $j'\leq j$. If it is stable, no sender would add any more information, and thus the Algorithm~\ref{alg: npostcalc} would have returned $\Sigma'$ instead of $\Sigma^*$. Hence, we reach a contradiction and there does not exist a stable posterior $\Sigma'$ that is preferred by all senders compared to $\Sigma^*$.
\end{Proof}

As a result of Proposition~\ref{Remark_3_alt}, Algorithm~\ref{alg: npostcalc} can identify at least one partially informative posterior. 
Note that Algorithm \ref{alg: npostcalc} takes an additional ordering parameter $N=\{k_1,k_2,\dots,k_m\}\in\bar{\mathcal{N}}$ as an input which is a particular ordering of $V_i$'s in the algorithm iterations. Since there are at most $m!$ possible orderings, there can be $m!$ partially informative posteriors that can be achieved by Algorithm \ref{alg: npostcalc}. By following a similar procedure as in the 2-sender case, we provide the structure of equilibrium posterior in the following corollary. 
\begin{corollary}\label{cor: npoststruct}
Stable posteriors obtained from Algorithm \ref{alg: npostcalc} are of the form $\Sigma^*=\Sigma_m$, where $\Sigma_m$ can be obtained by the recursion
\begin{align}
    \Sigma_j=\Sigma_{j-1}+(\Sigma_x-\Sigma_{j-1})^{\frac{1}{2}}P_j(\Sigma_x-\Sigma_{j-1})^\frac{1}{2},
\end{align}
where $2\leq j \leq m $, and $\Sigma_1=\argmin_{\Sigma_x\succeq S\succeq O}Tr(V_1S)$ for some projection matrix sequence $P_1,\hdots, P_n$.
\end{corollary}
\begin{Proof}
This can be proven by induction. From the structure of Algorithm \ref{alg: npostcalc}, and Corollary \ref{cor: stabpost1}, the statement is already true for iterations $j=1,2$. Assume that the statement is true for $j=k> 2$, which returns $\Sigma_k$. For the next iteration $k+1$, the optimization problem solved by the algorithm is $\Sigma_{k+1}=\argmin_{\Sigma_x\succeq S\succeq\Sigma_k}Tr(V_{k+1}S)$. By using Lemma~\ref{lem: transf}, this optimization problem can be written as 
\begin{align}\label{eqn:corrolary_2}
    \min_{I\succeq Z\succeq O} Tr(W_{k+1}Z)+Tr(V_{k+1}\Sigma_k).
\end{align}
Since the second term ($Tr(V_{k+1}\Sigma_k)$) is constant and does not affect the optimization problem, the problem in (\ref{eqn:corrolary_2}) is equivalent to solving $\min_{I\succeq Z\succeq O} Tr(W_{k+1}Z)$. Since the solution to this SDP has a projection matrix form, the above recursion is true for $j=k+1$, and hence by induction, the statement is true for any $m\geq 2$.
\end{Proof}
We note that the posterior achieved by Algorithm \ref{alg: npostcalc} is stable. This is because $S_i$ obtained after loop $j$ in Algorithm \ref{alg: npostcalc} satisfies $(\Sigma_x-S_i)^{\frac{1}{2}}V_i(\Sigma_x-S_i)^{\frac{1}{2}}\succeq O$. Since from Lemma \ref{lem: greatersig} any $S\succeq S_i$ satisfies $(\Sigma_x-S)^{\frac{1}{2}}V_i(\Sigma_x-S)^{\frac{1}{2}}\succeq O$, we have that any $S\succeq S_i$ is also stable for sender $i$. 
\begin{lemma}\label{lem: npostproj}
The matrix $S_i'$ obtained after each iteration of Algorithm \ref{alg: npostcalc} is of the form $\Sigma_x^{\frac{1}{2}}\tilde{P}_i\Sigma_x^{\frac{1}{2}}$ for some projection matrix $\tilde{P}_i$. In particular, this is also true for the final posterior returned by Algorithm \ref{alg: npostcalc}.
\end{lemma}
\begin{Proof}
We prove this by induction. It can be seen from Theorem \ref{th: Nasheq} that the statement is true for the first 2 iterations, i.e., $j=1,2$, of Algorithm \ref{alg: npostcalc}. Assume that the statement is true for $j=k>2$, which implies the matrix $S_k'=\Sigma_x^\frac{1}{2}\tilde{P}_k\Sigma_x^{\frac{1}{2}}$. From Corollary \ref{cor: npoststruct}, we can see that $S_{k+1}'=S_k'+(\Sigma_x-S_k')^\frac{1}{2}P_{k+1}(\Sigma_x-S_k')^\frac{1}{2}$ where $P_{k+1}$ is a projection matrix. Using the structure of $S_k'$ and Theorem \ref{th: Nasheq}, it can be seen that $S_{k+1}'=\Sigma_x^{\frac{1}{2}}\tilde{P}_{k+1}\Sigma_x^\frac{1}{2}$. Hence the statement is true for $j=k+1$ which completes the proof.
\end{Proof}
Next, we state a theorem to achieve such a stable posterior by using linear policies.
\begin{theorem}\label{Thm_3}
Let $\Sigma^*$ be a stable posterior covariance obtained from Algorithm \ref{alg: npostcalc}. The policies $\eta_i(\boldsymbol{x})=L\boldsymbol{x}$, $\forall i$, where $L^\top=\Sigma_x^{-\frac{1}{2}}U_{P'}\Lambda_{P'}$ and $P'=\Sigma_x^{-\frac{1}{2}}\Sigma^*\Sigma_x^{-\frac{1}{2}}=U_{P'}\Lambda_{P'}U_{P'}^\top$ are Nash equilibrium policies for the $m$-sender communication game.
\end{theorem}
\begin{Proof}
From Lemma \ref{lem: npostproj} it can be seen that the structure of stable posterior matrix obtained from Algorithm \ref{alg: npostcalc} is of the form $\Sigma^*=\Sigma_x^\frac{1}{2}P'\Sigma_x^\frac{1}{2}$, and thus $\Sigma^*$ can be reached by $\eta_i(\boldsymbol{x})=L\boldsymbol{x}=\Lambda_{P'}U_{P'}\Sigma_x^{-\frac{1}{2}}\boldsymbol{x}$.
\end{Proof}
As in the 2-sender game, senders can use different policies obtained by scaling of the policy given in Theorem~\ref{Thm_3} with full rank matrices that have orthonormal columns, but still reach the same posteriors and hence are NE policies.

As a summary, we have generalized above the results for the 2-sender strategic communication game setting in Subsections~\ref{subsect:3_A} and \ref{subsect:3_B} to a general $m>2$ sender case. We can obtain at most $m!$ different stable posteriors by using Algorithm~\ref{alg: npostcalc}. Further, these posteriors form partially informative equilibria, and hence cannot be dominated by any other more informative posteriors. Finally, since these posteriors can be achieved by using linear strategies, they are both stable and achievable posteriors, and thus they are equilibrium posteriors. 

 Now that we have characterized possible equilibrium posteriors, we return below to a natural question of interest 
which is the effect of multiple senders on information revelation. 
\subsection{Competition in Information Revelation}
We show here that competition increases information revelation in the following sense: Let $\Sigma^c$ be a posterior covariance obtained when the senders collaborate and minimize a (possibly weighted) sum of their payoffs.\footnote{One such collaborative minimization problem is provided as Example~6 in Section~\ref{Sect:Num_result}.} Let $\sigma^c$ be a set of all such possible Pareto optimal posterior covariances from the senders' perspective. Similarly, we denote by $\sigma^*$ as the set of all NE posteriors.
\begin{remark} If there is a posterior covariance 
obtained by a collaborative minimization of a common objective, and it is more informative than any NE solution, then it must also be a NE. In other words, letting $\Sigma^*\in\sigma^*$ and $\Sigma^c\in\sigma^c$, we have that if $\Sigma^c\succeq\Sigma^*$ then $\Sigma^c\in\sigma^*$.
\end{remark}

The proof of this remark follows directly by noting that if $\Sigma^*$ is an equilibrium posterior, any $\Sigma$ with $\Sigma_x\succeq\Sigma\succeq\Sigma^*$ is also a stable posterior. Further, such $\Sigma^c$ can be obtained by utilizing linear noiseless policies because it boils down to a single sender optimization problem provided in (\ref{eqn:lemma_4}), and thus solution techniques from \cite{tamura2018bayesian} can be used to obtain such policies. Hence, collaborative posterior set $\sigma^c$ cannot be more informative than the NE set $\sigma^*$. Further, adding senders is weakly beneficial for the receiver in the following sense.
\begin{prop}\label{Prop: eqshrink}
    Let $\sigma^*$ and $\sigma^{*'}$ be the sets of NE posteriors obtained when the sets of senders are $I$ and $I'$, respectively. If $I\subset I'$, then we have $\sigma^{*'}\subseteq \sigma^*$. 
\end{prop}
\begin{Proof}
    Let $\Sigma^{*'}\in\sigma^{*'}$. Then, $(\Sigma_x-\Sigma^{*'})^{\frac{1}{2}}V_i(\Sigma_x-\Sigma^{*'})^{\frac{1}{2}}\succeq0$ for all $i\in I'$. This implies that it is also true for all $i\in I$ since $I\subset I'$ which can be verified for all $\Sigma^{*'}\in\sigma^{*'}$, and as a result we have $\sigma^{*'}\subseteq \sigma^*$. 
\end{Proof}
By Proposition \ref{Prop: Fullinf}, full information revelation is always an equilibrium outcome and Proposition \ref{Prop: eqshrink} implies that the equilibrium set shrinks on adding another sender, hence intuitively we can see that the equilibrium shrinks towards full revelation in line with similar observation in the discrete state case \cite{gentzkow2017bayesian}. In the following section, we extend the single-stage game setting for the $m$-sender case to the multi-stage game setting. 

\section{Dynamic Information Disclosure}\label{Sect:Dynamic}

In this section we extend the formulation in the previous section to analyze a finite horizon hierarchical signaling game between multiple senders and a single receiver. 

Consider a dynamic state $\boldsymbol{x_{k}}\in\mathbb{R}^p$ that evolves according to an uncontrolled Markov process described by
\begin{align}
\boldsymbol{x_{k}}=A\boldsymbol{x_{k-1}}+\boldsymbol{w_{k-1}},
\end{align}
for each stage $k=1,\hdots,n$, where $\boldsymbol{x_k}$ is a zero-mean Gaussian process with $\boldsymbol{x_0}\sim \mathbb{N}(0,\Sigma_0)$ and $\boldsymbol{w_{k}}$ is a Gaussian noise vector sampled independently at every time instant from a distribution given by $\mathbb{N}(0,\Sigma_w)$. Thus, for each $k=1,\hdots,n$, the covariance of $\boldsymbol{x_k}$ is generated by the Lyapunov equation $\Sigma_k=A\Sigma_{k-1}A^\top+\Sigma_w$. After observing the state realization $ \boldsymbol{x_k}$, sender $i$ sends signal $ \boldsymbol{y^i_{k}}\in\boldsymbol{Y}_k^{i}$ to the receiver. Based on the received signals, at stage $k$, the receiver forms an estimate of $\boldsymbol{x_k}$ denoted by $\boldsymbol{u_k}\in \boldsymbol{U}_k$ where $\boldsymbol{Y}_k^{i}\subseteq \mathbb{R}^p$ and  $\boldsymbol{U}_k\subseteq \mathbb{R}^t$ are respectively the action spaces of sender $i$ and the receiver at time instant $k$.  

In this dynamic game,  the information structure of sender $i$ and the receiver at stage $k$ are $I_i(k)=\{\boldsymbol{x_{[k]}}\}$ and  $I_r(k)=\{ \boldsymbol{y^i_{[k]}} \}_{i\in[m]}$, respectively. At each stage $k$, each player $j$ has a possibly different finite-horizon quadratic loss function which is given by
\begin{align}
L_j(\boldsymbol{x_k};\boldsymbol{u_k}=\gamma_k(I_r(k)))=||Q_{j,k}\boldsymbol{x_k}+R_{j,k}\boldsymbol{u_k}||^2.
\end{align}
The game proceeds as follows. At stage $k$, each sender $i$ simultaneously commits to a policy $\eta^i_{[k]}:I_i(k)\rightarrow \boldsymbol{Y}^i_{k}$ such that $\boldsymbol{y^i}=\eta^i_{[k]}(\boldsymbol{x_{[k]}})$. Following this, receiver $r$ selects a strategy $\gamma_k:I_r(k)\rightarrow \boldsymbol{U_k}$ such that $\boldsymbol{u_k}=\gamma_k(\{ \boldsymbol{y^i_{[k]}} \}_{i\in[m]})$. Thus, at each stage, the game follows a solution concept of Nash equilibrium among senders followed by a Stackelberg response by the receiver. 
Explicitly, a tuple of signaling rules $(\{\eta^{i*}_{[n]}\}_{i\in[m]},\gamma^*_{[n]})$ achieve multistage Nash hierarchical equilibrium if
\begin{align}\label{eqn_nash_defn_mult_m}
    J_{i}(\eta^{i*}_{[k]},\eta^{-i*}_{[k]},\gamma_{[k]}^*(\eta^{i*}_{[k]},\eta^{-i*}_{[k]}))&\leq J_{i}(\eta^{i}_{[k]},\eta^{-i*}_{[k]},\gamma_{[k]}^*(\eta^{i}_{[k]},\eta^{-i*}_{[k]})),\quad\forall i\in[m],\hspace{2mm}\forall k\in[n],\nonumber\\
    \gamma_{[k]}^*(\{\eta^{i}_{[k]}\}_{i\in[m]})&=\argmin_{\gamma_k\in\Gamma_k} J_r (\{\eta^i_{[k]}\}_{i\in[m]},\gamma_{[k]}(\{\eta^i_{[k]}\}_{i\in[m]})),\quad\forall k\in[n],
\end{align}
where we assume in the second line of (\ref{eqn_nash_defn_mult_m}) that the minimization leads to a unique solution for the $m$-tuple of $\eta_i$'s. In (\ref{eqn_nash_defn_mult_m}), $J_j$ for $j=1,\hdots, m, r$ denote the players' respective expected cost functions, $\eta^{-i*}_{[k]}$ denotes strategy of all senders except sender $i$ at stage $k$ and $\Gamma_k$ denotes the receiver's strategy space which is all Borel measurable functions from $\mathbb{R}^{kp}$ to $\mathbb{R}^p$. As in this single-stage case, we consider finite horizon expected cost functions for all players $j=1,\hdots, m, r$ given by
\begin{align}\label{Eq: dynobjective}
    J_j(\{\eta^i_{[n]}\}_{i\in[m]};\gamma_{[n]})=\mathbb{E}\left[\sum_{k=1}^n L_j(\boldsymbol{x_k};\boldsymbol{u_k}=\gamma_k(I_r(k)))\right]=\mathbb{E}\left[\sum_{k=1}^n||Q_{j,k}\boldsymbol{x_k}+R_{j,k}\boldsymbol{u_k}||^2\right],
\end{align}
where $Q_{j,k}\in\mathbb{R}^{r\times p}$, $R_{j,k}\in\mathbb{R}^{r\times t}$ are cost matrices at stage $k$.  

We assume that $R_{r,k}'R_{r,k}$ is invertible. This implies that as in the single-stage case, the receiver's best response is linear in its posterior estimate at that stage, $\boldsymbol{\hat{x}_k}=\mathbb{E}[\boldsymbol{x_k}|\{\boldsymbol{y^i_{[k]}}\}_{i\in[m]}]$. Thus, the receiver's best response is given by
\begin{align}\label{eqn:dynamic_rec_cost}
    \gamma^*(\{\boldsymbol{y^i_{[k]}}\}_{i\in[m]})=-(R'_{r,k}R_{r,k})^{-1}R'_{r,k}Q_{r,k}\boldsymbol{\hat{x}_k}.
\end{align}
Incorporating (\ref{eqn:dynamic_rec_cost}) into sender $i$'s objective function, we then have to solve for an $n$-horizon $m$-player dynamic game for the Nash equilibrium polices $\eta_{[k]}^{i*}$. In particular, the objective function of sender $i$ at stage $k$ as the best response to the other senders' policies would involve the minimization problem
\begin{align}\label{Eq: simp1}
    \min_{\eta^i_k\in\Omega_k(\eta^{-i}_{[k]})}\sum_{k=1}^n\mathbb{E}[\|Q_{i,k}\boldsymbol{x_k}-R_{i,k}(R'_{r,k}R_{r,k})^{-1}R'_{r,k}Q_{r,k}\boldsymbol{\hat{x}_k}\|^2],
\end{align}
where $\Omega_{k}(\cdot)$ is the space of all Borel measurable senders' strategies at stage $k$ in the defined sense earlier. As in the single-stage case, (\ref{Eq: simp1}) is a functional optimization problem, and thus finding the NE among $m$ senders can be quite complex. To alleviate the difficulty encountered in any direct approach, we follow the path in the single-stage case and simplify the best response optimization by bounding it with a finite-dimensional optimization problem. For that, we rewrite (\ref{Eq: simp1}) as
\begin{align}\label{Eq: simp2}
\min_{\eta^i_k\in\Omega_k(\eta^{-i}_{[k]})}\mathbb{E}[\boldsymbol{x_k}^\top Q_{i,k}^\top Q_{i,k}\boldsymbol{x_k}]-2\mathbb{E}[\boldsymbol{\hat{x}_k}^\top\Lambda_k^{i\top}Q_{i,k}\boldsymbol{x_k}]+\mathbb{E}[\boldsymbol{\hat{x}_k}^\top\Lambda_k^{i\top}\Lambda_k^i\boldsymbol{\hat{x}_k}], 
\end{align}
where $\Lambda_k^i=R_{i,k}(R'_{r,k}R_{r,k})^{-1}R'_{r,k}Q_{r,k}$. The first term in (\ref{Eq: simp2}) does not depend on senders' strategies, and thus we can remove it from the objective function in (\ref{Eq: simp2}). By using the law of iterated expectations, we can rewrite the optimization problem in (\ref{Eq: simp2}) as follows
\begin{align}\label{eq: bestrespdynamic}
    \min_{\eta_k\in\Omega_k(\eta^{-i}_k)}\sum_{k=1}^n\mathbb{E}[\boldsymbol{\hat{x}_k}'V_k^i\boldsymbol{\hat{x}_k}],
\end{align}
where 
\begin{align}\label{eq: Vik}
V_k^i=\Lambda_k^{i\top}\Lambda_k^i-\Lambda_k^{i\top}Q_k^i-Q_k^{i\top}\Lambda_k^{i}.
\end{align}
We note that since $\mathbb{E}[\boldsymbol{\hat{x}_k}]=\mathbb{E}[\boldsymbol{x_k}]=0$, the posterior covariance at stage $k$, $\Sigma_k$, is given by $\mathbb{E}[\|\boldsymbol{\hat{x}_k}-\mathbb{E}[\boldsymbol{\hat{x}_k}]\|^2]=\mathbb{E}[\boldsymbol{\hat{x}_k}\boldsymbol{\hat{x}_k}^\top]$. Next, we provide an example to illustrate the above simplification in order to arrive at $V_k^i$ for the special class of  dynamic games involving only two senders and a receiver. This would be the counterpart of Example \ref{ex1} in the multi-stage case.
\begin{example} Consider a dynamic multi-stage communication game between two senders $1$, $2$ and receiver $r$, where $\boldsymbol{x_k}$ evolves as 
\begin{align*}
    \boldsymbol{x_{k+1}}\!=\!\begin{bmatrix}\boldsymbol{z_{k+1}}&\boldsymbol{\theta_{A_{k+1}}}&\boldsymbol{\theta_{B_{k+1}}}\end{bmatrix}\!^\top\!\!=\!\begin{bmatrix}A_Z &0&0\\ 0&A_{\theta_A}&0\\0&0&A_{\theta_B}\end{bmatrix}\begin{bmatrix}\boldsymbol{z_{k}}&\boldsymbol{\theta_{A_{k}}}&\boldsymbol{\theta_{B_{k}}}\end{bmatrix}^\top\!\!\!+\begin{bmatrix}\boldsymbol{w_{z_k}}&\boldsymbol{w_{z\theta_A}}&\boldsymbol{w_{z\theta_B}}\end{bmatrix}^\top\!\!\!\!.
\end{align*}
 Receiver $r$ aims to track process $\mathbf{z_k}$. On the other hand, sender $1$ wants the receiver to track a linear combination of $\mathbf{z_k}$ and a bias parameter $\boldsymbol{\theta_A}$, i.e., $\boldsymbol{z}_k+D_k\boldsymbol{\theta_{A_k}}$, and sender $2$ wants the receiver to track a linear combination of $\mathbf{z}_k$ with a different bias parameter $\boldsymbol{\theta_B}$, i.e.,  $\mathbf{z}_k+E_k\boldsymbol{\theta_{B_k}}$ where $D_k$ and $E_k$ are matrices representing the linear combinations with proper dimensions. Thus, the finite horizon cost functions for the senders are given by $J_{1}=\mathbb{E}[\sum_{k=1}^n\|\boldsymbol{z_k}+D_k\boldsymbol{\theta_{A_k}}-\boldsymbol{u_k}\|^2$, $J_{2}=\mathbb{E}[\sum_{k=1}^n\|\boldsymbol{z_k}+E_k\boldsymbol{\theta_{B_k}}-\boldsymbol{u_k}\|^2]$, and that of the receiver is given by $\mathbb{E}[\sum_{k=1}^n\|\boldsymbol{z_k}-\boldsymbol{u_k}\|^2]$. Then, we have $Q_k^1=\begin{bmatrix}I&D_k&0\end{bmatrix}$, $Q_k^2=\begin{bmatrix}I&0&E_k\end{bmatrix}$, and $R_{1,k}=R_{2,k}=R_{r,k}=-I$ for $k=1,\hdots,n$. By using the definition of $V_k^i$ in (\ref{eq: Vik}), we obtain
\begin{align*}
    V_k^1=\begin{bmatrix}-I_k &-D_k&0\\-D_k&0 &0\\0 &0 &0\end{bmatrix},\qquad V_k^2=\begin{bmatrix}-I_k &0 &-E_k\\0 &0 &0\\-E_k &0 &0\end{bmatrix}.
\end{align*}
\end{example}

Next, by using tools developed in Section~\ref{Sect:Strategic_com}, we identify a recursive structure satisfied by a set of stable posteriors in the dynamic game described in this section. The key insight that we provide in the following lemma is that greedily computing equilibrium posteriors using their static stage cost alone for all senders will result in posteriors that are stable for the cost-to-go value functions for all senders.
\begin{algorithm}[t]
\caption{Finding an Equilibrium Posterior for the $m$-Sender Multi-stage Game via SDP}\label{alg: dynpostcalc}
\begin{algorithmic}
\State{\textbf{Parameters:} $\Sigma_1,A,\Sigma_w$, $M=\{k_1,k_2,\dots,k_m\}\in\mathcal{M}$}
\State{\textbf{Compute} $V_k^i$ $\forall i\in[m],k\in[n]$ by using (\ref{eq: Vik})}
\State{\textbf{Variables:} $S_i\in\mathbb{S}^p$ with $S_i\succeq O$ $\forall i\in[n]$}
\State{$\Sigma_{-i}=O,S_0=O$}
\For{$k=1,\hdots, n$}
\State{\textbf{Run} Algorithm \ref{alg: npostcalc} with parameters: $\Sigma_x=\Sigma_k-AS_{k-1}A^\top$, and $M$ then \textbf{return:} $\Sigma^*$}
\State $S_k^*\gets\Sigma^*$
\EndFor
\State{\textbf{Return:} $S_1^*,\hdots,S_n^*$}
\end{algorithmic}
\end{algorithm}
\begin{lemma}\label{lem: dynstabpost}
There exist projection matrices $P_k$ for $k=1,\hdots,n$ such that the equilibrium posteriors for the $n$ stage dynamic game ($\{S^*_k\}_{k\in[n]}$) with players' objectives given by (\ref{Eq: dynobjective}) satisfy the recursion 
\begin{align}\label{eq: dynstablepost}
    S_{k+1}^*=AS_{k}^*A^\top+(\Sigma_k-AS_k^*A^\top)^{\frac{1}{2}}P_k(\Sigma_k-AS_k^*A^\top)^{\frac{1}{2}}.
\end{align}
\end{lemma}
\begin{Proof}
We note that as in the single-stage case, at each stage, for fixed strategies of the other senders $-i$, sender $i$ can only add more information to the receiver's estimation process. In other words, at stage $1\leq k'\leq n$, the equilibrium computation problem for sender $i$ is to minimize the total remaining cost denoted by $ \tilde{V}_{k'}^{i}(\{\{\eta^i_{k''}\}_{i\in[m]}\}_{k''\in[k'-1]})$\footnote{For ease of notation the dependence on the receiver's best response in the cost-to-go functions is suppressed.}, which can be expressed as 
\begin{align}\label{eqn:min_problem}
    \tilde{V}^i_{k'}(\{\{\eta^i_{k''}\}_{i\in[m]}\}_{k''\in[k'-1]})=\min_{\eta^i_{k'}\in\Omega_{k'}(\eta^{-i}_{k'})}\left[Tr(V^i_{k'}\boldsymbol{\hat{x}_{k'}}\boldsymbol{\hat{x}_{k'}}^\top)+\sum_{k=k'+1}^{n} Tr(V^i_{k}\boldsymbol{\hat{x}_k}\boldsymbol{\hat{x}_k}^\top)\right].
\end{align}
Further, we note that a fixed set of senders' strategies in all previous time steps, can be mapped into a posterior at present stage. Thus, (\ref{eqn:min_problem}) can be equivalently written as
\begin{align}\label{eqn:min_problemsimp}
    \tilde{V}^i_{k'}(\Sigma'_{k'-1})=\min_{\eta^i_{k'}\in\Omega_{k'}(\eta^{-i}_{k'})}\left[Tr(V^i_{k'}\boldsymbol{\hat{x}_{k'}}\boldsymbol{\hat{x}_{k'}}^\top)+\sum_{k=k'+1}^{n} Tr(V^i_{k}\boldsymbol{\hat{x}_k}\boldsymbol{\hat{x}_k}^\top)\right], 
\end{align}
where $\Sigma'_{k'-1}$ is the posterior induced by strategies $\{\{\eta^i_{k''}\}_{i\in[m]}\}_{k''\in[k'-1]}$. Although the strategies at previous time steps do not explicitly appear in the minimization problem stated in (\ref{eqn:min_problemsimp}), they implicitly affect the cost function through $\Omega_{k'}(\cdot)$
The optimization problem in (\ref{eqn:min_problemsimp}) needs to minimize not only the cost at stage $k'$ but also the cost at the future stages $k= k'+1,\dots,n$. Further, given the strategies of the other senders at the present stage, $\eta^{-i}_{k'}$, for reasons similar to Lemma \ref{lem: bestresplbound}, the value function $\tilde{V}_{k'}^i(\Sigma'_{k'-1})$ in (\ref{eqn:min_problemsimp}) is bounded below by 
\begin{align}
    \tilde{V}_{k'}^i(\Sigma'_{k'-1})\geq \min_{\Sigma_{k'}\succeq S_{k'}\succeq \Sigma_{k'}^{'-i}} [Tr(V^i_{k'}S_{k'})+\tilde{V}_{k'}^i(S_{k'})], 
\end{align}
where $\Sigma_{k'}^{'-i}$ is the posterior induced by senders $-i$ as a result of their strategies at the present stage, $\eta_{k'}^{-i}$. The explicit dependence of this bound on $\Sigma'_{k'-1}$ can be seen by observing that $\Sigma_{k'}^{'-i}\succeq A\Sigma'_{k'-1}A^\top$. This is because, even if all the other senders do not reveal any information at stage $k'$, posterior in stage $k'-1$, $\Sigma'_{k'-1}$, can be used for estimation at stage $k'$ as shown in \cite{sayin2019hierarchical}. Further, if $\Sigma'_{k'-1}\succeq \Sigma_{k'-1}''$ then we have $A\Sigma'_{k'-1}A^\top\succeq A\Sigma_{k'-1}''A^\top$ which implies that $\tilde{V}_{k'}^i(\Sigma_{k'-1}')\geq \tilde{V}_{k'}^i(\Sigma_{k'-1}'')$ for all senders $i\in[m]$. This is the key step which allows us to prove stability of posteriors obtained through greedy policies. Formally, let $S_{k'}^*$ be a stable posterior obtained considering the single-stage game at stage $k'$ where each sender $i\in[m]$ greedily computes its strategy considering only the cost at stage $k'$,  $Tr(V_{k'}^i\boldsymbol{\hat{x}_{k'}}\boldsymbol{\hat{x}_{k'}^\top})$. This posterior $S_{k'}^*$ is stable at stage $k'$. To prove this, consider at stage $k'$ the best response of player $i$ to all other players' $-i$ strategies which result in the posterior $S_{k'}^*$. This is bounded below by
\begin{align*}
    \tilde{V}^i_{k'}(\Sigma_{k'-1}')\geq \min_{\Sigma_{k'}\succeq S_{k'}\succeq S_{k'}^*} [Tr(V^i_{k'}S_{k'})+\tilde{V}_{k'}^i(S_{k'})].
\end{align*}
For any $S_{k'}\succeq S_{k'}^*,$ we have $ Tr(V^i_{k'}S_{k'})+\tilde{V}_{k'}^i(S_{k'})>Tr(V^i_{k'}S_{k'}^*)+\tilde{V}_{k'}^i(S_{k'}^*)$ from Lemma \ref{lem: nsenstable} and the discussion above. Thus, $S_{k'}^*$ is stable for sender $i$. Since this is true for all senders $i,\forall i\in[m]$ and all posteriors $\Sigma_{k'-1}'$, this is a stable posterior for the multi-stage game.

 Thus, a set of stable posteriors for the $n$-stage game can be written as
\begin{align*}
S^*_k=AS_{k-1}^*A^\top+(\Sigma_k-AS_{k-1}A^\top)^{\frac{1}{2}}P_k(\Sigma_k-AS_{k-1}A^\top)^{\frac{1}{2}},
\end{align*}
where $P_k$ is the projection matrix at stage $k$ obtained by solving the single-stage game for NE as in Lemma \ref{lem: npostproj}. These posteriors $S_1,\hdots,S_n$ are stable since no sender wants to reveal any more information at any stage.
\end{Proof}
The set of stable posteriors described in Lemma \ref{lem: dynstabpost} can be obtained through Algorithm~\ref{alg: dynpostcalc}. In the following lemma, we prove that these posteriors can be achieved by using linear strategies.
\begin{lemma}\label{Lemma_11}Let $S_1^*,S_2^*,\hdots,S_n^*$ and $P_1,\hdots,P_n$ be the stable posteriors and corresponding projection matrices as provided in (\ref{eq: dynstablepost}). Then, linear strategies given by $L_k^i=(\Sigma_k-AS_{k-1}^*A^\top)^{-\frac{1}{2}}U_k\Lambda_k$, where $P_k=U_k\Lambda_kU_k^\top$ is the eigen-decomposition of the projection matrix $P_k$, achieve the dynamic multistage NE. 
\end{lemma}
\begin{Proof}
We show that the equilibrium can be attained when all the senders commit to the same linear policy at each stage, leading to the stable posterior.\footnote{This linear policy can easily be extended to the scenario where senders can follow full rank scaling of these policies.} Let us assume that all senders $\{T_i\}_{i\in[m]}$ commit to memoryless linear strategies $\{L_k^i\}_{k\in[n]}$ at each stage $k$. Further, at each stage, all senders commit to the same policy, $L_k^i=L_k,\forall i\in[m]$ for some $L_k$, which implies that $\boldsymbol{y}_k^i=\boldsymbol{y}_k,\forall i\in[m]$. Hence correspondingly, the posterior at stage k, $\boldsymbol{\hat{x}_k}=\mathbb{E}[\boldsymbol{x}_k|\{\{\boldsymbol{y_k^i}\}_{k\in[k]}\}_{i\in[m]}]=\mathbb{E}[\boldsymbol{x}_k|\{\{\boldsymbol{y_k}\}_{k\in[k]}\}]$, can be calculated as
\begin{align*}
    \hat{\boldsymbol{x}}_k=A\hat{\boldsymbol{x}}_{k-1}+(\Sigma_k-A\Sigma'_{k-1}A^\top)L_k(L_k^\top(\Sigma_k-A\Sigma'_{k-1}A^\top)L_k)^\dagger L_k^\top(\boldsymbol{x}_k-A\hat{\boldsymbol{x}}_{k-1}),
\end{align*}
where $\hat{\boldsymbol{x}}_1=\Sigma_1L_1(L_1^\top\Sigma_1L_1)^\dagger\boldsymbol{x}_1$. Further, by using the fact that $\mathbb{E}[\hat{\boldsymbol{x}}_{k-1}(\boldsymbol{x}_k-A\hat{\boldsymbol{x}}_{k-1})^\top]=O$, we obtain the posterior covariance for stages $k\geq 2$ as
\begin{align*}
    \Sigma'_{k}=A\Sigma'_{k-1}A^\top+(\Sigma_k-A\Sigma'_{k-1}A^\top)^\frac{1}{2}P'_k(\Sigma_k-A\Sigma'_{k-1}A^\top)^\frac{1}{2},
\end{align*}
where $P'_k=Z_k(Z_k^\top Z_k)^\dagger Z_k^\top$ with $Z_k=(\Sigma_k-A\Sigma'_{k-1}A^\top)^\frac{1}{2}L_k$ which implies that $P'_k$ is a projection matrix. Further $\Sigma'_1=\Sigma_1^\frac{1}{2}P_1\Sigma_1^\frac{1}{2}$ where $P_1=Z_1(Z_1^\top Z_1)^\dagger Z_1^\top$ is also a projection matrix with $Z_1=\Sigma_1^\frac{1}{2}L_1$. Thus by considering the eigen-decomposition of projection matrices obtained from Algorithm \ref{alg: dynpostcalc} where $P^*_k=U_k\Lambda_kU_k^\top$, we can complete the proof.
\end{Proof}

Similar to the single-stage multiple-sender case in Section~\ref{Sect:Strategic_com}, as noted in Proposition~\ref{Remark_3_alt}, the posteriors obtained through Algorithm \ref{alg: dynpostcalc}, although greedy, are also partially informative in the defined sense, i.e., all senders prefer the posteriors obtained through Algorithm \ref{alg: dynpostcalc} to any set of posteriors that are more informative at any stage of the multi-stage game.



As a recap, in this section, we have derived a set of stable and achievable posteriors by minimizing the current stage cost in the optimization problem given in (\ref{eqn:min_problem}). We note that this solution might be sub-optimal in the sense that it does not account for the cost at future stages. Thus, a tighter lower bound for the senders' optimization problem at each stage can be obtained by solving the exact dynamic programming, i.e., Backward induction. However, such a lower bound might not be achievable using linear noiseless policies. We leave finding such admissible equilibria for the dynamic game setting as a future research direction.       

\section{Multi Sender Multi Receiver Strategic Communication}\label{Sec: Multireceiver}
\begin{figure}[t]
\centerline{\includegraphics[width=0.65\columnwidth]{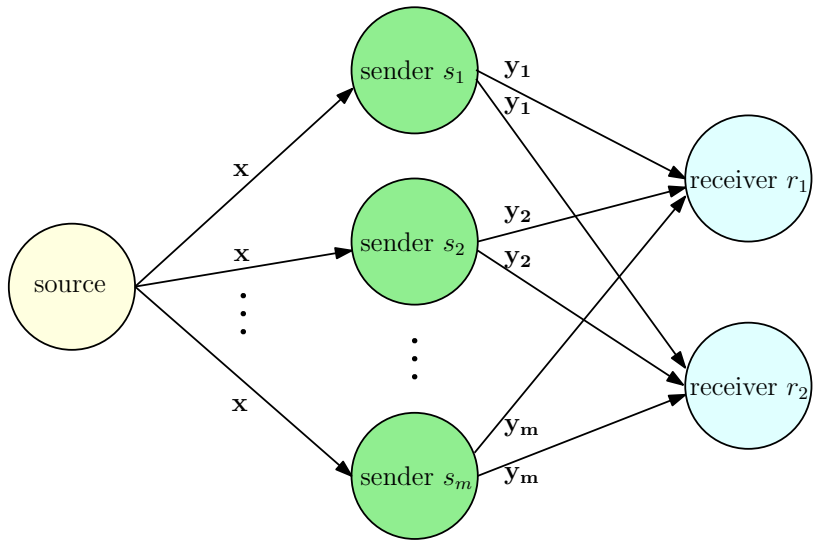}}
	\caption{The strategic communication system consisting of $m$ senders labeled as $s_1,\!\cdots\!,s_m$ and $2$ receivers labeled as $r_1$ and $r_2$. }
	\vspace{-0.35cm}
	\label{Fig:system_model_2}
	\vspace{-0.3cm}
\end{figure}
In this section we illustrate the extension of techniques from previous sections to a game with multiple receivers as shown in Fig.~\ref{Fig:system_model_2}. We derive our results to the case with 2 receivers and note that our results can be extended to $d>2$ receivers easily.


More precisely, consider an $m+2$ player static game with $m$ senders and 2 receivers each with their own objectives that depend on a random state $\boldsymbol{x}$ sampled from a zero mean Gaussian distribution with positive-definite covariance $\Sigma_x$, i.e., $\boldsymbol{x}\sim\mathbb{N}(0,\Sigma_x)$, and actions of the receivers $r_1$ and $r_2$ are denoted by $\boldsymbol{u}_1$ and $\boldsymbol{u}_2$, respectively. 
Here, each player $i$ where $i\in\{s_1,\hdots,s_{m}, r_1,r_2\}$ has its own expected cost function given by
\begin{align}\label{cost_msmr}
    J_i(\{\eta_{s_j}\}_{j\in[m]};\gamma_{r_1},\gamma_{r_2})=\mathbb{E}[L_i(\boldsymbol{x};\boldsymbol{u_1}=\gamma_{r_1}(I_{r_1});\boldsymbol{u_2}=\gamma_{r_2}(I_{r_2}))]=\mathbb{E}\left[\left\|Q_i\boldsymbol{x}+R_{i1}\boldsymbol{u}_1+R_{i2}\boldsymbol{u}_2\right\|^2\right].
\end{align}
Here, similar to the previous sections, we denote by $L_i(\boldsymbol{x};\boldsymbol{u_1}=\gamma_{r_1}(I_{r_1});\boldsymbol{u_2}=\gamma_{r_2}(I_{r_2}))=\left\|Q_i\boldsymbol{x}+R_{i1}\boldsymbol{u}_1+R_{i2}\boldsymbol{u}_2\right\|^2$ the cost function of player $i$ where $I_{r_1} =I_{r_2} = \{\boldsymbol{y}_1,\hdots,\boldsymbol{y}_m\}$ are the information structure of the receivers $r_1$ and $r_2$, and $\eta_{s_j}(\cdot)$ is the strategy of sender $s_j$ for $j\in[m]$ and $\gamma_{r_1}(\cdot)$, $\gamma_{r_2}(\cdot)$ are strategies of receivers $r_1$ and $r_2$, respectively. The game proceeds as in the previous sections where all the senders simultaneously commit to their own strategy followed by the receivers' best responding to the committed senders' strategies. Due to the presence of multiple receivers whose actions possibly affect each other as reflected in utilities given by (\ref{cost_msmr}), 
the receivers in turn play a Nash game among themselves after observing the senders' commitments. Thus, a policy tuple $(\{\eta^*_{s_j}\}_{j\in[m]},\gamma_1^*, \gamma_2^*)$ is said to be in an equilibrium if
\begin{align}
    &J_i(\eta_i^*,\eta_{-i}^*,\!\{\gamma_{r_k}^*(\eta_i^*,\eta_{-i}^*)\}_{k\in[2]})\!\leq\! J_i(\eta_i,\eta_{-i}^*,\!\{\gamma_{r_k}^*(\eta_i,\eta_{-i}^*)\}_{k\in[2]}),~\forall\eta_i\in\Omega,~\forall i\in\{s_1,\hdots,s_m\},\!\!\\
    &J_i(\{\eta_{s_j}\}_{j\in[m]},\gamma_i^*(\{\eta_{s_j}\}_{j\in[m]}),\gamma_{-i}^*(\{\eta_{s_j}\}_{j\in[m]}))\leq J_i(\{\eta_{s_j}\}_{j\in[m]},\gamma_i(\{\eta_{s_j}\}_{j\in[m]}),\gamma_{-i}^*(\{\eta_{s_j}\}_{j\in[m]})),
\end{align}
where $\gamma_i\in\Gamma$ for $i\in\{r_1, r_2\}.$ In order to construct such an equilibrium tuple, we first need to analyze the Nash game played among the receivers for fixed senders' strategies $\{\eta_{s_j}\}_{j\in[m]}$. 

For fixed senders' strategies, receivers' play a Nash game with utilities which are strictly convex in their action variables, since $R_{{r_\ell}\ell}^\top R_{{r_\ell}\ell}$ is assumed to be of full rank for all $\ell\in[2]$. Thus, in order to determine the NE of this strictly convex quadratic game among receivers, by differentiating each receiver's cost function, we find the best response dynamics of each receiver for given senders' strategies and the action of the other receiver.  
This leads to a set of necessary and sufficient conditions for $\boldsymbol{u}_i's$ to be in equilibrium due to strict convexity \cite{bacsar1998dynamic}. Thus, we obtain 
\begin{align}\label{eqn_mult_sen_mult_rec}
    R\boldsymbol{u}=-\boldsymbol{q},
\end{align}
where $\boldsymbol{u}=
    [\boldsymbol{u}_1^\top, \boldsymbol{u}_2^\top
]^\top$, $\hat{\boldsymbol{x}}=\mathbb{E}[\boldsymbol{x}|\boldsymbol{y}_1,\hdots,\boldsymbol{y}_m]$, and

\begin{align}
    R=&\begin{bmatrix}
        R_{r_11}^\top R_{r_11} & R_{r_11}^\top R_{r_12}\\
        R_{r_22}^\top R_{r_21} & R_{r_22}^\top R_{r_22}\\
    \end{bmatrix},  \label{eq: Rmatrix} \\
    \boldsymbol{q}=&\begin{bmatrix} \boldsymbol{\hat{x}}^\top Q_{r_1}^\top 
    R_{r_11}, & \boldsymbol{\hat{x}}^\top Q_{r_2}^\top R_{r_22} 
\end{bmatrix}^\top.\label{eq: Qmatrix}
\end{align}
\begin{prop}\label{prop: receievrbestresp}
    The $2$-receiver quadratic game admits a unique solution when $R$ given in (\ref{eq: Rmatrix}) is invertible. This unique NE solution is given by
    \begin{align*}
        \boldsymbol{u}^*=-R^{-1}\boldsymbol{q}:=\begin{bmatrix}
            -K_1\boldsymbol{\hat{x}}, \\ -K_2\boldsymbol{\hat{x}}
        \end{bmatrix},
    \end{align*}
where $\boldsymbol{q}$ is provided in (\ref{eq: Qmatrix}).
\end{prop}
We note from (\ref{eq: Qmatrix}) and Proposition \ref{prop: receievrbestresp} that at NE, each receiver's best response is linear in the posterior estimate $\hat{\boldsymbol{x}}$. This specific structure allows us to use tools similar to those in previous sections to find the NE among the senders.
Considering the receivers' NE policy given in Proposition~\ref{prop: receievrbestresp}, senders play a Nash game in the space of posteriors. 

Next, we find the NE policies for senders. 
Senders want to minimize their own cost functions given by    
\begin{align}  \label{eqn:cost_msmr}  \min_{\eta_i\in\Omega}\mathbb{E}\left[\left\|Q_i\boldsymbol{x}+\sum_{k=1}^2 R_{ik}K_k\hat{\boldsymbol{x}}\right\|^2\right], \quad\forall i\in \{s_1,\dots, s_m\}.
\end{align}
Thus, utilizing techniques from Section \ref{Sect:Strategic_com}, 
we can rewrite the objective function of each sender in (\ref{eqn:cost_msmr}) as
\begin{align}\label{eqn_senders_obj}
    \min_{\eta_i\in\Omega} Tr(V_iS),
\end{align}
where $S=\mathbb{E}[\hat{\boldsymbol{x}}\hat{\boldsymbol{x}}^\top]$ and $V_i$ is given by 
\begin{align}
    V_{i}=&K_1^\top R_{i 1}^\top R_{i2}K_2+K_2^\top R_{i2}^\top R_{i1}K_1+K_1^\top R_{i1}^\top R_{i1}K_1+K_2^\top R_{i2}^\top R_{i2}K_2\nonumber\\
    &-K_1^\top R_{i1}^\top Q_{i}-K_2^\top R_{i2}Q_{i}-Q_{i}^\top R_{i1}^\top K_1-Q_{i}^\top R_{i2}K_2, \quad i\in \{s_1,\dots, s_m\}.
\end{align}
We note that even in the presence of multiple receivers, senders' Nash game can be written in terms of a trace minimization. Thus, we can utilize Algorithm~\ref{alg: npostcalc} to identify stable posteriors which are achievable by using linear strategies as proved in Theorem~\ref{Thm_3}.   

\section{Numerical Results}\label{Sect:Num_result}
In this section, we provide simulation results to analyze the effects of multiple senders in the communication game across various settings. In the first 3 numerical results, we consider two senders and a single receiver and a state of the world given by $\boldsymbol{x}=\begin{bmatrix}z &\theta_A &\theta_B\end{bmatrix}^\top$. For all these examples, we consider quadratic cost functions for the players as noted below:
\begin{align}
 J_{1} &=\mathbb{E}[(z+\beta \theta_A+\alpha\theta_B-u)^2], \\
 J_{2} &=\mathbb{E}[(z+\alpha\theta_A+\beta\theta_B-u)^2], \\
 J_{r} &=\mathbb{E}[(z-u)^2],
\end{align}
where we specify $\alpha$ and $\beta$ values for each example.\footnote{Here, the actions are generated by policies/strategies as in Section~\ref{Sect:system_model}, but we suppress this representation throughout this section.} We note that both senders having access to the full state allows us to have such objectives for the sender which depend on both $\theta_A,$ $\theta_B$.\footnote{In the first 3 numerical results, one could consider a more general model where $\alpha$ and $\beta$ could also be sender specific such that the expected cost functions are given by $J_{1} =\mathbb{E}[(z+\beta_1 \theta_A+\alpha_1\theta_B-u)^2]$ and $J_{2} =\mathbb{E}[(z+\alpha_2\theta_A+\beta_2\theta_B-u)^2]$.}

\subsubsection{Construction of Nash equilibrium policies}
We first consider $\alpha=0$, and $\beta=1$, i.e., sender $1$ has a quadratic cost to minimize $\mathbb{E}[(z+\theta_A-u)^2]$ while sender $2$ wants to minimize $\mathbb{E}[(z+\theta_B-u)^2]$ and the receiver wants to minimize $\mathbb{E}[(z-u)^2]$. We consider a zero-mean Gaussian prior on the state of the world with the covariance matrix $\Sigma_x$ given by
\begin{align}
    \Sigma_x=\begin{bmatrix}1 &0.5 &0.7\\0.5 &1.5 &0.2\\ 0.7 &0.2 &1\end{bmatrix}.
\end{align}
By using Algorithm \ref{alg: postcalc} to compute stable equilibrium posterior, we get 
\begin{align}\label{eq: expost}
\Sigma^*=\begin{bmatrix}
    0.9715  &  0.5571 &   0.7793\\
    0.5571  &  1.3859    &0.0413\\
    0.7793  &  0.0413   & 0.7794
   \end{bmatrix},
\end{align}
which achieves a NE expected cost values of $J_1=1.4144$, $J_2=0.8699$, and $ J_r=0.0285$ for senders 1 and 2, and the receiver, respectively. 
From Theorem \ref{th: Nasheq}, we can construct NE policies to be $\eta_1(\boldsymbol{x})=\eta_2(\boldsymbol{x})=L\boldsymbol{x}$, where 
\begin{align*}
L=\begin{bmatrix}
   0&   0  &  0\\
    0.1046&   -0.7421   & 0.5714\\
   -0.7807  & -0.1868 &  -0.1464
\end{bmatrix} .  
\end{align*}
As a sanity check we can see that such policies would return the equilibrium posterior $\Sigma_L=\Sigma_xL^\top(L\Sigma_x L^\top)^{\dagger}L\Sigma_x=\Sigma^*$ as given by (\ref{eq: expost}). 

\subsubsection{Correlation between the state of the world parameters}
\begin{figure}[t]
	\begin{center}
	    \subfigure[]{			\includegraphics[scale=0.4]{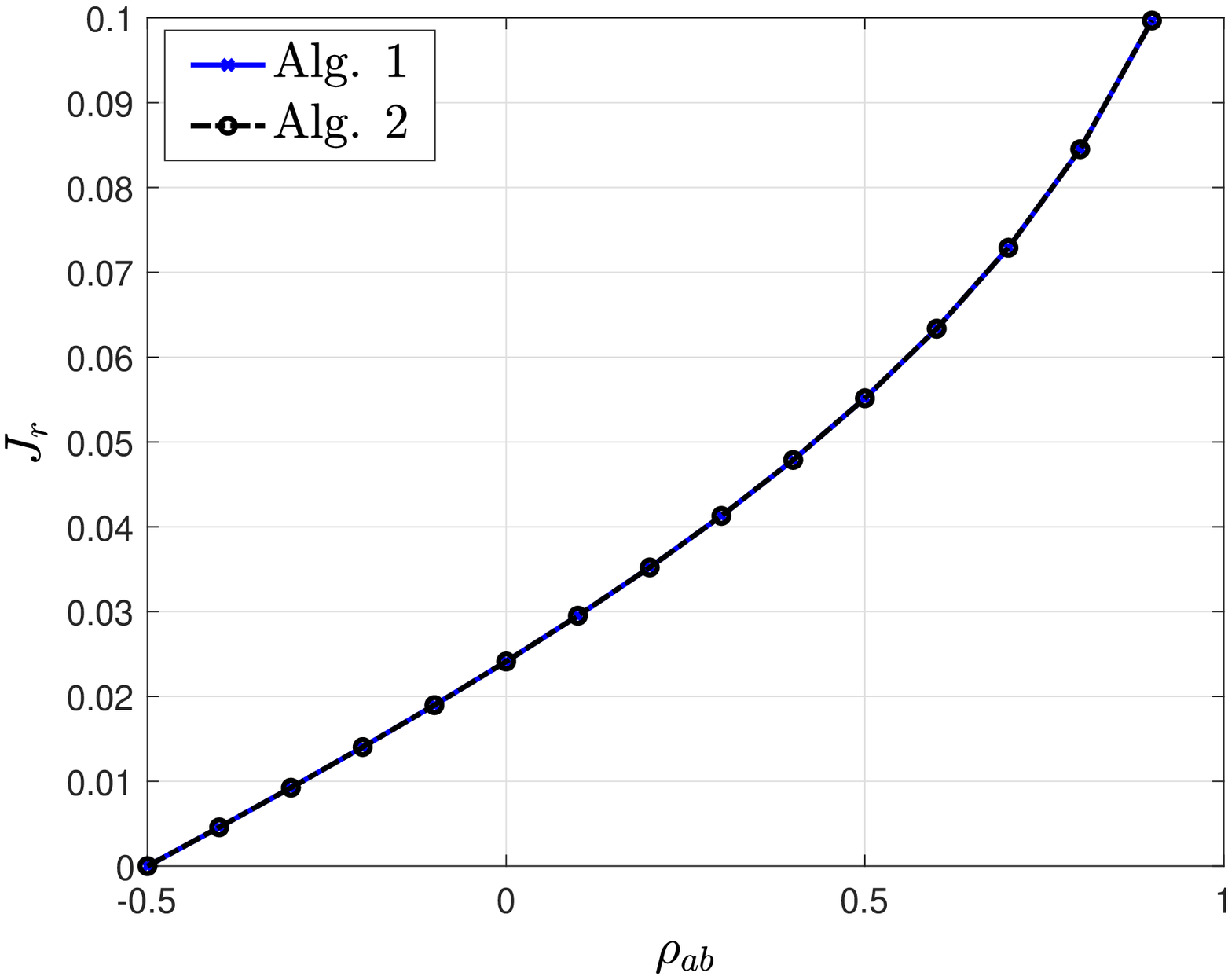}}
		\subfigure[]{%
			\includegraphics[scale=0.4]{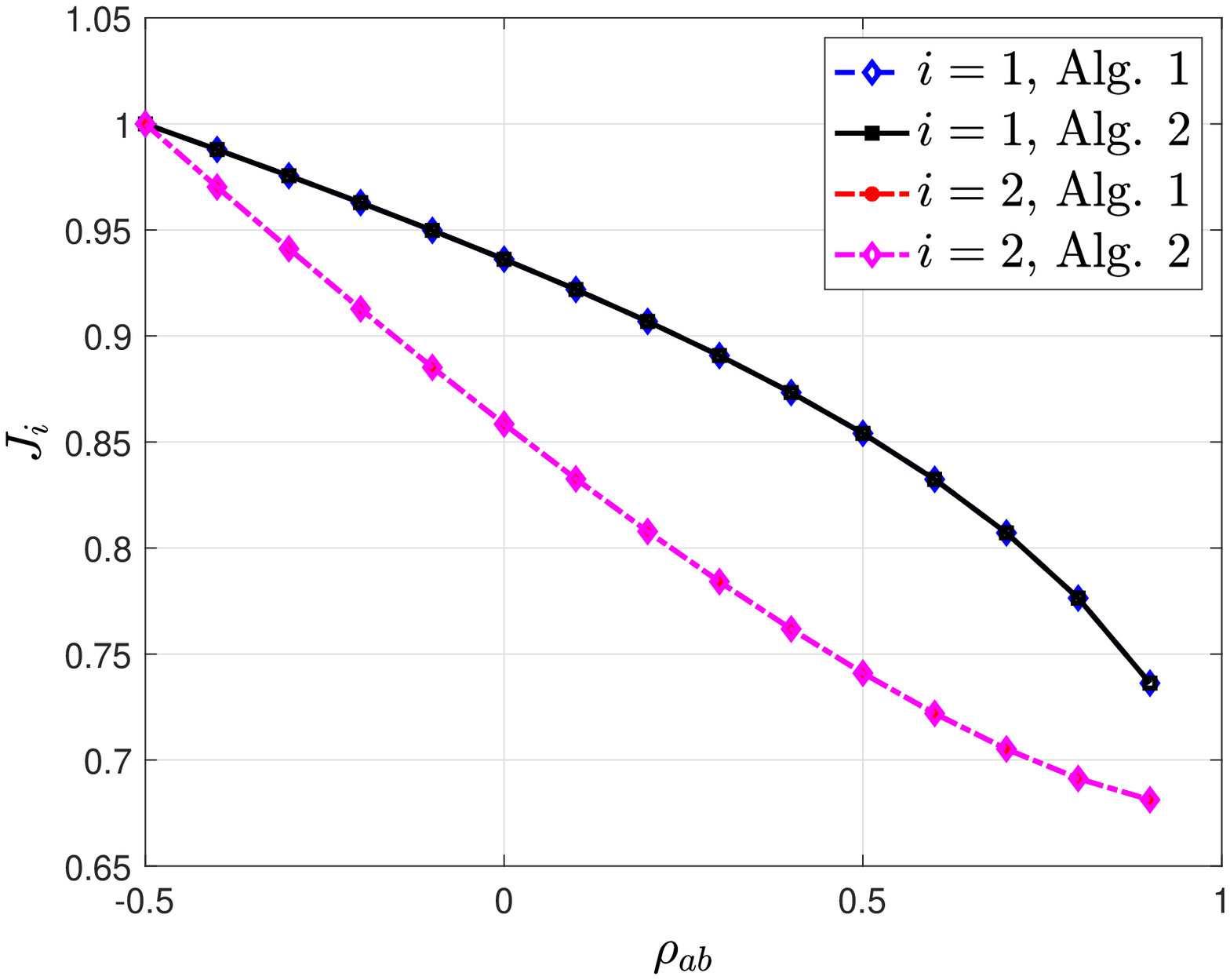}}
	\end{center}\vspace{-0.5cm}
	\caption{The expected costs of the (a) receiver, (b) sender $i$ for $i=\{1,2\}$ with respect to correlation coefficient $\rho_{ab}$.}
	\label{fig:sim1}
	\vspace{-0.5cm}
\end{figure}

In the second numerical example, we consider the same setting as in the first example, i.e., $\alpha=0$, and $\beta=1$, but with a different covariance matrix $\Sigma_x$. Here, we take 
\begin{align}\label{eqn:example_1}
\Sigma_x=\begin{bmatrix}1 & 0.5 &0.5\\0.5 & 1 &\rho_{ab}\\0.5 &\rho_{ab} &1\end{bmatrix},
\end{align}
where $\rho_{ab}$ is the correlation coefficient between $\theta_A$ and $\theta_B$ which is given by $\rho_{ab} = \frac{Cov(\theta_A,\theta_B)}{\sqrt{Var(\theta_A)Var(\theta_B)}}$. In this example, we vary $\rho_{ab} = \{-0.5, -0.4, \cdots, 0.9\}$ and find the expected costs of the receiver and the senders obtained by Algorithms~\ref{algorithm2} and \ref{alg: postcalc}. In Fig.~\ref{fig:sim1}(a), we see how the error covariance of the receiver, i.e., $J_{r} = \mathbb{E}[({z}-{u})^2]$, changes with respect to $\rho_{ab}$. When $\rho_{ab}<0$, $\theta_A$, and $\theta_B$ are negatively correlated, and as a result, senders reveal more information to the receiver about the state of the world $\boldsymbol{z}$. As the correlation $\rho_{ab}$ increases, the information revealed to the receiver decreases and the expected cost of the receiver $J_{r}$ increases. In Fig.~\ref{fig:sim1}(b), we see how the expected costs of the senders changes with respect to $\rho_{ab}$. As the senders' goals become aligned, they can manipulate the receiver more, and decrease their expected costs effectively. In Figs.~\ref{fig:sim1}(a)-(b), as both Algorithms~\ref{algorithm2} and \ref{alg: postcalc} give the same equilibrium posteriors, we see that they achieve the same expected costs for the receiver and for the senders.         

\subsubsection{Correlation between senders' objectives} 
\begin{figure}[t]
	\begin{center}
	    \subfigure[]{%
			\includegraphics[scale=0.4]{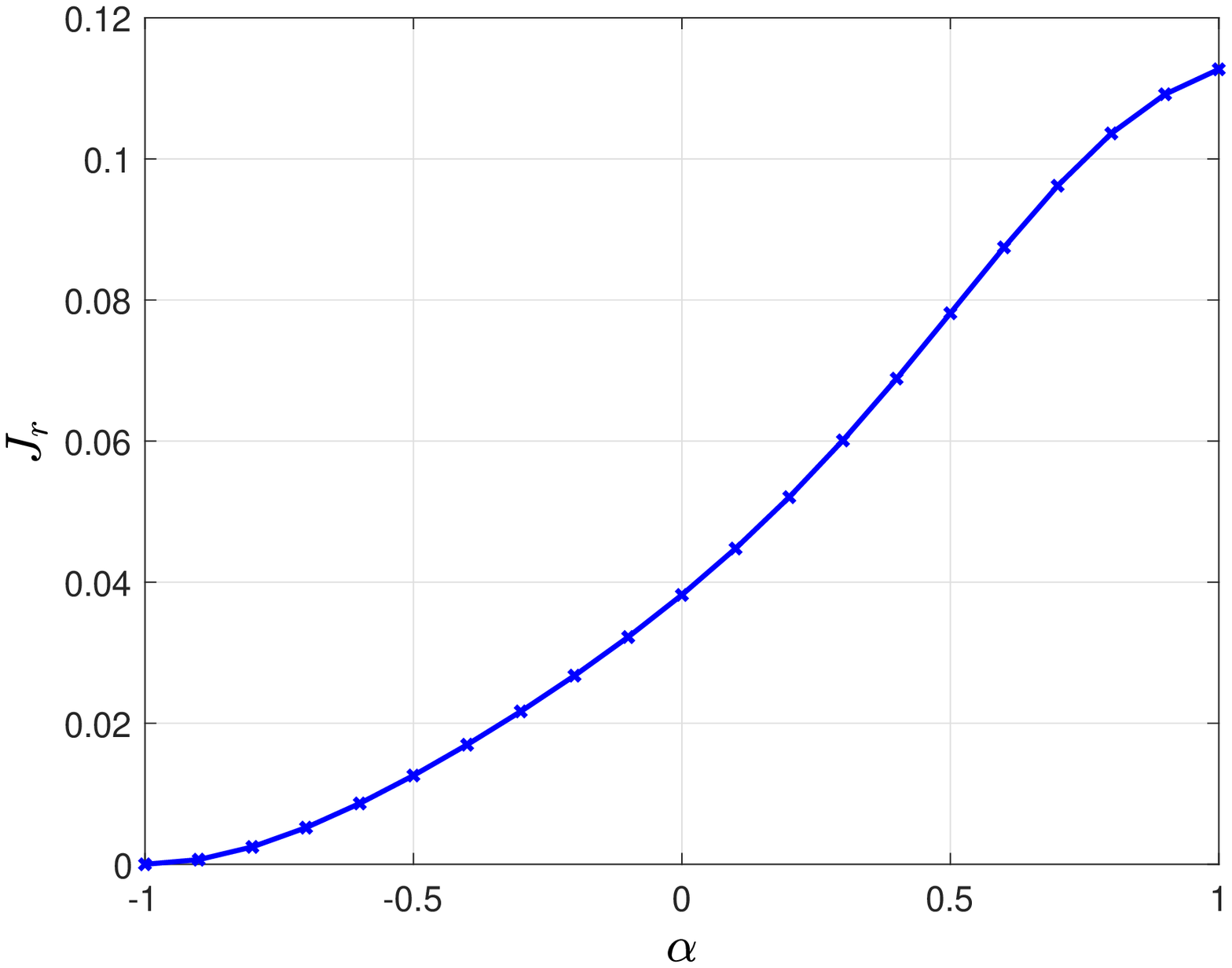}}
		\subfigure[]{%
			\includegraphics[scale=0.4]{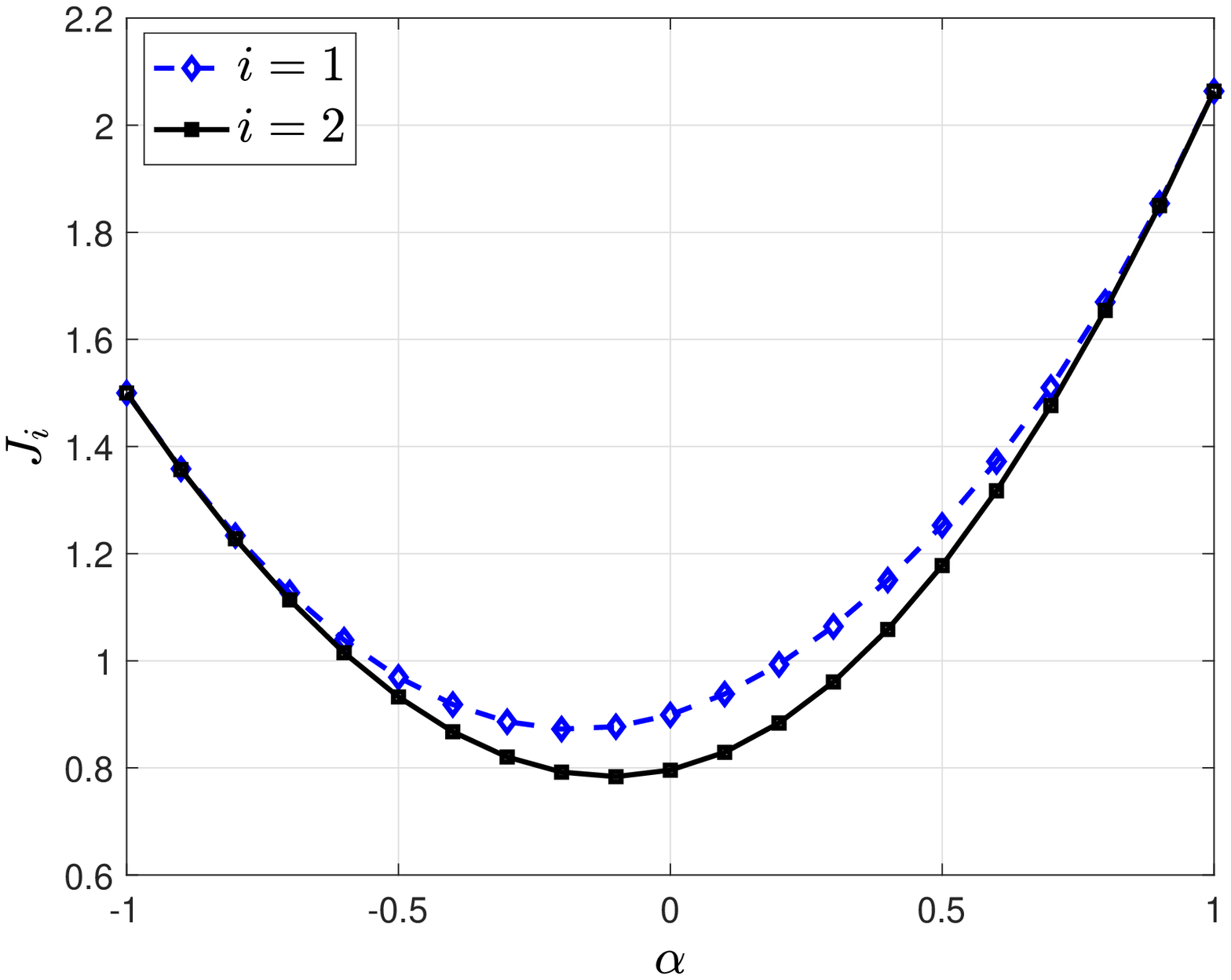}}
	\end{center}\vspace{-0.5cm}
	\caption{The expected costs of the (a) receiver, (b) sender $i$ for $i=\{1,2\}$ with respect to $\alpha$.}
	\label{fig:sim3}
	\vspace{-0.5cm}
\end{figure}

In the third numerical example, we consider $\Sigma_x$ as in (\ref{eqn:example_1}) with $\rho_{ab} = 0.25$. In this example, we take $\beta=1$ and vary $\alpha \in [-1,1]$. Thus, the senders' objective functions include both the parameters ${\theta_A}$ and ${\theta_B}$ where we have $J_{1}=\mathbb{E}[({z}+{\theta_A} + \alpha{\theta_B} -{u})^2]$ and $J_{2}=\mathbb{E}[({z}+\alpha{\theta_A} + {\theta_B} -{u})^2]$. In this example, $\alpha$ represents the alignment of interests in $\theta_A$, $\theta_B$ for the senders. As in the previous examples, the receiver is only interested in $z$ and thus, it has the objective function  $J_{r} = \mathbb{E}[({z}-{u})^2]$. We see in Fig.~\ref{fig:sim3}(a) that when $\alpha = -1$, i.e., when the senders' interests are exactly mismatched, the receiver can use this to its advantage and recover ${z}$ perfectly and thus, $J_r = 0$ when $\alpha=-1$. As $\alpha$ increases, the correlation between the senders' objectives increases. As a result, the receiver gets less information about the parameter ${z}$, and its expected cost $J_r$ increases with $\alpha$. When there is an exact mismatch between the senders' interests, i.e., when $\alpha=-1$, we see in Fig.~\ref{fig:sim3}(b) that the senders' expected costs are high, as the receiver can recover ${z}$ perfectly. As $\alpha$ gets closer to 0, the senders' objective functions become less dependent, and as a result, senders can achieve their minimum expected costs. As the senders' objective functions are positively correlated, i.e., when $\alpha>0$, the expected cost is increasing for both senders.

\subsubsection{The effect of increasing the number of senders} 
\begin{figure}[t]
	\begin{center}
		\subfigure[]{%
			\includegraphics[scale=0.4]{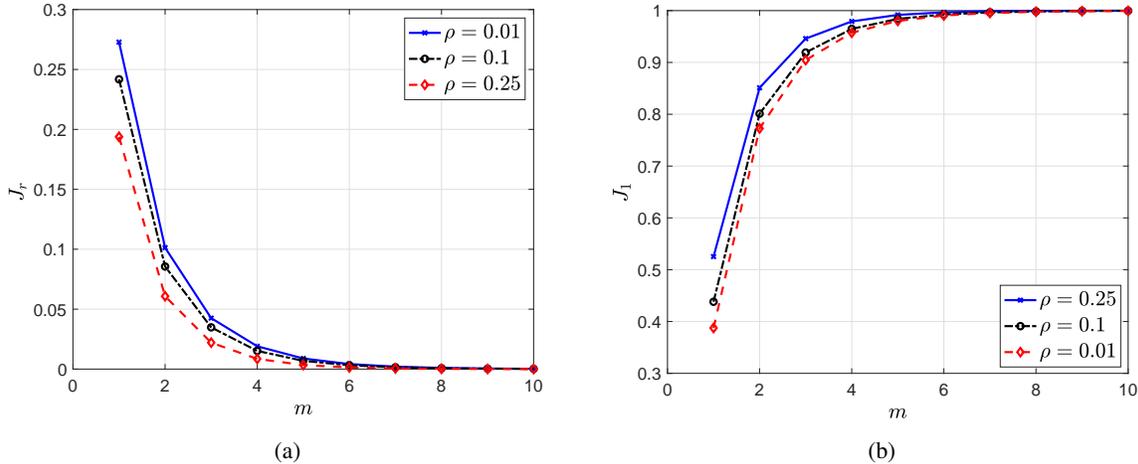}}
			\subfigure[]{%
			\includegraphics[scale=0.4]{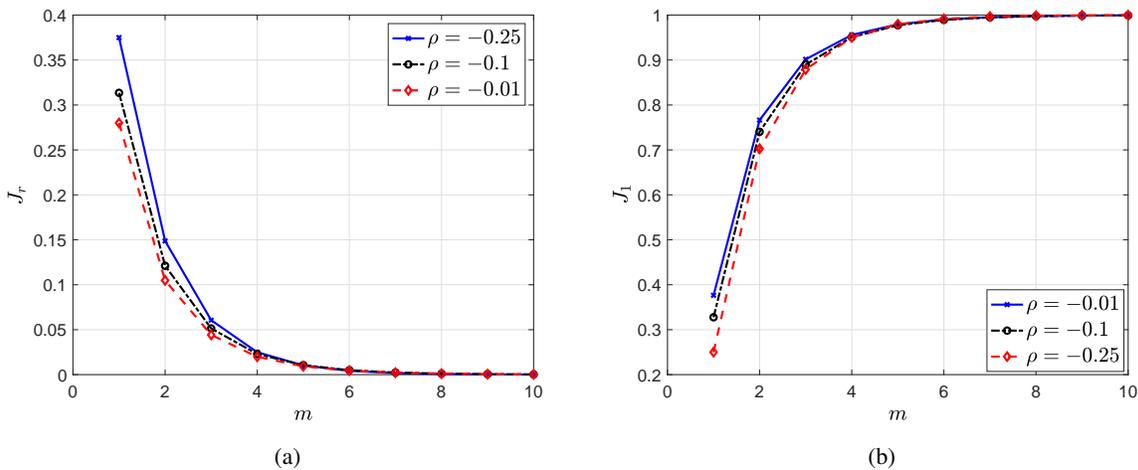}}
	\end{center}\vspace{-0.5cm}
	\caption{The expected costs of the (a) receiver, (b) sender $1$ with respect to the number of senders when $\rho= \{0.01, 0.1, 0.25\}$.}
	\label{fig:sim2}
	\vspace{-0.3cm}
\end{figure}

\begin{figure}[t]
	\begin{center}
		\subfigure[]{%
			\includegraphics[scale=0.4]{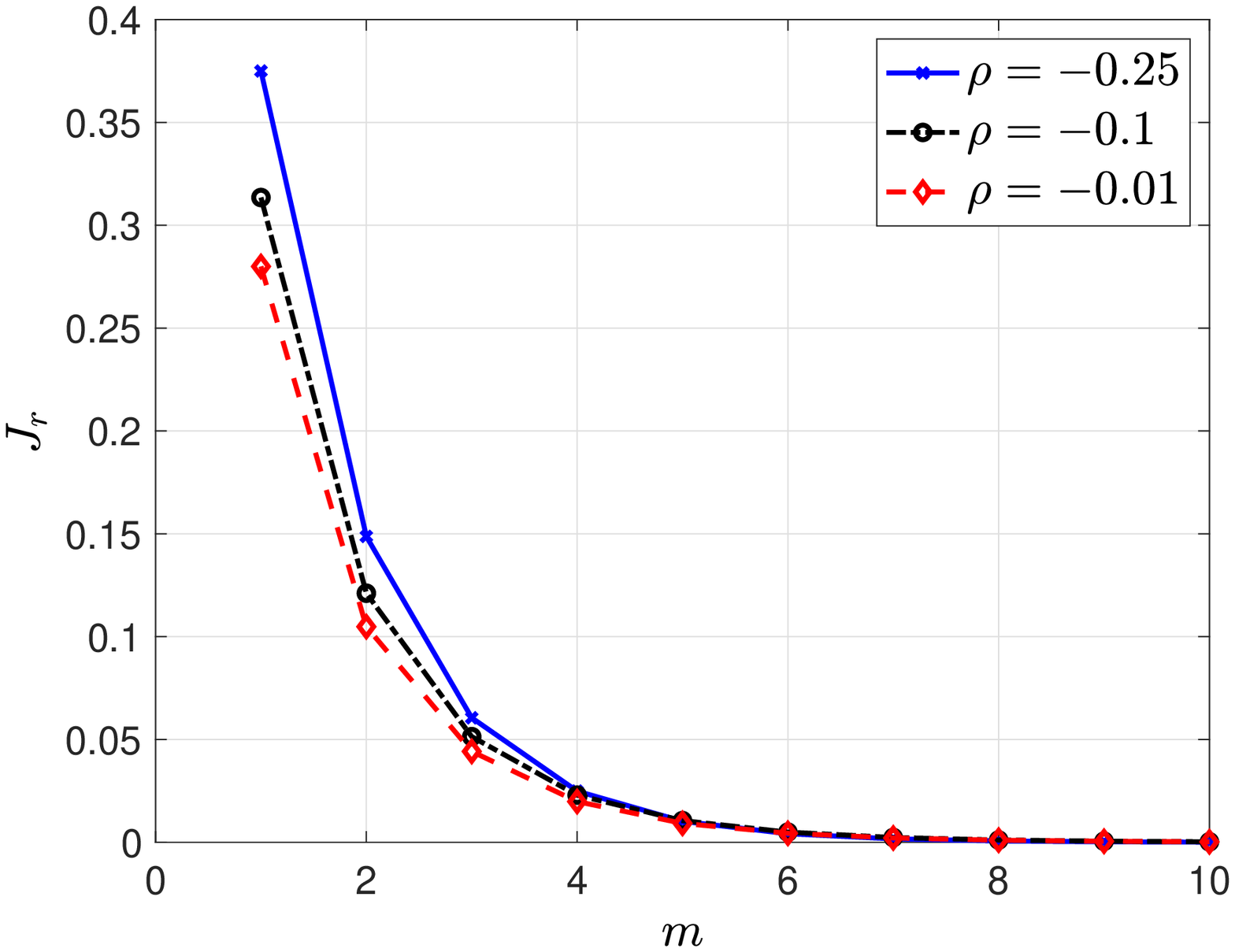}}
			\subfigure[]{%
			\includegraphics[scale=0.4]{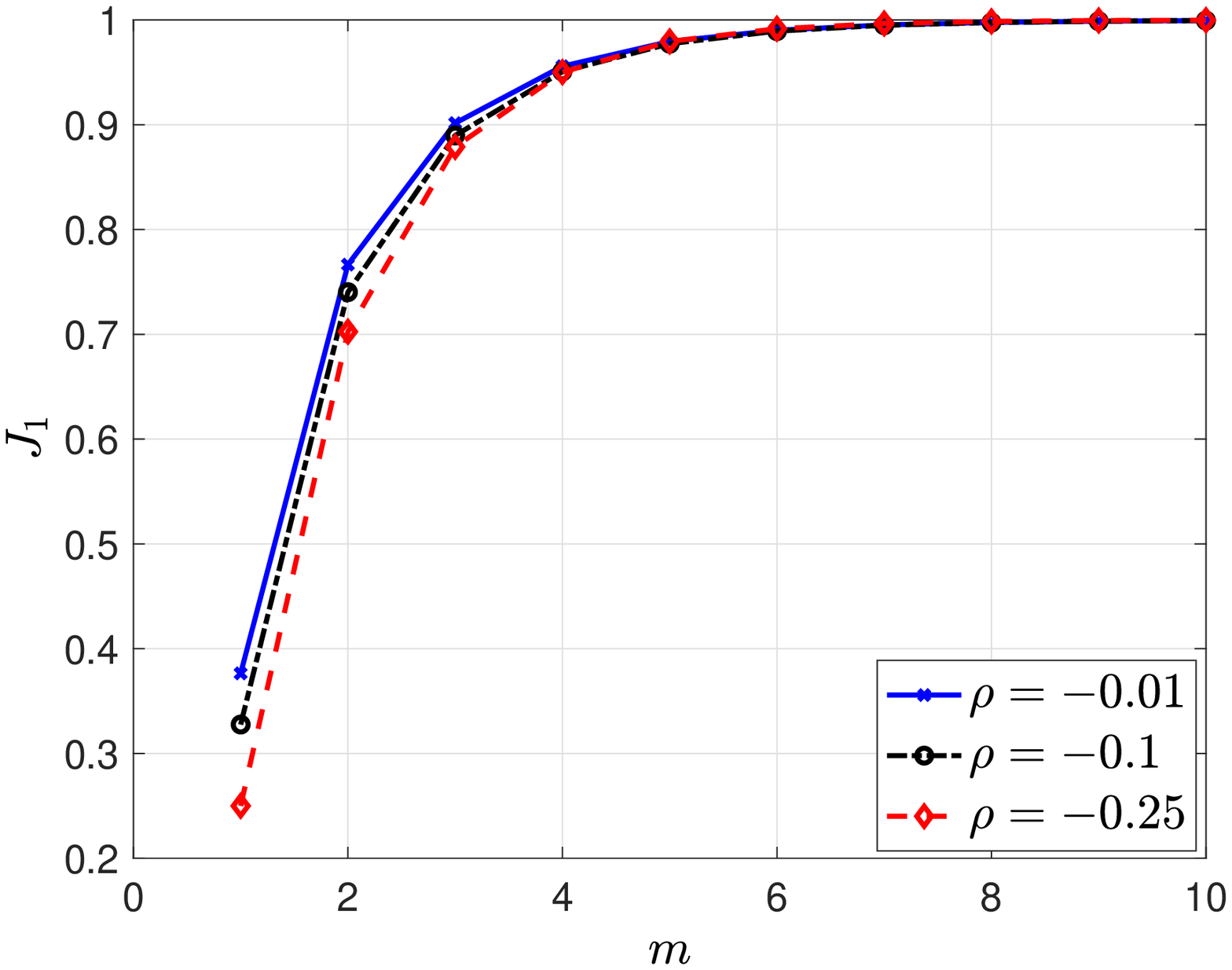}}
	\end{center}
  \vspace{-0.5cm}
	\caption{The expected costs of the (a) receiver, (b) sender $1$ with respect to the number of senders when $\rho= \{-0.01, -0.1, -0.25\}$.}
	\label{fig:sim4}
 \vspace{-0.5cm}
\end{figure}

In the fourth numerical example, we see the effect of increasing the number of senders $m$ on the expected costs of the receiver and sender $1$. Let $\boldsymbol{x}=\begin{bmatrix}{z}, & {\theta_1},\cdots, &{\theta_m}\end{bmatrix}^\top$. The sender $i$ has the objective $J_{i}=\mathbb{E}[({z}+{\theta_i}-{u})^2]$ for $i=1,\cdots, m$ and the receiver is only interested in ${z}$, and thus has the objective function  $J_{r} = \mathbb{E}[({z}-{u})^2]$. With these objective functions, we have the following matrices 
\begin{align*}
    Q_1&=\begin{bmatrix}1 &1 &0&\cdots&0\end{bmatrix},\quad R_1=-1,\\ Q_2&=\begin{bmatrix}1 &0 &1&\cdots&0\end{bmatrix},\quad R_2=-1,\\
    \vdots &\qquad \qquad\quad\vdots \qquad\qquad \quad \hspace{1mm} R_i=-1, \\
    Q_m&=\begin{bmatrix}1 &0 &0&\cdots&1\end{bmatrix},\quad R_m=-1,\\Q_r&=\begin{bmatrix}1 &0&0&\cdots &0\end{bmatrix},\quad R_r=-1.
\end{align*}
Then, we can find the corresponding $V_i$ matrices by using (\ref{eqn:Vi}). In this example, we consider the covariance matrix to be
\begin{align}\label{eqn:example_2}
\Sigma_x= \begin{bmatrix} 
    1 & \rho & \rho &\dots &\rho\\
    \rho & 1 & 0 &\dots &0\\
    \vdots & \vdots& \vdots&\ddots &\vdots \\
    \rho & 0 & 0 &\dots &1\\
    \end{bmatrix},
\qquad
\end{align}
where each sender's private information is independent from the other senders' private information but they are correlated with the receiver's target information ${z}$ with correlation coefficient $\rho$. 
For this example, we increase the number of senders $m$ from 1 to 10. In Fig.~\ref{fig:sim2}, we plot the receiver's and the first sender's expected costs as functions of $m$ when $\rho= \{0.01,0.1,0.25\}$. For all $\rho$ values, as we increase the number of senders $m$, more information is revealed by the senders and as a result the receiver is able to fully recover the desired information $\boldsymbol{z}$ and the receiver's expected cost $J_r$ reduces down to 0 as seen in Fig.~\ref{fig:sim2}. We see that as $\rho$ gets higher, the senders reveal more information to the receiver, and thus $J_r$ gets lower. Similarly, as we increase the number of senders, the receiver gets more information from the other senders about its desired information ${z}$, it gets less manipulated by each sender, and as a result $J_{1}$ increases. As the receiver fully recovers ${z}$, the expected cost of sender $1$ approaches to $J_{1}=\mathbb{E}[({z}+{\theta_1}-{z})^2] = \mathbb{E}[\theta_1^2] =1$ for all $\rho$ values. We see from Fig.~\ref{fig:sim2}(b) that the receiver is able to decrease its cost efficiently when the correlation coefficient between  ${z}$ and its private information ${\theta_i}$, i.e., $\rho$, decreases. In Fig.~\ref{fig:sim4}, we repeat the same example for $\rho= \{-0.01, -0.1, -0.25\}$, i.e., when ${z}$ and ${\theta_i}$ are negatively correlated. We see in Fig.~\ref{fig:sim4}(a) that the receiver can fully recover its desired information ${z}$ and also the expected cost of sender $1$, $J_{1}$, decreases as the correlation between ${z}$ and ${\theta_i}$, i.e., $\rho$, decreases. 

 \begin{table}[t]
\small
\centering
	\begin{center}
		\begin{tabular}{ | c | c | c | c |c |c |}
			\hline
			 $N \in \bar{\mathcal{N}}$ in Algorithm~\ref{alg: npostcalc} & $J_{1}$  & $J_{2}$ & $J_{3}$& $\sum_{i=1}^{m}J_{i}$ & $J_{r}$\\ \hline
			$N= \{3	,2,1\}$
			 & \textbf{0.7590} & $0.8527$ & $0.9191$ & $2.5308$ & $0.0348$  \\ \hline
			$N= \{3	,1,2\}$
			 & $0.8527$ & \textbf{0.7590} & $0.9191$ & $2.5308$ & $0.0348$ \\ \hline 
			 $N= \{2	,3,1\}$
			 & \textbf{0.7590} & $0.9191$ & $0.8527$   & $2.5308$ & $0.0348$\\ \hline 
			 $N= \{2	,1,3\}$
			 & $0.8527$ & $ 0.9191$ & \textbf{0.7590} & $2.5308 $
			 & $0.0348$\\ \hline 
			 $N= \{1	,3,2\}$
			 & $0.9191$ & \textbf{0.7590} & $0.8527$ & $2.5308$ 
			 & $0.0348$\\ \hline 
			 $N= \{1	,2,3\}$
			 & $0.9191$ & $0.8527$ & \textbf{0.7590} & $2.5308$& $0.0348$ \\
			 \hline 
			 Full revelation
			 & $1$ & $1$ & $1$ & $3$& $0$ \\\hline
		\end{tabular}
	\end{center}
	\caption{The expected costs of the senders and receiver when we apply Algorithm~\ref{alg: npostcalc} with different ordering of senders. The minimum value in each column is indicated with bold text. }
	\label{table:results}
	\vspace{-0.5cm}
\end{table}

\subsubsection{Nash equilibrium posteriors obtained from Algorithm~\ref{alg: npostcalc}}
In the fifth numerical example, we compare the NE posteriors obtained by solving the sequential optimization problem with varying order of $V_i's$ for optimization considered in Algorithm~\ref{alg: npostcalc}. First, we consider $\Sigma_x$ in (\ref{eqn:example_2}) with $\rho = 0.1$ for $m=3$. The senders' and the receiver's objective functions are given by $J_{i} =\mathbb{E}[({z}+{\theta_i}-u)^2]$ for $i =1,2, 3$ and $J_{r} =\mathbb{E}[({z}-u)^2]$, respectively. We run Algorithm~\ref{alg: npostcalc} for $6$ different possible orderings and find the NE posteriors. We summarize the results by considering expected costs of individual senders, total sum of their costs and the receiver's expected cost in Table~\ref{table:results}. We see that since $\theta_i$ variables are independent and identically distributed (i.i.d.), changing the ordering of $V_i's$ does not change the senders' total and the receiver's expected costs, i.e., $\sum_{i=1}^m J_{i} = 2.5308$ and $J_r = 0.0348$ for all $N\in\bar{\mathcal{N}}$, respectively. On the other hand, the order in Algorithm~\ref{alg: npostcalc} affects the sender's individual cost. We see in the first row of Table~\ref{table:results} that when the sequential optimization problem is solved in the order of $N= \{3,2,1\}$, then sender $3$ has the highest cost ($J_{3} = 0.9191$) whereas sender $1$ has the lowest cost ($J_{1} = 0.7590$). As we solve for SDP involving $V_1$ at the last stage, 
it can achieve the lowest cost compared to other senders. We re-emphasize that this sequence of ordering $V_i's$ is only to solve the optimization problem in Algorithm~\ref{alg: npostcalc} but any of the resulting posteriors above are in NE and are partially informative.
\begin{table}[t]
\small
\centering
	\begin{center}
		\begin{tabular}{ | c | c | c | c |c |c |}
			\hline
			 $N \in \bar{\mathcal{N}}$ in Algorithm~\ref{alg: npostcalc} & $J_{1}$  & $J_{2}$ & $J_{3}$& $\sum_{i=1}^{m}J_{i}$ & $J_{r}$\\ \hline
			$N= \{3	,2,1\}$
			 & $0.9255$ & $0.9746$ & $0.9923$ & $ 2.8925$ & \textbf{0.0066}  \\ \hline
			$N= \{3	,1,2\}$
			 & $0.9663$ & $0.9424$ & $0.9898$ & $2.8985$ & $ 0.0088$ \\ \hline 
			 $N= \{2	,3,1\}$
			 & \textbf{0.9196} & $0.9913$ & $0.9798$   & \textbf{2.8907} & $0.0078$\\ \hline 
			 $N= \{2	,1,3\}$
			 & $0.9660$ & $0.9866$ & \textbf{0.9513} & $2.9040 $
			 & $0.0120$\\ \hline 
			 $N= \{1	,3,2\}$
			 & $0.9898$ & \textbf{0.9407} & $0.9799$ & $2.9105$ 
			 & $0.0096$\\ \hline 
			 $N= \{1	,2,3\}$
			 & $0.9880$ & $0.9737$ & $0.9531$ & $2.9147$& $0.0113$\\ \hline 
			 Full revelation
			 & $1$ & $1$ & $1$ & $3$& $0$  \\ \hline
		\end{tabular}
	\end{center}
	\caption{The expected costs of the senders and the receiver when we apply Algorithm~\ref{alg: npostcalc} with different orderings for senders. The minimum value of each column is indicated with bold text. }
	\label{table:results_2}
	\vspace{-0.5cm}
\end{table}

Next, we consider the following prior $\Sigma_x$:
\begin{align}\label{eqn:example_3}
\Sigma_x= \begin{bmatrix} 
    100 & 2.5 & 5 &7.5\\
    2.5 & 1 & 0 & 0\\
    5 & 0& 1&0  \\
    7.5 & 0 & 0 &1\\
    \end{bmatrix},
\qquad
\end{align}
where the correlation coefficients are $\rho_{z \theta_1} =0.25 $, $\rho_{z \theta_2} =0.5 $, and $\rho_{z \theta_3} =0.75 $. In other words, the correlation between $z$ and $\theta_i$ is increasing with $i$. Since $\theta_i$s are not i.i.d. anymore, the order in Algorithm~\ref{alg: npostcalc} changes the total expected costs of the senders and the receiver. We see in Table~\ref{table:results_2} that the receiver's expected cost is minimized (maximized) when $V_3$ is solved first (last) in Algorithm~\ref{alg: npostcalc}, i.e., when $N = \{3,2,1\}$ ($N = \{2,1,3\}$), respectively. Additionally, for the senders, we observe that each one's individual cost is minimized when its order is last in Algorithm~\ref{alg: npostcalc}. Thus, there is no equilibrium among those calculated by our algorithm that is preferred by all senders. However, their total expected cost, i.e., $\sum_{\ell= 1}^{m}J_{\ell}$, is minimized when $N= \{2,3,1\}$ where the NE posterior also minimizes sender $1$'s individual cost whereas it maximizes sender $2$'s individual cost compared to other NE posteriors.

In both examples, we see in Tables \ref{table:results} and \ref{table:results_2} that NE policies that we obtain in this paper are nontrivial in the sense that compared to the full revelation type, which is also an equilibrium, we attain equilibria that are preferred by all senders.

\subsubsection{Cooperative vs non-cooperative games among senders}
\begin{figure}[t]
	\begin{center}
		\subfigure[]{	\includegraphics[scale=0.4]{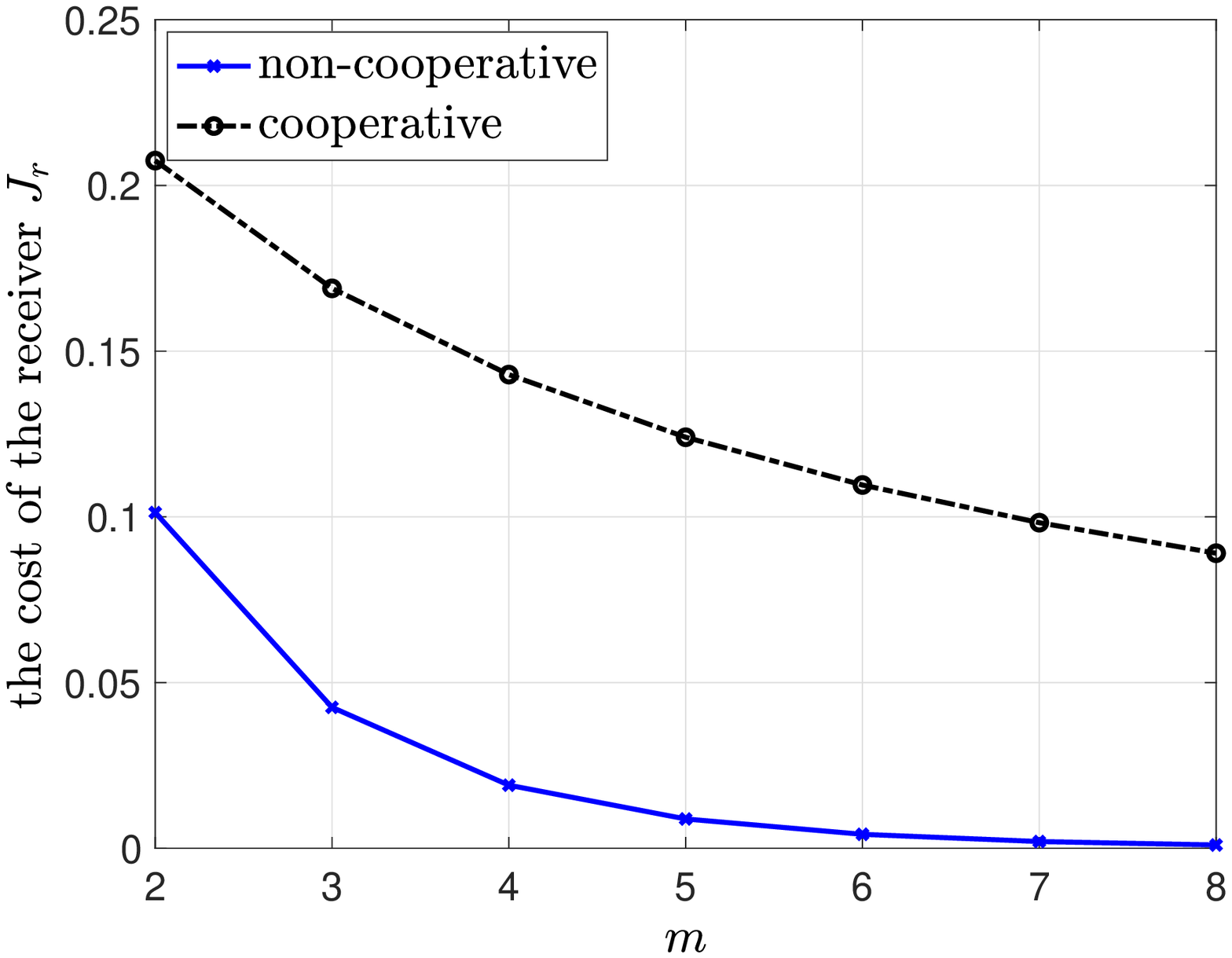}}
			\subfigure[]{%
			\includegraphics[scale=0.4]{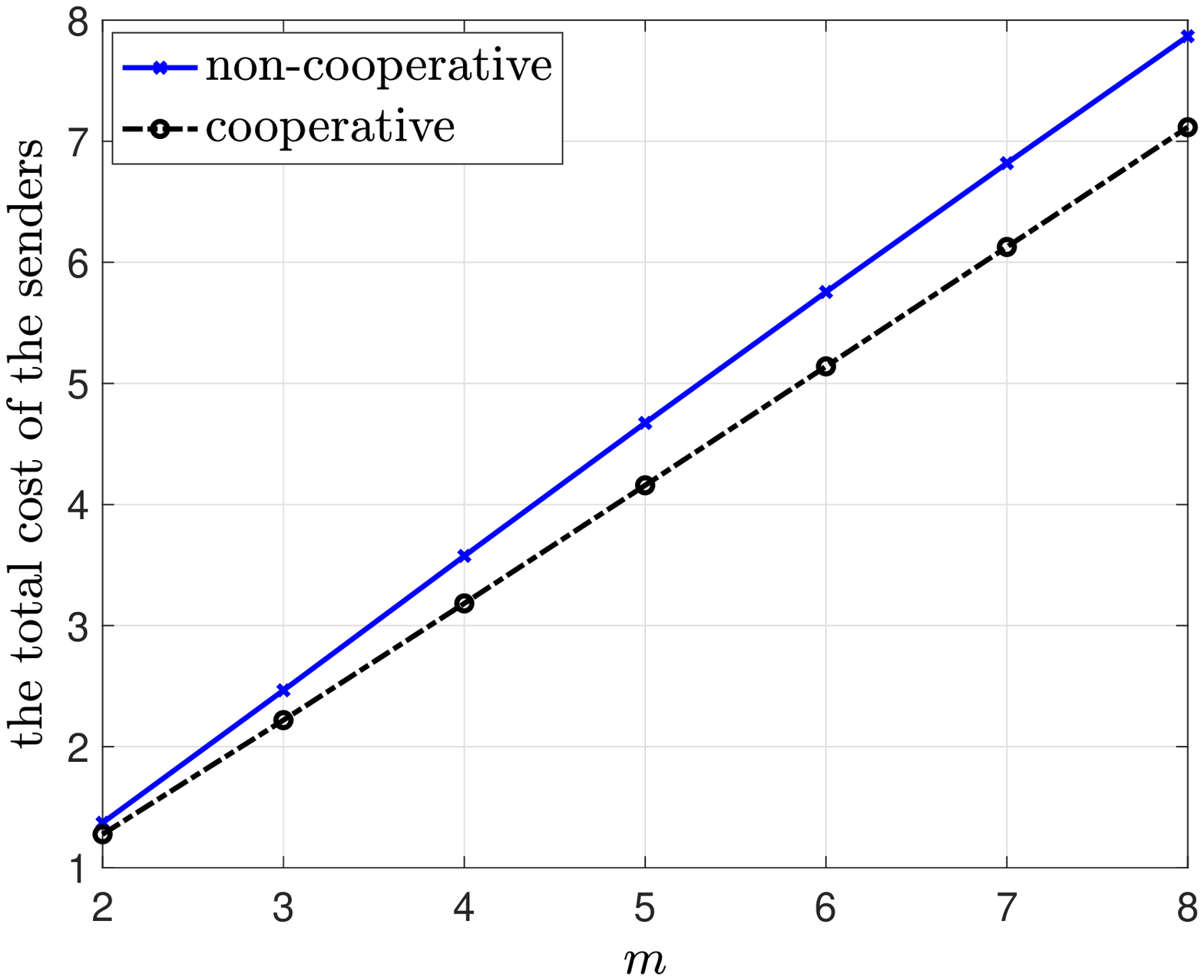}}
	\end{center}\vspace{-0.5cm}
	\caption{(a) The expected costs of the receiver, (b) the total cost of the senders when the senders are cooperative and non-cooperative.}
	\label{fig:sim5}
\end{figure}

In the sixth numerical example, we consider the $\Sigma_x$ in (\ref{eqn:example_2}) for $\rho = 0.01$. Our goal is to compare the expected costs of all the senders and the receiver when the senders are cooperative, i.e., each sender wants to minimize a global objective function which is the summation of individual cost functions of the senders, and when the senders are non-cooperative, and hence playing a Nash game, which is the setting in our work. In the cooperative game, the cost function for each sender is given by $J_{i} = \sum_{\ell=1}^{m} \mathbb{E}[({z}+{\theta_\ell}-u)^2]$. Similar to the derivations in (\ref{eqn:Vi}), the objective function of each sender can be written as  
\begin{align} \label{eqn:num_res_5}
    \min_{\Sigma_x\succeq \Sigma\succeq 0} Tr\left( \left(\sum_{\ell=1}^{m} V_\ell\right) \Sigma\right).
\end{align}
Here, as the senders have the same global objective function, the optimal solution to (\ref{eqn:num_res_5}) will minimize the expected cost of each sender. Thus, it is sufficient to solve the optimization problem in (\ref{eqn:num_res_5}) just for a single sender. Then, we compare the cooperative game cost with the non-cooperative game setting where the senders have $J_{i} =\mathbb{E}[({z}+{\theta_i}-u)^2]$ for $i =1,\dots, m$, as in the fourth numerical example. In order to make a fair comparison, we compute $\sum_{i=1}^m J_{i}$ for the non-cooperative game, and compare it with the expected cost of the single sender objective function $J_{i} = \sum_{\ell=1}^{m} \mathbb{E}[({z}+{\theta_\ell}-u)^2]$ in the cooperative game scenario. We see in Fig.~\ref{fig:sim5}(b) that cooperation among the senders helps reduce their total expected cost compared to the non-cooperative game setting. We note that the total expected cost of the senders increases linearly with $m$. We see in Fig.~\ref{fig:sim5}(a) that the cooperation between the senders is not beneficial for the receiver as it gets less informative signals from the senders. However, the receiver's expected cost decreases as the number of senders, $m$, increases. Thus, even if the senders cooperate, having multiple senders can still be helpful for the receiver.  

\subsubsection{Dynamic information disclosure} In the seventh numerical example, we consider the multi-stage game with $m=2$ senders and $n = 10$ stages where the $3$-dimensional state of the world $\boldsymbol{x_k} = [z_k, \theta_{A_k},\theta_{B_k} ]$ evolves over time as follows:
\begin{align}
    \boldsymbol{x_k} = A\boldsymbol{x_{k-1}}+ \boldsymbol{w_{k-1}}, \quad k\in\{1,\dots,n\},
\end{align}
where $A = I$, and $\boldsymbol{x_0}$ has a zero-mean Gaussian distribution with variance $\Sigma_0$, i.e., $\mathbb{N}(0,\Sigma_0)$, where 
\begin{align*}
    \Sigma_0 =\begin{bmatrix}
    1 & 0.5 &0.5\\
    0.5&1 &0\\
    0.5 &0 &1
    \end{bmatrix},
\end{align*}
and, $w_{k}$ has $\mathbb{N}(0,\Sigma_w)$ with $\Sigma_w = I$. For this example, we consider the case where the objective functions of the senders are changing over each stage $k$, denoted by $J_{i}^k$ and are given by 
\begin{align}\label{eqn:ex_7}
    J_{1}^k =\mathbb{E}\left[ \left({z_k}+\frac{k}{n}{\theta_{A_k}}-u_k\right)^2\right], \quad J_{2}^k =\mathbb{E}\left[ \left({z_k}+\frac{n-k}{n}{\theta_{B_k}}-u_k\right)^2\right], \quad \forall k \in [n].
\end{align}
The receiver's expected cost function at stage $k$ is given by $J_{r}^k =\mathbb{E}[ (z_k-u_k)^2]$. With these expected cost functions, we find $V_k^i$ in (\ref{eq: Vik}) as follows:
\begin{align*}
    V_k^1=\begin{bmatrix}-1 & -\frac{k}{n}&0\\-\frac{k}{n}&0 &0\\0 &0 &0\end{bmatrix},\qquad V_k^2=\begin{bmatrix}-1 &0 &-\frac{n-k}{n}\\0 &0 &0\\-\frac{n-k}{n} &0 &0\end{bmatrix}.
\end{align*}

\begin{figure}[t]
	\begin{center}
		\subfigure[]{	\includegraphics[scale=0.4]{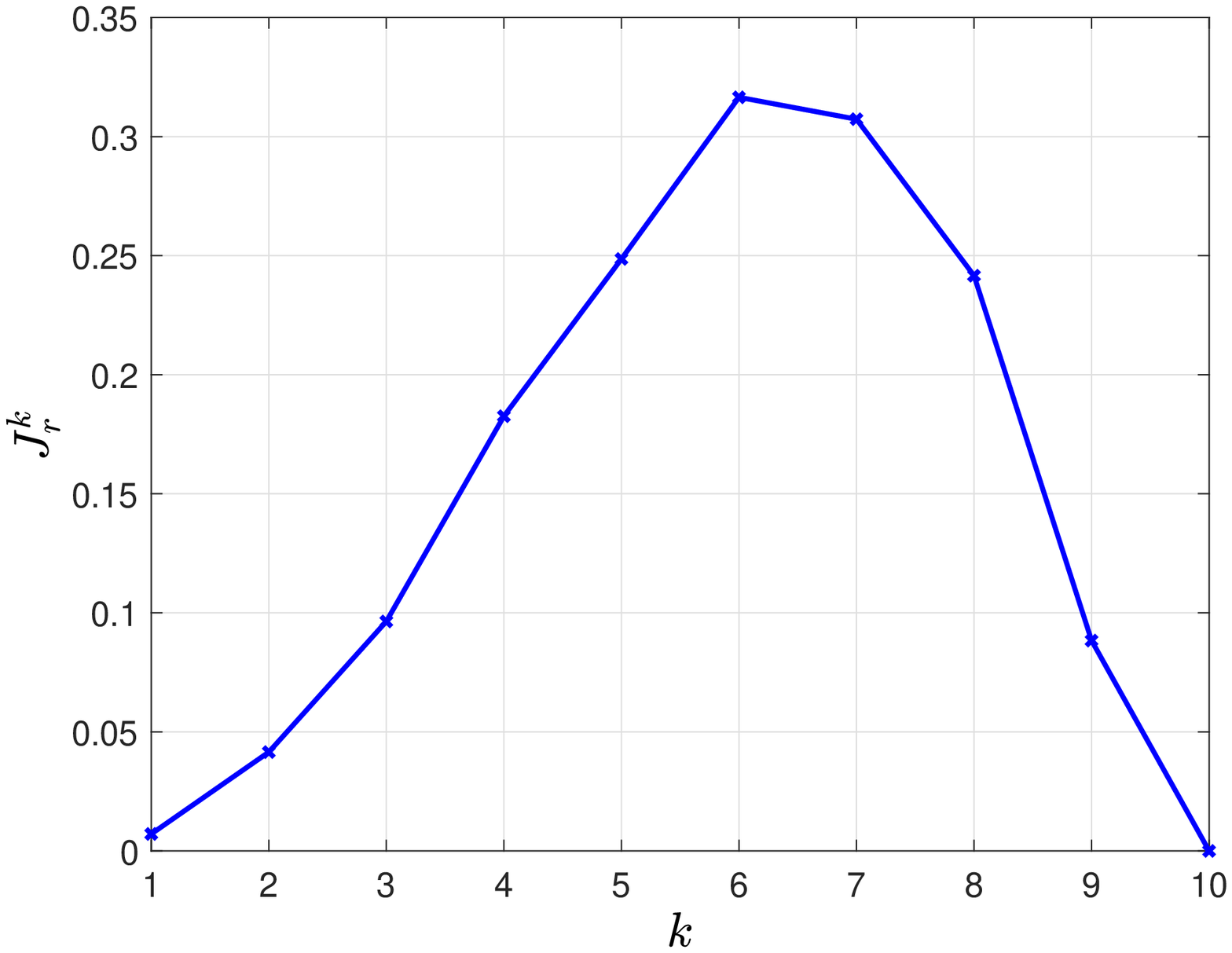}}
			\subfigure[]{%
			\includegraphics[scale=0.4]{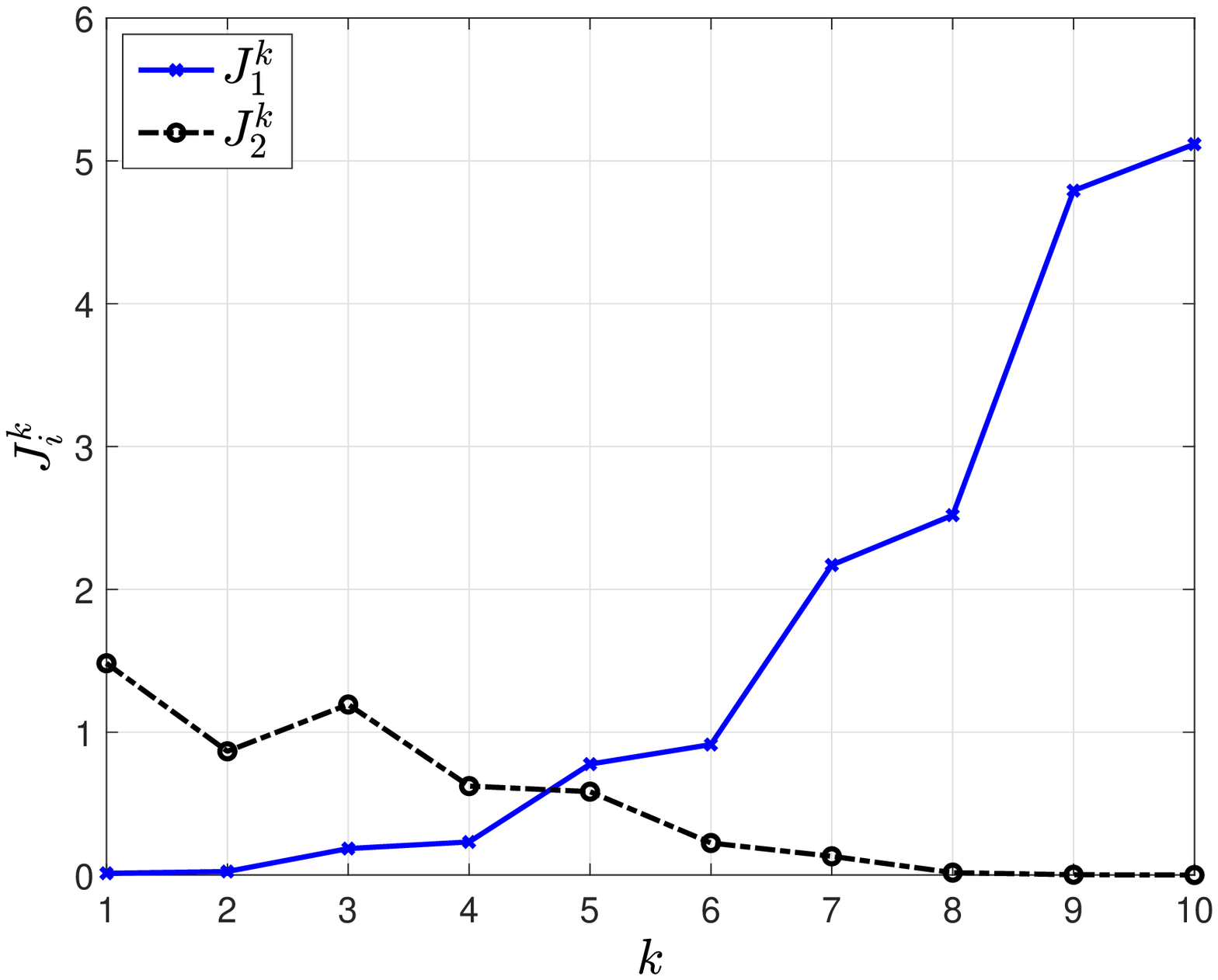}}
	\end{center}\vspace{-0.5cm}
	\caption{Dynamic information disclosure games with time horizon $n=10$ for $m=2$ senders the expected costs of (a) the receiver, (b) the senders.}
	\label{fig:sim7}
\end{figure}

In this example, based on (\ref{eqn:ex_7}), the objective function of sender $1$ is initially aligned with that of the receiver whereas sender $2$ has a more misaligned objective. As the time horizon $k$ gets closer to $n$, we see that the objective of sender $1$ becomes misaligned whereas sender $2$'s objective function becomes aligned with that of the receiver. We plot the senders' and the receiver's cost functions at stage $k$ in Fig.~\ref{fig:sim7}. As seen in Fig.~\ref{fig:sim7}(a), the receiver is able to minimize its cost well when at least one sender has aligned objective functions with the receiver, which happens when $k$ is small and when $k$ is close to $n$. We observe from Fig.~\ref{fig:sim7}(b) that sender $1$'s expected cost is increasing with $k$. As sender $1$'s objective function diverges from the receiver's objective, the receiver gets more reliable information from sender $2$. Here, sender $2$ is able to minimize its cost by revealing more information to the receiver as its objective function becomes more aligned as $k$ grows.  

\subsubsection{Multi-Sender Multi-Receiver}
\begin{figure}[t]
\centerline{\includegraphics[width=0.65\columnwidth]{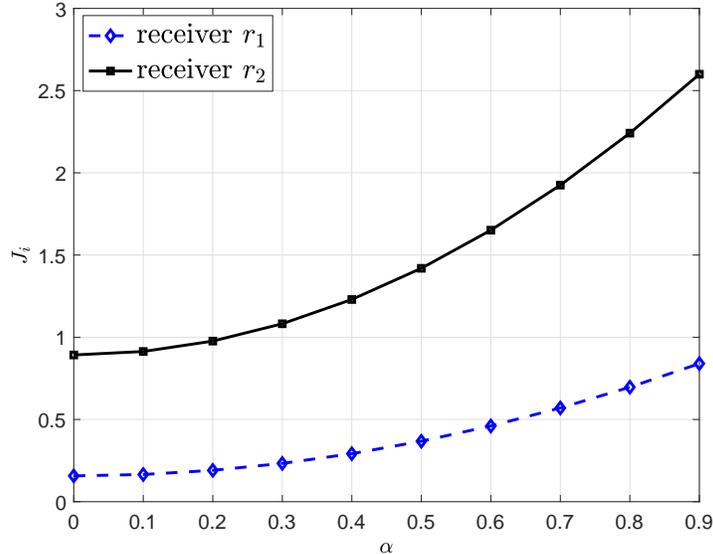}}
	\caption{The expected costs at the receivers for a system with a single sender and 2 receivers. }
	\vspace{-0.35cm}
	\label{Fig:num_res_8}
	\vspace{-0.3cm}
\end{figure}

In the eighth numerical example, we first consider a single sender and 2 receivers, and a state of the world $\boldsymbol{x} = [z, \theta_A, \theta_B ]^\top$,  $\boldsymbol{x} = \begin{bmatrix}
    z & \theta_A &\theta_B 
\end{bmatrix}^\top$which has a zero-mean Gaussian prior with covariance matrix $\Sigma_x$ given by 
\begin{align}
    \Sigma_x=\begin{bmatrix}1 &0.5 &0.5\\0.5 &1 &0.25\\ 0.5 &0.25 &1\end{bmatrix}.
\end{align}
The sender's and the receivers' cost functions are given by 
\begin{align}
 J_{s_1} =\mathbb{E}\left[\left(z+ \theta_B-\frac{u_1+u_2}{2}\right)^2\right], ~
 J_{r_1} \!=\!\mathbb{E}\left[\left(z- u_1-\alpha u_2\right)^2\right],~
 J_{r_2} \!=\!\mathbb{E}\left[\left(z+\theta_A\!-\!\alpha u_1\!-\!u_2\right)^2\right].
\end{align}
By using Proposition~\ref{prop: receievrbestresp}, we found the optimum receivers' best response as
    \begin{align*}
        \boldsymbol{u}^*=\begin{bmatrix}
            -K_1\boldsymbol{\hat{x}} \\ -K_2\boldsymbol{\hat{x}}
        \end{bmatrix},
    \end{align*}
where $K_1 = [-1, \frac{\alpha}{1-\alpha}, 0]$ and $K_2 = [-1, -\frac{1}{1-\alpha}, 0]$. Then, we write the objective function of the sender in (\ref{eqn_senders_obj}) as
\begin{align}
    \min_{\eta\in\Omega} Tr(VS),
\end{align}
where $S=\mathbb{E}[\hat{\boldsymbol{x}}\hat{\boldsymbol{x}}^\top]$ and 
\begin{align}
   V=\begin{bmatrix} 
   -1 &0 &-1\\0 &\frac{1}{4} &-\frac{1}{2}\\ -1 &-\frac{1}{2} &0 
   \end{bmatrix}.
\end{align}
In this example, we vary $\alpha =\{0,0.1,\dots, 0.9\}$, and find the optimum posterior induced by the sender $S$, and then calculate the corresponding expected costs of players. We note that the sender's objective function does not depend on the variable $\alpha$. By using Algorithm~\ref{alg: npostcalc}, we find $S$ as 
\begin{align}
   S^*=\begin{bmatrix} 
   0.843 &0.49 & 0.73\\0.49 &0.29 &0.42\\ 0.73 &0.42 &0.63 
   \end{bmatrix},
\end{align}
the corresponding expected cost at the sender is given by  $Tr(VS^*) = 0.3464$. 
From Theorem \ref{th: Nasheq}, we can construct equilibrium policy to be $\eta_1(\boldsymbol{x})=L\boldsymbol{x}$, where 
\begin{align*}
L=\begin{bmatrix}
   -0.6452  & -0.0993&   -0.4466 \\
    0&   0   & 0\\
   0  & 0 &  0
\end{bmatrix} .  
\end{align*}
The receivers' expected costs are given in Fig.~\ref{Fig:num_res_8}. As we increase $\alpha$, the receivers' objective functions start to couple with each other. As a result, their expected costs increases with $\alpha$ as shown in Fig.~\ref{Fig:num_res_8}.

\section{Discussion on Sequential Persuasion}\label{Sect:Discussion}
In this section, we highlight an interesting property of the equilibrium obtained in the static multi-sender communication game. First, note that there can be multiple equilibrium definitions based on the times of commitment by the senders. Two of them considered in this paper include the NE (Theorem \ref{th: Nasheq}) due to simultaneous commitment among senders as stated in (\ref{eq: Nash}) and a sequential commitment with second sender entering the game between an existing sender and receiver as derived in Theorem \ref{th: seqcommit}. Although the posteriors obtained in both cases through Algorithm \ref{alg: postcalc} are the same, the way that they are achieved is different due to the commitment order. Further, as stated in the text, NE in general for a multi-sender communication problem is not unique, and thus can result in multiple equilibrium posteriors. On the other hand, in the case of sequential commitment when the second sender enters an existing communication game as described in Theorem \ref{th: seqcommit}, the second sender faces an optimization problem which has essentially a unique solution, and thus resulting in a unique equilibrium posterior for this game. The fact that these two equilibrium posteriors can be obtained from the same algorithm boils down to the partially informative Nash equilibrium refinement as defined earlier.

Alternatively, one can consider a strictly sequential game between the senders where sender $1$ moves first knowing that sender $2$ would join the game in the next stage before the receiver takes an action. Since both senders have complete access to the state, any policy profile $(\eta_1,\eta_2)$ of the senders can have an equivalent policy profile $(\eta_1',\eta_2')$ such that $\eta_2'$ does not add any additional information about the state \cite{li2021sequential}. This can be done by adding incentive compatibility constraints for sender $2$ into sender $1$'s optimization problem. Thus without loss of generality, sender $1$'s optimization problem after following the simplification in (\ref{eq:senobj_v2}) and by utilizing Lemma~\ref{lem: bestresplbound} becomes
\begin{align*}
    \min_{\Sigma_x\succeq S\succeq 0}\quad& Tr(V_1S)\\
     \text{s.t.}\quad& (\Sigma_x-S)^\frac{1}{2}V_2(\Sigma_x-S)^\frac{1}{2}\succeq O.
\end{align*}
Due to the similarity between this objective and the steps in Algorithm~\ref{alg: postcalc}, similar techniques can be used to obtain an equilibrium posterior. We leave this problem as a future research direction.

To summarize one key difference between sequential games and NE is that in sequential commitment the order of the senders matters and the equilibrium reached is essentially unique. Further, it can be seen that NE posteriors are at least as informative as sequential equilibrium posteriors. This is in line with the observation made in \cite{li2021sequential} for discrete state and action spaces.

\section{Conclusion and Future Direction}\label{Sect:Conclusion}
In this work, we have considered a strategic communication game problem between multiple senders and a receiver with misaligned objective functions. First, by focusing on the 2-sender case, we have posed the senders' optimization problems in terms of the receiver's posterior covariance. Then, by relaxing the constraint, we have proposed a sequential optimization technique to find the stable posteriors which can be achieved by using linear noiseless policies, and as a result they are in fact the equilibrium policies for the senders. Next, we have extended our solution to multiple senders ($m>2$) and characterized partially informative equilibrium posteriors, i.e., there exists no other more informative equilibrium which is beneficial for all the senders to deviate. Further, we have added dynamics to the problem by considering a Markovian evolution process of the underlying state with finite-horizon quadratic costs for the players. In this case we have shown that the equilibrium obtained by considering single-stage games is in fact stable and nontrivial. Finally, we extended the analysis to account for multiple receivers. 
Through extensive simulations, we have shown that having multiple senders is beneficial to the receiver as its estimation error cannot get worse by increasing the number of senders. It is also beneficial to the receiver to have senders with maximum misalignment between their objectives.

One immediate future direction of research is to explore the multi-stage case further, and characterize all possible 
equilibrium strategies. Further, we note that there might be other procedures (inspired by the approach in \cite{tamura2018bayesian}) that can achieve equilibrium posteriors not obtained by our algorithms. We plan to investigate such procedures for achieving admissible equilibria\cite{bacsar1998dynamic} from which deviation to any other equilibrium is not profitable for any of the senders. Another promising direction is to consider the senders with private information where the private information is statistically known but its realization is not observable by the other players. It would be interesting to explore the effect of misalignment between sender objectives and verify if it still would be beneficial for the receiver to have multiple senders in all scenarios. Finally, it is crucial to understand the importance of equilibrium in linear policies. More precisely, it would be interesting to investigate if all the stable posteriors are achievable (or equivalently, identifying a necessary and sufficient condition for a posterior covariance to be an equilibrium posterior) using senders policies in some other policy class. In contrast with persuasion in the presence of a single sender, it is not obvious to achieve all stable posteriors using linear plus noise policies as in \cite{9222530}. 
\bibliographystyle{unsrt}
\bibliography{IEEEabrv,refs}
\end{document}